\documentclass[nofootinbib,superscriptaddress]{revtex4}

\usepackage{amsfonts,amssymb,amsthm,bbm,amsmath}
\usepackage{graphicx}
\usepackage{subfigure}
\usepackage{hyperref}
\usepackage{color}

\newcommand{\C}{{\mathbb C}}

\newcommand{\R}{{\mathbb R}}

\newcommand{\cG}{{\mathcal G}}

\newcommand{\cD}{{\mathcal D}}
\newcommand{\cC}{{\mathcal C}}

\newcommand{\cU}{{\mathcal U}}

\newcommand{\SU}{\mathrm{SU}}
\newcommand{\SB}{\mathrm{SB}}

\newcommand{\SL}{\mathrm{SL}}
\newcommand{\SO}{\mathrm{SO}}
\newcommand{\U}{\mathrm{U}}
\newcommand{\ISO}{\mathrm{ISO}}
\newcommand{\ISU}{\mathrm{ISO}}

\renewcommand{\sb}{{\mathfrak sb}}
\newcommand{\so}{{\mathfrak so}}
\newcommand{\su}{{\mathfrak su}}

\renewcommand{\sl}{{\mathfrak sl}}
\newcommand{\isu}{{\mathfrak{iso}}}

\newcommand{\be}{\begin{equation}}
\newcommand{\ee}{\end{equation}}
\newcommand{\beq}{\begin{eqnarray}}
\newcommand{\eeq}{\end{eqnarray}}
\newcommand{\bes}{\begin{eqnarray}}
\newcommand{\ees}{\end{eqnarray}}
\newcommand{\bea}{\begin{eqnarray}}
\newcommand{\eea}{\end{eqnarray}}
\newcommand{\nn}{\nonumber}

\newcommand{\mat} [2] {\left ( \begin{array}{#1}#2\end{array} \right ) }
\newcommand{\tabl} [2] {\begin{array} {#1} #2 \end{array}}

\newcommand{\la}{\langle}
\newcommand{\ra}{\rangle}

\newcommand{\f}{\frac}

\def\nn{\nonumber}

\def\arr{\rightarrow}
\def\dr{\rightarrow}
\def\ov{\overline}
\def\ot{\otimes}

\newcommand{\id}{\mathbb{I}}
\def\ka{\kappa}

\def\eps{\epsilon}

\DeclareMathOperator{\tr}{Tr}

\def\vsigma{\vec{\sigma}}
\def\veps{{\varepsilon}}
\def\vveps{{\vec \varepsilon}}

\def\vb{\vec{b}}
\def\hb{\hat{b}}

\def\tell{\tilde{\ell}}
\def\tu{\tilde{u}}
\def\tL{\tilde{L}}

\def\tX{\tilde{X}}

\def\tlambda{\tilde{\lambda}}
\def\tz{\tilde{z}}
\def\bz{\bar{z}}
\def\balpha{\bar{\alpha}}
\def\bbeta{\bar{\beta}}

\def\poi#1{\{ #1 \}}
\def\com#1{[ #1 ]}
\def\one{{\bf 1}}
\def\UQ{{\cU_{q}(\su(2))}}
\def\mone{^{{-1}}}

\def\bX{{\bf X}}
\def\bR{{\bf R}}
\def\demi{\f{1}{2}}
\def\vT{{\vec T}}
\def\tT{{\tilde T}}

\def\btT{{\bf{\tilde T}}}
\def\bT{{\bf{T}}}
 \def\bee{{\bf{e}}}
\def\btz{{{\ov{\tilde z}}}}
\def\tz{{\tilde{{z}}}}
\def\tlambda{{\tilde{{\lambda}}}}
\def\tu{{\tilde u}}
\def\talpha{{\tilde \alpha}}
\def\tbeta{{\tilde \beta}}
\def\bTop{{\bf{T}}^{op}}
\def\Top{{{T}}^{op}}
\def\Lop{{{L}}^{op}}
\def\tLop{{\tilde{L}}^{op}}
\def\tbTop{{\tilde{\bf{T}}}^{op}}
\def\tTop{{\tilde {T}}^{op}}
\def\tLop{{\tilde {L}}^{op}}

\def\ovz{{\overline{z}}}

\def\fB{{\mathfrak{B}}}
\def\fg{{\mathfrak{g}}}
\def\fd{{\mathfrak{d}}}

\def\hfb{\hat{\mathfrak{b}}}

\newtheorem{theorem}{Theorem}[section]

\newtheorem{prop}[theorem]{Proposition}

\begin{document}

\title{Deformed phase space for 3d loop gravity and hyperbolic discrete geometries}

\author{{\bf Valentin Bonzom}}\email{bonzom@lipn.univ-paris13.fr}
\affiliation{LIPN, UMR CNRS 7030, Institut Galil\'ee, Universit\'e Paris 13,
99, avenue Jean-Baptiste Cl\'ement, 93430 Villetaneuse, France, EU}

\author{{\bf Ma\"it\'e Dupuis}}\email{m2dupuis@uwaterloo.ca}
\affiliation{Department of Applied Mathematics, University of Waterloo, Waterloo, Ontario, Canada}
%\affiliation{University Erlangen-Nuremberg, Institute for Theoretical Physics III, Erlangen, Germany}

\author{{\bf Florian Girelli}}\email{fgirelli@uwaterloo.ca}
\affiliation{Department of Applied Mathematics, University of Waterloo, Waterloo, Ontario, Canada}
%\affiliation{University Erlangen-Nuremberg, Institute for Theoretical Physics III, Erlangen, Germany}

\author{{\bf Etera R. Livine}}\email{etera.livine@ens-lyon.fr}
\affiliation{Laboratoire de Physique, ENS Lyon, CNRS-UMR 5672, 46 all\'ee d'Italie, Lyon 69007, France}
%\affiliation{Perimeter Institute, 31 Caroline St N, Waterloo ON, Canada N2L 2Y5}

\date{\today}

\begin{abstract}
We revisit the loop gravity space phase for 3D Riemannian gravity by algebraically constructing the phase space $T^*\SU(2)\sim\ISO(3)$ as the Heisenberg double of the Lie group $\SO(3)$ provided with the trivial cocyle. Tackling the issue of accounting for a non-vanishing cosmological constraint $\Lambda \ne 0$ in the canonical framework of 3D loop quantum gravity, $\SL(2,\C)$ viewed as the Heisenberg double of $\SU(2)$ provided with a non-trivial cocyle is introduced as a phase space. It is a deformation of the flat phase space $\ISO(3)$ and reproduces the latter in a suitable limit. The $\SL(2,\C)$ phase space is then used to build a new, deformed LQG phase space associated to graphs. It can be equipped with a set of Gauss constraints and flatness constraints, which form a first class system and Poisson-generate local 3D rotations and deformed translations. We provide a geometrical interpretation for this lattice phase space with constraints in terms of consistently glued hyperbolic triangles, i.e. hyperbolic discrete geometries, thus validating our construction as accounting for a constant curvature $\Lambda<0$. Finally, using ribbon diagrams, we show that our new model is topological.

%The $\SL(2,\C)$ phase space, defined with the classical $r$-matrix of $\sl_{2}$, is as a deformation of the flat phase space $\ISO(3)$ and reproduces the latter in a suitable limit. On this new deformed LQG phase space defined as many copies of the $\SL(2,\C)$ phase space, we introduce closure constraints and flatness constraints, which form a first class system and Poisson-generate deformed equivalent of 3d rotations and translations. We provide a geometrical interpretation for this phase space with constraints in terms of consistently glued hyperbolic triangles - hyperbolic discrete geometries- thus validating our construction as accounting for a constant curvature $\Lambda<0$. Finally, using ribbon diagrams, we show that our new models are topological.%, similarly to BF theory with the moduli space of flat $\SL(2,\C)$ connections.

\end{abstract}

\maketitle

%%%%%%%%%%%%%%%%%%%%%%%%%%%%%%%%%%%%%%%%%%%%%%%%%%%%%%%%%%%%%%%%%%%%%%%%%%%%%%%%%%%%%%%
\setcounter{tocdepth}{0}

\tableofcontents
%\newpage

%%%%%%%%%%%%%%%%%
\section*{Introduction}
%%%%%%%%%%%%%%%%%

3D gravity is an interesting laboratory to explore the quantization of gravity. It retains all the main features of 4d gravity but is much simpler since it is a topological field theory and  admits a finite number of (topological) degrees of freedom \cite{carlip}, which can be solved and quantized exactly. In the main different quantization frameworks, Chern-Simons, spinfoam and loop quantum gravity (LQG), the cosmological constant $\Lambda$ is taken as a coupling constant on the same grounds as Newton's constant for gravity $G_{N}$.

When $\Lambda=0$, all of these frameworks can be used to describe the quantum gravity regime. It is possible to construct bridges between each of them so that they are indeed consistent \cite{louapre, karim, karim-cat, ale-karim}. An interesting feature is that in all three frameworks, a quantum group -the Drinfeld double- appears as the natural symmetry structure \cite{louapre, karim-cat}.

When $\Lambda\neq0$, things get more intricate. The Chern-Simons and spinfoam approaches show that the use of a quantum group is necessary to encode  $\Lambda\neq 0$ \cite{witten, viro}. Furthermore both approaches can be related explicitly \cite{CS-TV}. Note that in the cases of a Lorentzian signature or a negative cosmological constant  $\Lambda<0$, these two approaches do not yet provide a definite answer since the representations of the associated non-compact quantum groups are not yet entirely understood. The LQG case is rather different and trickier. Indeed $\Lambda$ appears as a coupling constant in the quantum Hamiltonian constraint, and one does not know how to solve it on the kinematical Hilbert space if $\Lambda\neq0$. Moreover, these kinematical states of geometry are standard spin network states, which are not defined in terms of a quantum group. This discrepancy is surprising since
one would expect the dynamics of 3D loop quantum gravity for $\Lambda \ne 0$ to be given by the Turaev-Viro spinfoam model (similarly to 3D loop quantum gravity for $\Lambda=0$ defined by the Ponzano-Regge spinfoam model) and thus described in terms of quantum group spin networks. Nevertheless, observables for 3D LQG were recently defined  in terms of  a quantum gauge group ($\UQ$ with $q$ real)  and studied in details \cite{ours1, ours2}. This analysis showed that the theory for $\Lambda <0$ can indeed be interpreted as describing quantum hyperbolic geometries, as one could expect.

\medskip

The key question is then: \textit{how does such quantum group structure appear in the LQG framework}?  On the one hand, we have the classical action for gravity with $\Lambda\neq0$, with its canonical analysis, constraint algebra and geometrical interpretation as constant curvature geometries. On the other hand, we have a candidate quantum theory defined in terms of a quantum gauge group and associated spin network observables and admitting a good geometrical interpretation. The goal is now to connect the two frameworks. See \cite{pranzetti} for an early attempt to tackle this issue.
In the present article, we do not fully answer this question yet, but  we provide an essential step towards the final answer of this question, by introducing a classical phase space for 3D loop quantum gravity with $\Lambda <0$ which can be interpreted as a classical representation of a quantum gauge group symmetry and whose quantization is expected to lead to the observables of quantum hyperbolic geometries as introduced in \cite{ours1,ours2}. %Indeed, recent works have showed that $\SU(2)$ spin networks states describe the quantization of discrete geometries when $\Lambda=0$. Furthermore,

More precisely, we know very well how to discretize the Hamiltonian formulation of the $BF$ action, with $\Lambda=0$, using standard lattice gauge theory tools \cite{ale-karim,valentin1, valentin2}. Such discrete theory can then be quantized to give rise to LQG and the Ponzano-Regge spinfoam model. Hence,  we would like to construct the analogue of such discrete theory  and deal with  discrete  hyperbolic/spherical geometries, which upon quantization would lead to spin network defined in terms of quantum groups and hopefully the dynamics given by the Turaev-Viro model.

We provide here such a model. The key technical innovation is the use of the Heisenberg double which describes the general framework for phase spaces based on groups \cite{frolov1, frolov2}. In particular, using this formalism, one can show that $T^*\SU(2)$ can be described by the Euclidian group $\ISO(3)$, equipped with a symplectic structure. This allows to deform $T^*\SU(2)$ to the case where momentum space (i.e. the flux space) is curved. In this case\footnote{We shall focus for simplicity on the Euclidian case, with $\Lambda<0$.}, the relevant phase space becomes $\SL(2,\C)$. We can then deform the usual  lattice gauge construction using the deformed Heisenberg double. We have constructed a set of two first class constraints, the Gauss constraint which encode the $\SU(2)$ invariance at the vertices of the lattice as well as the flatness constraint which encodes the invariance under some deformed translations (since we are dealing with homogeneously curved geometries).  In the $\SL(2,\C)$ case, the symmetries are non-trivial in the sense that they are equipped with non-trivial Poisson brackets. The symmetry groups are actually Poisson-Lie groups, which are the classical version of the notion of quantum group \cite{chari}.    We have showed that the model we have constructed  is topological and explored the geometrical meaning of the constraints. Just as in the flat case, the Gauss constraint is geometrically equivalent to the (hyperbolic) cosine law and the flatness constraint is equivalent to evaluating the extrinsic curvature in terms of the dihedral angles. By construction, we have the classical Drinfled double as symmetry structure, both in the flat case ($\ISO(3)$) and the curved case ($\SL(2,\C)$).

The discrete lattice gauge theory we have constructed resembles therefore very much  to what we would obtain by discretizing a 3D $BF$ theory with a cosmological constant. However the precise link between such a continuum theory and our discrete theory still has to be identified. The quantization of our model will be discussed in the outlook.

\medskip

The scheme of the article goes as follows. In Section I, we recall the Heisenberg double formalism and in particular how the standard $T^*\SU(2)$ phase space can be described by the group $\ISO(3)$. We recall then how $\SL(2,\C)$ can also be seen as a phase space and how $T^*\SU(2)$ is recovered from it in a suitable limit.
In Section II, we recall the notion of Poisson symmetries, which are given here by Poisson Lie groups, the classical version of a quantum group. In particular we explain how the classical Drinfeld double appears as the symmetry structure of the Heisenberg double. In the $\SL(2,\C)$ case, a deformed notion of translation is obtained. We also provide the infinitesimal realization of the different symmetries (rotations and (deformed) translations), in terms of Poisson brackets.
In Section III, we define the lattice gauge theory model by putting on each edge of the lattice the relevant Heisenberg double. A useful graphical representation for this is to use ribbons instead of lines.  We provide the definition of a set of first class constraints and show how they relate to the infinitesimal symmetries discussed in Section II.
In Section IV, we describe the geometric meaning of the constraints.
In Section V, we show that the discrete models we have defined, either in terms of $\ISO(3)$ or $\SL(2,\C)$ are topological.
We also provide, in appendix, the details of the phase space structure for $\SL(2,\C)$ as well as the proofs for the realization of the infinitesimal symmetries in terms of the Poisson brackets.

%%%%%%%
\section{LQG phase space and its deformation}
%%%%%%%%

\subsection{Definitions}
A Poisson Lie group is a Lie group equipped with an additional structure, a Poisson bracket satisfying a compatibility condition with the group multiplication. Such a Poisson Lie group is associated with a unique Lie bialgebra, the  Lie algebra of the Lie group together with an additional structure, a cocycle $\delta$. The cocycle is derived from the Poisson bracket associated to the Poisson Lie group \cite{chari, yvette}.

\medskip

Consider the Lie bialgebra $(\fg, \delta_{\fg})$ with generators $e_i$, structure constants $\alpha_{ij}^k$ and cocycle $\delta_\fg$ which defines a second family of structure constants $\beta^{ij}_k$,  such that
$$
[e_i, e_j]=\alpha_{ij}^k e_k, \qquad \delta_\fg(e_k)= \beta^{ij}_k e_i \otimes e_k.
$$
The classical double $\fd(\fg)$ is a Lie bialgebra defined from the following structures.
\begin{itemize}
\item \textit{The dual Lie bialgebra $\fg^*$}, with generators $f^i$.  The dual map is built from the bilinear map $\la f^i,e_j\ra =\delta^i_{j}$. The Lie bracket on $\fg^*$ is built from the cocycle on $\fg$.
$$ \la [ f^i,  f^j], e_k\ra = \la f^i\ot f^j, \delta(e_k)\ra  \quad \leadsto \quad  [ f^i,  f^j] = \beta^{ij}_kf^k.
$$
The cocycle $\delta_*$ on $\fg^*$ is built by duality from the Lie bracket on $\fg$.
$$
\la \delta_*(f^i),  e_j \ot e_k \ra = \la f^i, [ e_j,  e_k]\ra  \quad \leadsto \quad  \delta_*(f^i)=  \alpha^i_{jk} f^j\ot f^k.
$$
$(\fg^*, \delta_*)$ is thus the Lie bialgebra with "interchanged" structure constants with respect to $(\fg, \delta_{\fg})$.

\item \textit{As a Lie algebra $\fd(\fg)$} with generators $e_i,f_j$ and commutators
$$
[e_i,e_j]= \alpha _{ij}^k e_k, \quad [f^i,f^j]= \beta ^{ij}_k f^k, \quad [e_i,f^j]= \beta _{j}^{ik} e_k-  \alpha _{jk}^i f^k.
$$
Hence, we have that $\fd(\fg) = \fg^* \bowtie \fg$ as a Lie algebra.  The bilinear form constructed above, supplemented by $\la e_i,e_j \ra =\la f^i,f^j \ra =0 $ is then $\fd(\fg)$-invariant.

\item \textit{The cocycle structure} on  $\fd(\fg)$ is given by
$$ \delta_\fd(d) = [d\ot d,r], \quad d\in \fd, \quad r= \sum f^i\ot e_i, $$
where $r= \sum f^i\ot e_i $ is the classical $r$-matrix.
When restricted respectively to $\fg$ and $\fg^*$, we have
$$
\delta_\fd(e_i) =- \delta(e_i), \quad \delta_\fd(f^i) = \delta_*(f^i).
$$
\item \textit{The $r$-matrix} can be split into the antisymmetric and symmetric parts.  We note  $r^t\equiv e_i\ot f^i$.
$$
\textbf{a}= \demi (r-r^t)= \demi \sum _i ( f^i\ot e_i - e_i\ot f^i), \quad \textbf{s}= \demi (r+r^t).
$$
 We shall consider the case where $\textbf{s}$ is non-degenerate and an invariant of $\fd$.   In this case, we say that the classical double $\fd$ is factorizable  and quasi-triangular. Note that $\textbf{s}$ can be interpreted as a (quadratic) Casimir for $\fd$.

\end{itemize}

We can integrate $\fd$ to get the  Lie group $\cD\sim G^*\bowtie G$. It can be equipped with two typical Poisson structures, characterized by the Poisson bivectors $\pi_\pm$.
\beq
\pi_\pm(D) = - \com{\textbf{a},D\ot D}_\pm, \quad D\in \cD,
\eeq
where $[ \; , \;]_-$ denotes the usual commutator and $[ \; , \;]_+$ the anticommutator. \\
The group $\cD$ equipped with $\pi_-$ is called the \textit{Drinfeld double} of $G$. It is a Poisson Lie group and as such is not a symplectic space. \\
The group $\cD$ equipped with $\pi_+$ is called the \textit{Heisenberg double}. It is a symplectic space. We can determine the Poisson brackets between the matrix elements of $D\in \cD$ in terms of the $r$-matrix, using the expression of $\pi_+$ and the expression of $\textbf{a}$ in terms of $r$ and $r^t$. We get
\be\label{poisson structure}
\{D\stackrel{\ot}{,}D\}_{\pi_+}
\,=\,
%-rG_1G_2-G_1G_2r^\dagger=
 - \com{\textbf{a},D\ot D}_{+} = -r (D\otimes D) + (D\otimes D)r^t,
\,
\ee
where the subscript $\pi_+$ refers to the fact that we are using the Poisson bracket structure coming from the bivector $\pi_+$. Since we are going to deal only with this Poisson bracket in this Section, we shall omit it for simplicity. We will come back to the Drinfeld double structure implemented by $\pi_-$ in  Section \ref{symmetries} when addressing  the symmetries. We use the standard notation  $D_1 = D\ot \one$ and  $D_2 = \one \ot D$.
$$
\{D\stackrel{\ot}{,}D\}\equiv\{D_1, D_2\}\equiv  \left(\begin{array}{ccccccc}
\poi{D_{11},D_{11}}&\poi{D_{11},D_{12}}& \hdots& \poi{D_{12},D_{11}}&\poi{D_{12},D_{12}} &\hdots\\
\poi{D_{11},D_{21}}& \poi{D_{11},D_{22}}&\hdots& \poi{D_{12},D_{21}} &\poi{D_{12},D_{22}}&\hdots  \\
\vdots&\vdots &\vdots &\vdots&\vdots\\
\poi{D_{21},D_{11}}&\poi{D_{21},D_{12}} &\hdots& \poi{D_{22},D_{11}}&\poi{D_{22},D_{12}}&\hdots\\
\poi{D_{21},D_{21}}& \poi{D_{21},D_{22}}&\hdots& \poi{D_{22},D_{21}} &\poi{D_{22},D_{22}} &\hdots \\
\vdots&\vdots &\vdots &\vdots&\vdots
 \end{array}  \right)
$$
%We have then, using the expression of $\textbf{a}$ in terms of $r$ and $r^t$,

%In the following we shall omit the subscript $+$ for simplicity.

\medskip

We have constructed the classical double of the Lie algebra $\fg$, with cocycle $\delta$. However, we could also construct the classical double of $\fg^*$ with cocycle $\delta_*$. In this case, we  obtain the Lie algebra $\fg\bowtie \fg^*$ with cocycle generated by the $r$-matrix $\tilde r= \sum e_i\ot fi = r^t$. The resulting group $D$ will be the same. The difference from the previous construction is how this group is factorized and the symplectic structure on it. We  have now $D\sim G\bowtie G^* $. The symplectic structure is given by
\be\label{poisson structure 2}
\{D_1 , D_2\}
\,=\,
%-rG_1G_2-G_1G_2r^\dagger=
-\tilde r (D\otimes D) + (D\otimes D)\tilde r^t = -r^t (D\otimes D) + (D\otimes D)r.
\,
\ee

\begin{figure}[h]
\begin{center}
\includegraphics{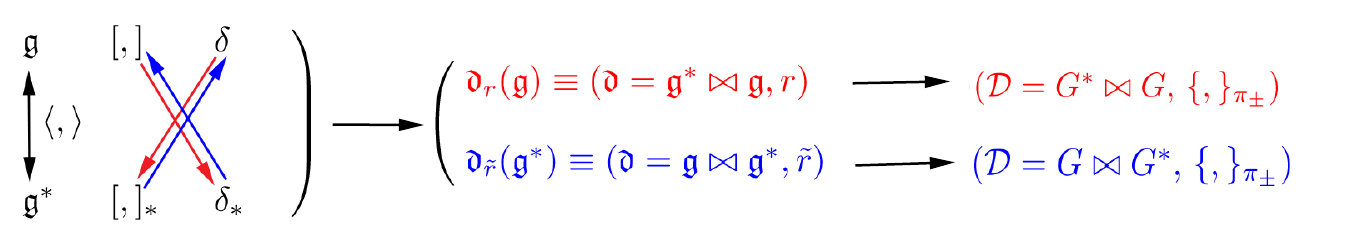}
\end{center}
\end{figure}

\medskip

%$$
%\left. \begin{array}{lll}
%\fg \hspace{.5cm}& [,] \hspace{.5cm}& \delta\hspace{.5cm} \\
% \\\hspace{.2cm} \la,\ra&&\\
% \\
%\fg^* & [,]_* & \delta_*\\
%\end{array}
%\right) \hspace{1cm} \left( \begin{array}{lll}  \textcolor{red}{\fd_r(\fg)\equiv (\fd= \fg^*\bowtie \fg ,r)} & \hspace{1cm} &\textcolor{red}{(\cD=G^*\bowtie G,\{,\}_\pm)}  \\ \\
% \textcolor{blue}{\fd_{\tilde r}(\fg^*)\equiv (\fd= \fg\bowtie \fg^* ,\tilde r)} & \hspace{1cm} &\textcolor{blue}{(\cD=G\bowtie G^*,\{,\}_\pm)}
%\end{array}\right.
%$$

Let us illustrate the full construction for  $\cD=\ISU(3)$ and $\SL(2,\C)$. At the infinitesimal level, we start with the Lie algebra $\fg=\so(3)$ (resp. $\su(2)$) with generators $J_i$, (resp. $\sigma_i$) satisfying the commutation relations  $
\com{J_i,J_j}= \epsilon_{ij}^k\,  J_k$, (resp. $\com{\sigma_i,\sigma_j}= 2i\epsilon_{ij}^k\,  \sigma_k$) . But we consider two possible cocycles $\delta_\fg$: $\ISO(3)$ comes from a choice of trivial cocycle for $\so(3)$ whereas $\SL(2, \C)$ emerges as the double for $\su(2)$ equipped with a non-trivial cocycle. Indeed, a given choice of  cocycle entirely determines the associated dual Lie bialgebra $\fg^*$. The details are given in the next two sections.

\subsection{$\ISO(3)$ as  phase space}
We first consider $\so(3)$  equipped with the simplest cocycle,  the trivial cocycle $\delta_{0}\equiv 0$.  The classical double $\fd_0(\so(3))$ is then given by the following structures.
\begin{itemize}
\item We note $E^i$ the dual of $J_i$. The $\fd_0$-invariant bilinear form is $\la E^i,J_j\ra =\delta^i_{j}, \, \la J_i,J_j\ra=\la E^i,E^j\ra=0, \; \forall \, i, j \in \{1,2, 3\} $.
\item The dual Lie algebra $\so(3)^*$, generated by the $E's$ is Abelian. Indeed,  $[E^i,E^j]=0$ since the $\so(3)$ cocycle $\delta_0$ is trivial. Hence $\so(3)^*\sim \R^3$. Its associated cocycle is given by the constant structure defined by the Lie algebra structure of $\so(3)$, that is, $\delta_*(E^k)= \epsilon^{k}_{ij}\,E^i\otimes E^j$.
\item As a Lie algebra $\fd_0(\so(3)) \sim \isu(3)=\R^3\rtimes \so(3)$.
\item The (coboundary) cocycle $\delta_{\fd_0}$ is generated by the $r$-matrix $r=\sum_i  E^i\ot J_i$.
\end{itemize}
We  construct now the Heisenberg double $\cD_{\pi_+}$. It is given by the Lie group $\ISO(3)$ while the Poisson structure is specified by the $r$-matrix, as in \eqref{poisson structure}. A group element $D\in I\SO(3)$ is characterized by a rotation $\textbf{R}\in\SO(3)$ and a translation $\bX\in\R^3$, with product,
\beq
(\textbf{R}_1,\bX_1)(\textbf{R}_2,\bX_2)= (\textbf{R}_1\textbf{R}_2, \bX_1 + \textbf{R}_1\rhd \bX_2),
\eeq
where the rotation $\textbf{R}_1$ acts in the vector representation on the 3-vector $\bX_2$. A general element  $D=(\textbf{R},\bX)$, can be factorized into a pure translation and a pure rotation. %in two different ways, respectively the left and the right decompositions.
\bes
D=(\textbf{R},\bX)&=& \ell u,  \quad \ell = (\textbf{1},\bX), \quad  u = (\textbf{R}, \textbf{0}).
%&=& \tu \tell,  \quad \tell = (\textbf{1}, \vec {\tilde X})=(1, \textbf{R}\mone \rhd \bX), \quad  \tu = (\textbf{R},\vec 0).
\eeq
The Poisson bracket structure for $D$ is given by \eqref{poisson structure} in terms of the classical double $r$-matrix, $r=\sum_i  E^i\ot J_i$.
%We have indeed $u\tell =\ell u $ so $\tell = u\mone \ell u$.
We can then deduce the Poisson structure for the variables $u$ and $\ell$,
\bes \label{poisson euclidian}
\poi{\ell_1,\ell_2}=-\com{r, \ell_1\ell_2}, & \quad \poi{\ell_1,u_2}=-\ell_1 r u_2, \quad  \poi{u_1,\ell_2}=\ell_2 r^t u_1, & \quad \poi{u_1,u_2}=-\com{r^t,u_1u_2}.
\ees
When using the explicit parametrization of $u$ and $\ell$, these Poisson brackets become the standard  $T^*\SU(2)$ Poisson structure:
\beq\label{left TSU2}
\poi{X_i,X_j}=\epsilon_{ij}^kX_k, \quad \poi{X_i,\textbf{R}}= -J_i \textbf{R}, \quad \poi{\textbf{R},\textbf{R}}=0,
\eeq
Hence $\ISO(3)$ equipped with the Poisson structure \eqref{poisson euclidian} is an equivalent description of the phase space $T^*\SU(2)$, which is commonly used in LQG.

\medskip

Instead of considering the left decomposition of $\ISO(3)$, $D=\ell u$, we could consider the right decomposition  $D= \tilde{u} \tilde{\ell}$. Since, $\tu\tell =\ell u $ we have that
\bes
D=(\textbf{R},\bX)
&=& \tu \tell,  \quad \textrm{ with } \tell = (\textbf{1},  {\tilde \bX})=(1, \textbf{R}\mone \rhd \bX), \quad  \tu = (\textbf{R},\vec 0).
\eeq
 So $\tu= u$ and  $\tell = u\mone \ell u$. \\
  This factorization comes from building the classical double of the Abelian Lie algebra $\R^3$ equipped with the non trivial cocycle $\delta_*(E^i)= \epsilon^{k}_{ij} E^i\otimes E^j$.  The dual bialgebra is then $\so(3)$ with a trivial cocycle.  The relevant $r$-matrix for the classical double $\fd_0(\R^3)$ is then $\tilde r=\sum_i  J^i\ot E_i=r^t$. The symplectic structure on $\ISO(3)= \SO(3)\ltimes \R^3$ is given by \eqref{poisson structure 2} which implies the following Poisson brackets for the $\tu$ and $\tell$ variables,
\bes \label{poisson euclidian bis}
\poi{\tell_1,\tell_2}=-\com{ r , \tell_1\tell_2}, & \quad \poi{\tell_1,\tu_2}=-\tell_1  r \tu_2, \quad  \poi{\tu_1,\tell_2}=\tell_2  r^t \tu_1, & \quad \poi{\tu_1,\tu_2}=-\com{ r^t,\tu_1\tu_2}.
\ees
In terms of the $T^*\SU(2)$ variables, we have then
\beq\label{right TSU2}
\poi{\tX_i,\tX_j}=\epsilon_{ij}^k\tX_k, \quad \poi{\tX_i,\textbf{R}}=  \textbf{R}\, J_i, \quad \poi{\textbf{R},\textbf{R}}=0.
\eeq
We see that the two decompositions of the Heisenberg double built on $\ISO(3)$ correspond to considering the left and right invariant vectors on $\SO(3)$.

\medskip

An explicit realization of the generators of $\so(3)$ and $\R^3$ as well as of the $r$-matrices, $r$ and $r^t$, and of the variables $\ell$ and $u$ can be found in Appendix \ref{ISO(3)}.

\medskip

The explicit link between the variables of the left and of the right decompositions of $D \in \ISO(3)$ allows us to evaluate the Poisson brackets between $\tell$ and $\ell$ and $u$  and between $\tu$ and $\ell$ and $u$. Indeed, $\tu=u$ implies that
\be
\{ \tu_1, u_2 \}=0, \qquad \{\tu_1, \ell_2\}=\ell_2 r^t \tu_1,
\ee
and $\tell=u^{-1} \ell u$ implies that
\be
\{\tell_1, \ell_2\}=0, \qquad \{\tell_1, u_2\}=u_2 r \tell_1.
\ee

\subsection{$\SL(2,\C)$ as a phase space}
We  construct the Heisenberg double structure on $\SL(2,\C)$, following the same recipe as above. The factorization of $\SL(2,\C)$ into $\SU(2)$ and $\SB(2,\C)$ is given by the Iwasawa decomposition. As in the Euclidian group case, we have a left and a right decomposition. We construct the phase space structure for each if these decompositions. We are going to use the spinorial representation following the standard literature.  In particular, we  use the spinorial representation of $\su(2)$, so we use  as generators, the Pauli matrices with commutation relations $[\sigma_i,\sigma_j]=2i\,\eps_{ij}^k\, \sigma_k$ (see appendix \ref{poi formulae} for the expression of the Pauli matrices).

 %%%%%%%%
\subsubsection{Left Iwasawa decomposition}
%%%%%%%%%%%%
  Our starting point is now  $\su(2)$ equipped with a non trivial cocycle.
 \be \label{nontrivial cocycle}
 \delta(\sigma_k)=2 i\, \ka \left( \delta^i_k \delta^j_z-\delta^j_k \delta^i_z\right) \sigma_i \otimes \sigma_j.
 \ee
The classical double $\fd(\su(2))$ is then given by the following structures.
\begin{itemize}
\item We note $\tau^i$ the dual of $\sigma_i$. The $\fd$-invariant bilinear form is $\la \tau^i,\sigma_j\ra =\delta^i_{j}, \, \la \sigma_i,\sigma_j\ra=\la \tau^i,\tau^j\ra=0$.%, which is the Killing form on $\SL(2,\C)$.
\item The dual Lie algebra $\su(2)^*$ is not Abelian anymore since the $\su(2)$ cocycle $\delta$ in \eqref{nontrivial cocycle} is non trivial. We have
\be
[\tau^i,\tau^j]=2i\ka (\delta _{jz}\tau^i - \delta _{iz}\tau^j).
\ee
Therefore,  we have that  $\su(2)^*=\sb(2,\C)$, the Lie algebra which  generates the $2\times2$ lower triangular matrices. The generators $\tau^i$ can be expressed in terms of the Pauli matrices.
\be
\tau^i=i \ka (\sigma_i - \f 12 [\sigma_z, \sigma_i])= \ka( i\sigma_i + \epsilon_{zi}^k \sigma_k).
\ee
Their explicit expression as 2-by-2 matrices can be found in appendix \ref{poi formulae}.
The $\sb(2,\C) $ cocycle is induced by the Lie algebra structure of $\su(2)$, so $\delta_*(\tau^k)= 2i\epsilon^{k}_{ij}\,\tau^i\otimes \tau^j$.
\item As a Lie algebra $\fd =\sb(2,\C) \bowtie \su(2) \sim \sl(2,\C)$.
\item The (coboundary) cocycle $\delta_{\fd}$ is generated by the $r$-matrix
\be \label{r sl2c}
r=\f 14\sum_i  \tau^i\ot \sigma_i \,=\,
\frac{i \ka}{4}
\left(\begin{array}{cccc}
1&&&\\ &-1&&\\&4&-1&\\&&&1
\end{array}\right).\ee
\end{itemize}
We have constructed above the classical double $\fd(\su(2))$, which leads to the Lie algebra $\sb(2,\C) \bowtie \su(2) \sim \sl(2,\C)$. This corresponds to the "left Iwasawa decomposition" of $ \sl(2,\C)$. This decomposition is naturally extended to the group element  $D\in\SL(2,\C)$.
\be\label{def1}
D=\ell u,
\qquad
u=\mat{cc}{\alpha & -\bar{\beta} \\ \beta & \bar{\alpha}}\,\in\SU(2),
\quad
\ell=\mat{cc}{\lambda & 0 \\ z & \lambda^{-1}}\,\in\SB(2,\C),\quad
\lambda\in\R_+^*,\,\,z\in\C\,.
\ee
It is common in the literature to express the Poisson structure in terms of the Hermitian conjugated $r^\dagger$ instead of the transpose $r^t$ \cite{marmo}. Note that in this particular case, $r^\dagger= \sum_i (\tau^i)^\dagger \otimes \sigma_i=-\sum_i \sigma_i\otimes \tau^i=-r^t$. Hence the symplectic structure between the matrix elements of $\SL(2,\C)$ becomes
\be\label{SL(2,C) poi}
\{D_1,D_2\}
\,=\,
-rD_1D_2-D_1D_2r^\dagger, \quad D\in\SL(2,\C).
\ee
The full phase space structure for the configuration variables $u$ and momenta $\ell$ now is given by
\begin{eqnarray}
&&
\poi{\ell_1,\ell_2}=-\com{r, \ell_1\ell_2}, \quad
  \label{mm}   \label{lu poi}\quad
\poi{\ell_1,u_2}=-\ell_1 r u_2, \quad
\poi{u_1,\ell_2}=-\ell_2 r^\dagger u_1, \quad
\label{mc} \quad
\poi{u_1,u_2}=\com{r^\dagger,u_1u_2}.\label{cc}
\end{eqnarray}
Note that since $z$ is complex, we also need to consider its conjugated $\ov{z}$. To this aim, we introduce $l= (\ell^\dagger)\mone$,
\beq
l=\mat{cc}{\lambda\mone & - \ov{z} \\ 0 & \lambda}.
\eeq
To specify, the Poisson brackets of $l$ with $\ell$ and $u$, we use the fact that the $\SB(2,\C)$ group  and the algebra $\sb(2, C)$ are preserved under $ \ell \rightarrow l$, $\tau^i \rightarrow (\tau^i)^\dagger$. As in \cite{marmo}, we require that the Poisson brackets of $u$ and $\ell$ are preserved under this mapping as well.
\bes
 \poi{l_1,\ell_2}&=&-\f14 [ (\tau^i)^\dagger\otimes \sigma_i, l_1 \ell_2]=-\com{r^\dagger, l_1\ell_2},
\quad \poi{l_1,u_2}=-\f14 l_1((\tau^i)^\dagger\otimes \sigma_i)u_2=-l_1 r^\dagger u_2, \nn\\
\poi{l_1,l_2}&=& -\f14[(\tau^i)^\dagger\otimes \sigma_i, l_1 l_2]=-\com{r^\dagger, l_1l_2}. \label{lu poi 2}
\ees
The explicit expression of these Poisson brackets in terms of the matrix elements $\lambda, \, z, \, \bar{z}$ and $\alpha, \, \beta$ is given in Appendix \ref{poi formulae}.
\medskip

The Heisenberg double $\SL(2,\C)$ can be seen as a deformation of the Heisenberg double $\ISO(3)$. We shall discuss the limit in section \ref{sec:limit}. In this latter case,  angular momentum variables are given by  $\bX\in\R^3$. When moving to the $\SL(2,\C)$ case,  we have deformed $\R^3$ into the hyperboloid of radius $\ka\mone$.  Since we are using the spinorial representation, a point on the hyperboloid can be obtained from the combinations $D D^\dagger$ and $D^\dagger D$. We construct the two vector-like quantities out of these two combinations,
\be
DD^\dagger \equiv L=\ell\ell^\dagger
=\mat{cc}{\lambda^2 & \lambda\,\bz \\ \lambda\,z & \lambda^{-2}+|z|^2}\,,
\quad
T_0=\f1{2\ka}\tr L
=\f1{2\ka}\left(\lambda^2+\lambda^{-2}+|z|^2\right), \quad \bT \equiv \f1{2\kappa}\,\tr(\ell\ell^\dagger\,\vsigma),
\ee
with explicit components
\be
\qquad
T_z=\f1{2\kappa}\,\tr \ell\ell^\dagger\sigma_z
=\f1{2\kappa}\,(\lambda^2-\lambda^{-2}-|z|^2),
\quad
T_+=\f1{\kappa}\,\tr \ell\ell^\dagger\sigma_+
=\f1\kappa\,\lambda z,\quad
T_-=\overline{T_+}=\f1\kappa\,\lambda \bz\,.
\ee
The 4-vector $T^\mu$ defines indeed  a point on the hyperboloid since we have   $T^2=T_\mu T^\mu= T_0^2-\vT^2=T_0^2-(T_z^2+T_+T_-)=1/{\ka^2}$.
%This corresponds to $\det L=1$ with $L=\ka T_\mu\sigma^\mu=\ka(T_0\one+T_z\sigma_z+T_+\sigma_-+T_-\sigma_+)$.
Using either  the formula with the $r$-matrix or  the explicit commutators of $z$, $\ovz$ and $\lambda$, we can compute the Poisson brackets between the $T_\mu$ components.
\beq\label{poit}
\poi{T_0,\bT}=0,
\quad
\poi{T_z,T_+}=i \ka\, (T_0 + T_z) \,T_+,
\quad
\poi{T_z,T_-}=-i \ka\, (T_0 + T_z) \,T_-,
\quad
\poi{T_+,T_-}=2 i \ka \, (T_0 + T_z)\,T_z\,,
\eeq
where $\ka(T_0 + T_z)=\lambda^2$ plays a role of a re-scaling compared to the commutators of the standard $\su(2)$ algebra.

The other vector we can construct is the analogue of the vector $ {\tilde \bX}= \textbf{R}\mone \rhd \bX $.
\beq
 u D^\dagger D u\mone \equiv \Lop \equiv \ell^\dagger\ell \,=\,\mat{cc}{\lambda^2+|z|^2 & \lambda^{-1}\bz \\  \lambda^{-1}z & \lambda^{-2}},  \quad T^{op}_0=T_0, \quad \bTop \equiv \f1{2\kappa}\,\tr(\ell^\dagger\ell\,\vsigma)\,,
\eeq
with explicit expressions:
\be
\Top_z=\f{1}{2\kappa}\,(\lambda^2-\lambda^{-2}+|z|^2),\quad
\Top_+=\f1\kappa\,\lambda^{-1}z,\quad
\Top_-=\f1\kappa\,\lambda^{-1}\bz.
\ee

%%%%%%%%%%%%
\subsubsection{Right Iwasawa decomposition}
%%%%%%%%
Instead of building the classical double from $\su(2)$, we could have constructed it out of $\sb(2,\C)$, with the non trivial cocycle $\delta_*(\tau^k)=2i\epsilon^k_{ij} \tau^i \otimes \tau^j$. This  leads to the "right  Iwasawa decomposition" of $\sl(2,\C)\sim \su(2)\bowtie \sb(2,\C)\sim \fd_{\tilde r}(\sb(2,\C))$. The associated $r$-matrix is then  $\tilde r = \f14\sum \sigma_i\ot \tau_i = r^t=-r^\dagger$. This decomposition can be extended to   group elements.
\be\label{def2}
D=\tu\tell,
\qquad
\tu=\mat{cc}{\talpha & -\bar{\tbeta} \\ \tbeta & \bar{\talpha}}\,\in\SU(2),
\quad
\tell=\mat{cc}{\tlambda & 0 \\ \tz & \tlambda^{-1}}\,\in\SB(2,\C),\quad
\tlambda\in\R_+^*,\,\,\tz\in\C\,.
\ee
The phase space structure is then given by $\tilde r$.
$$
\{D_1,D_2\}
\,=\,
-\tilde rD_1D_2-D_1D_2\tilde r^\dagger \,=\,
r^\dagger D_1D_2+D_1D_2r.
$$
 The Poisson structure  on the variables $\tu, \tell$ and $\tilde{l}$, where $\tilde{l}$ is defined as previously by $\tilde{l}=(\tell^\dagger)^{-1}$ is the following,
\be
\begin{gathered}
\poi{\tell_1,\tell_2}=-\com{r, \tell_1\tell_2},
\qquad
\poi{\tell_1,\tu_2}=\tu_2 r \tell_1,
\qquad
\poi{\tu_1,\tu_2}=\com{r^\dagger, \tu_1\tu_2}\,,\\
\poi{\tilde{l}_1, \tell_2}=-\com{r^\dagger, \tilde{l}_1 \tell_2}, \qquad \poi{\tilde{l}_1, \tilde{l}_2}=-\com{r^\dagger, \tilde{l}_1 \tilde{l}_2}, \qquad \poi{\tu_1,\tell_2}=\tu_1r^\dagger \tell_2.
\end{gathered}
\ee
%For instance, this gives
%\be
%\poi{\tlambda, \tz} = -\f{i\ka}{2}\tlambda\tz\,, \qquad \poi{\bar{\tz}, \tz }=i \ka (\tlambda^2-\tlambda^{-2}).
%\ee

As earlier, we can  introduce the different vectors associated to this decomposition.
\bes
\tL&=&\tell\tell^\dagger=\,\mat{cc}{\tlambda^2 & \tlambda \btz \\  \tlambda \tz & \tlambda^{-2}+|\tz|^2},
\btT=\f1{2\kappa}\,\tr(\tell\tell^\dagger\vsigma), \quad
\tT_z=\f1{2\kappa}\,(\tlambda^2-\tlambda^{-2}-|\tz|^2),\quad
\tT_+=\f1\kappa\,\tlambda \tz,\quad
\tT_-=\f1\kappa\,\tlambda \btz\,,
\qquad
\,,
\ees
Note that the two decompositions describe the same group element $D= \ell u = \tu \tell$. Hence we have that $ \tL=\tu^{-1}\ell\ell ^\dagger \tu= \tu^{-1}L \tu$. As a consequence, the 3-vector $\btT$ is related to the 3-vector $\bT$ by the rotation $\tu^{-1}$,  $\btT=\tu^{-1}\rhd \bT$. \\
We can also construct the other vector $\tbTop$, given by the other combination $\tell^\dagger\tell$.
\beq
\tbTop \equiv \f1{2\kappa}\,\tr(\tell^\dagger\tell\,\vsigma)\,, \quad \tTop_z=\f1{2\kappa}\,(\tlambda^2-\tlambda^{-2}+|\tz|^2),\quad
\tTop_+=\f1\kappa\,\tlambda^{-1}\tz,\quad
\tTop_-=\f1\kappa\,\tlambda^{-1}\btz\,.
\eeq
Since $\tLop=\tell^\dagger\tell = u^{-1}\ell^\dagger\ell u= u^{-1}\Lop u$, the 3-vector $\tbTop$ is related to the 3-vector $\bTop$ by the rotation $u^{-1}$, $\tbTop= u^{-1} \rhd \bTop$.

\medskip

This explicit link between the variable of the left and right Iwasawa decomposition of $D \in \SL(2,\C)$ allows us to express the Poisson brackets between the $\ell$, $l$, $u$ variables and the $\tell$, $\tilde{l}$, $\tu$ variables.
By considering $\tell^\dagger \tell = u\mone \ell ^\dagger \ell u$, we deduce that
\be
\left\{ \tell^{-1}_1,u_2\right\} = u_2\,r\,\tell^{-1}_1, \qquad \left\{ \tell_1, \ell_2\right\} = 0.
\ee
From the first equality of the above equation, we can in particular infer that
\be
\{\tilde{l}_1, u_2\}=-\tilde{l}_1u_2 r^\dagger.
\ee
Finally, using $\tu = \ell u \tell^{-1}$, we get
\be
\left\{\tu_1, u_2 \right\} = 0,\qquad \left\{\tu_1, \ell_2 \right\} = -r^\dagger\,\tu_1 \ell_2.
\ee
More details about the computations of these commutation relations and the commutation relations in terms of the components of $\tell, \, \tilde{l}, \, \ell, \, u$ and $\tu$ can be found in the Appendix \ref{poi formulae}.

%%%%%%%
\subsection{$\SL(2,\C)$ phase space as a deformation of the  $\ISO(3)$ phase space }\label{sec:limit}
%%%%%%%%
We would like to show here how the $\ISO(3)$ phase space can be obtained in the limit $\ka\dr0$, so that $\SL(2,\C)$ as a phase space can be viewed as a deformation of the $\ISO(3)$ phase space.

As groups, we know that the Ionu-Wigner contraction allows to recover   $\ISO(3)$ from $\SL(2,\C)$.  This contraction is implemented through $\ka\dr 0$, such that $\tau_i\dr E_i$, the generators of the translations. Indeed the $\sb(2,\C)$ Lie bracket becomes the $\R^3$ Lie brackets.
\beq
\com{\tau_i,\tau_j}= 2i\ka (\delta_{iz}\tau_j- \delta_{jz}\tau_i)\dr 0= \com{E_i,E_j}.
\eeq
%Second as  phase spaces, we consider the Poisson brackets \eqref{lu poi} and \eqref{poisson euclidian}. In each case, they are characterized by the $r$-matrix. For the $\SL(2,\C)$ case, we have  $r=\f14 \sum \tau_i\ot \sigma_i$, whereas in the $\ISO(3)$ case, we have  $r= \sum E_i\ot J_i$. As above, since $\tau_i\dr E_i$ it is clear that the Poisson structures are related in each case. Note however that in one case we have used the spinor representation, whereas in the other case, we have used the vector representation. This was just for calculation convenience. The $\SL(2,\C)$ can also be treated in the vector representation.

Due to the difference of representation (spinor for $\SL(2,\C)$ versus vector for $\ISO(3)$), we might wonder what the relations between the different coordinates used to parameterize the two momentum spaces $\R^3$ and $\SB(2,\C)$ are.  By construction, the generators $\tau_i$ have dimension $\ka$, so that the coordinates on $\SB(2,\C)$ should have dimension $\ka\mone$, in agreement with the geometric interpretation of the momentum space as the hyperboloid of radius $\ka\mone$. If we consider the "Euler" parametrization of $\ell\in\SB(2,C)$, we consider ${\bf j} \in \R^3$ with dimension $\ka\mone$ such that
\be
\label{vecj}
\ell=e^{ i j_z\tau^z}e^{i j_x\tau^x}e^{ i j_y\tau^y},
\qquad
\lambda=e^{-{\f\kappa2 j_z}}, \quad z=-\ka e^{{\f\kappa2 j_z}} \left({j_x+ij_y}\right)=-e^{{ \f\kappa2 j_z}} \ka j_+\,, \quad \ov{z}=-e^{{\f\kappa2 j_z}} \ka j_-\, .
\ee
%Note that with this parametrization, $\lambda $ and $z$ have    dimension 1 and $\ka\mone$.

We would like to check  that the  Poisson brackets of the ${\bf j}$'s expressed in terms of $\lambda,\, z, \,\overline{z}$ give rise to the Poisson brackets between the usual angular momentum variables $\bX$ in the limit $\ka\dr0$. We get
\beq
\poi{j_z,j_+}=\f{i}\ka \lambda z = - i j_+, \quad   \poi{j_z,j_-}=\f{i}\ka \lambda z =  i j_-, \quad
\{j_+,j_-\}  %\f{i}\ka(-\lambda^2|z|^2+\lambda^4-1)
= \f{i}\ka(\ka^2 j_+j_- +e^{-2\ka j_z}-1) \underset{\ka\arr0}{\longrightarrow}\, -2ij_z.
\eeq
Hence we have recovered the standard angular momentum Poisson relations, modulo the fact  that $j_+$ and $j_-$ are swapped with respect to the usual commutation relations, i.e. $j_+\dr X_-$, $j_-\dr X_+$ and $j_z\dr X_z$.
%This is due to the fact that we have used the lower triangular matrix to encode $j_+$ and  the  upper triangular matrix for $j_-$. This is just a matter of convention.

For the sake of completeness, we can also take the limit $\ka\to0$ of the vector $\bf T$ in terms of ${\bf j}$,
\beq
T_+=-j_+, \quad T_-=-j_-,\quad T_z= \f{1}{2\ka}(e^{-\ka j_z} -  e^{\ka j_z}+\ka^2 e^{-\ka j_z} j_+j_-)\underset{\ka\arr0}{\longrightarrow}\, -j_z.
\eeq
The Poisson brackets \eqref{poit} between the components of $\bT$ also reproduce the angular momentum Poisson algebra,
\be
\ka T_0 \arr 1, \quad
\{T_z,T_\pm\}\arr \pm iT_\pm,\quad
\{T_+,T_-\}\arr 2iT_z\,,
\ee
meaning that $\bT$ and $\bX$ coincide in the limit $\ka\dr 0$.

The bottom line is that
$\SL(2,\C)$ as a phase space can be viewed as a deformation of  $\ISO(3)$ seen as a phase space.

%%%%%%%
\section{Poisson Lie symmetries}\label{symmetries}
%%%%%%%%
The introduction of Poisson Lie groups is motivated by the identification of the phase space symmetries. They must be compatible with the Poisson structure, meaning that the symmetry action is a \textit{Poisson map} \cite{yvette}. The symmetries are then called \textit{Poisson-Lie group symmetries}. Having such a symmetry  amounts to putting a Poisson structure\footnote{It will never be symplectic by construction.} on the symmetry group, which is compatible with the group product (i.e. the group multiplication is a Poisson map). In this case we have a Poisson Lie group as symmetry group. Often this Poisson structure  does not need to be specified, but when the phase space symplectic structure is non trivial,  the Poisson structure on the symmetry group also becomes non trivial and cannot be overlooked.
%A necessary condition to have a Poisson Lie group is that the Poisson structure is consistent w

In the previous section, we have considered some examples of Poisson Lie groups. We had for $G=\SU(2)$ or $\SO(3)$ and its dual $G^*$,
\beq
(G,\poi{,}_G) \leadsto \poi{u_1,u_2}_G=-[r^t,u_1u_2], \quad  (G^*,\poi{,}_{G^*})\leadsto \poi{\ell_1,\ell_2}_G=-[r,\ell_1\ell_2].
\eeq
From $G$ and $G^*$, we constructed $\cD=G^*\bowtie G$. As a phase space, $\cD$ was equipped with the symplectic structure $\pi_+$ of the Heisenberg double which is not a Poisson Lie group structure. $\cD$ can also be  equipped with the Poisson structure
\beq \label{drinfeld structure}
\pi_-(D)=-[\textbf{a}, D_1\ot D_2]= - [r,D_1D_2], \quad D\in \cD.
\eeq
Then, $\cD=\cD_{\pi_-}$ is a Poisson Lie group and is called a \textit{Drinfeld double}. The Drinfeld double $\cD_{\pi_-}$ can actually be seen as the symmetry structure behind the Heisenberg double $\cD_{\pi_+}$ as the following theorem shows.
\begin{theorem}\textbf{Drinfeld double as symmetry structure}\\
The multiplication maps
$$
\begin{array}{ccl}
\cD_{\pi_-}\times \cD_{\pi_+} &\dr&  \cD_{\pi_+} \\
(U,D) &\dr&  UD
\end{array} \qquad
\begin{array}{ccl}
 \cD_{\pi_+}\times  \cD_{ -\pi_-} &\dr&  \cD_{\pi_+} \\
(D,V) &\dr&  DV
\end{array}
$$
%$(\cD,\pi_-)\times (\cD, \pi_+) \dr  (\cD, \pi_+)$ and $(\cD,\pi_+)\times (\cD, -\pi_-) \dr  (\cD, \pi_+)$
 are Poisson maps, so they define left and right Poisson actions of respectively $\cD_{\pi_-}$ and $\cD_{-\pi_-}$ on $\cD_{\pi_+}$, \cite{lu}. Explicitly, we  have the symmetry transformations $U$ and $V$ that do Poisson commute with $D$ an element of the Heisenberg double and
\bes
 \poi{U_1D_1, U_2D_2}_{\pi_+}&=&\poi{U_1,U_2}_{\pi_-}D_1D_2 + U_1U_2 \poi{D_1,D_2}_{\pi_+}, \\  \poi{D_1V_1, D_2V_2}_{\pi_+}&=&\poi{D_1,D_2}_{\pi_+}V_1V_2 - D_1D_2 \poi{V_1,V_2}_{\pi_-}, \quad U,V, D \in \cD.
 \ees
%The maps
%$$
%\begin{array}{rcl}
%(G,\pi_G)\times (\cD, \pi_+) &\dr&  (\cD, \pi_+) \\
%(v,d) \dr  vd
%\end{array}
%\begin{array}{rcl}
% (\cD, \pi_+) \times  (G^*,\pi_{G^*}) &\dr&  (\cD, \pi_+) \\
%(d,m) \dr  dm
%\end{array}
%$$
%are Poisson, so they define left and right Poisson actions of $(G,\pi_G)$ and $ (G^*,\pi_{G^*})$ on $(\cD, \pi_+)$.
\end{theorem}

Therefore, $\cD_{\pi_-}$ equipped with \eqref{drinfeld structure} is the symmetry structure behind a phase space with symplectic structure defined by $\pi_+$. The symmetries of the phase spaces considered in the previous section, namely $\cD_{\pi_+}=\ISO(3)$ and $\SL(2,\C)$, are respectively $\cD_{\pi_-} = \ISO(3), \,\SL(2,\C)$ equipped with \eqref{drinfeld structure} which makes them Poisson Lie groups. The $r$-matrix of \eqref{drinfeld structure} is the one in the definition of the symplectic structure of $\cD_{\pi_+}$ given by \eqref{r iso3} for $\ISO(3)$ and \eqref{r sl2c} for $\SL(2,\C)$.
Explicitly, if we write an element $D$ seen as generating a symmetry transformation   in the left decomposition $D=mv$, with $m \in G^*$ and $v \in G$, the different Poisson brackets read%later on $m\in G^*$ will be a (deformed) translation and $v\in G$ a rotation, the different Poisson brackets read
\beq
\poi{m_1,m_2}_{\pi_-}=-[r,m_1m_2], \quad \poi{v_1,v_2}_{\pi_-}=-[r^t,v_1v_2], \quad \poi{m_1,v_2}_{\pi_-}=\poi{v_1,m_2}_{\pi_-}=0.
\eeq
If $G=\SU(2)$ or $\SO(3)$, $v$ is a rotation and $m$ is a translation in the case $G^*=\R^3$ or a deformed translation for $G^*=\SB(2, \C)$. From the previous commutation relations, we can see that in general the coordinates encoding the symmetry transformation will not Poisson commute. Upon quantization, they will then give rise to the notion of \textit{quantum group}. Indeed Poisson Lie groups are the classical version of a quantum group \cite{chari}.

%Hence the  symmetries are  equipped with a Poisson structure too, possibly trivial. More explicitly, the coordinates encoding the symmetry transformation, such as  the angles for example, might have some non-trivial Poisson bracket. %In the following, we shall focus on infinitesimal symmetries parameterized by $\vec \veps$, hence these coordinates $\vec \veps$ might have a non-trivial Poisson bracket. We shall not need to determine them for our purposes, so we do not dwell further on this.
%\beq
%\poi{\ell_1,\ell_2} = [r,\ell_1\ell_2], \quad \poi{\ell_1,u_2}=0, \quad \poi{u_1,u_2}= [r,u1_u_2].
%\eeq
%Note  that   these non-trivial Poisson brackets on $G$ and $G^*$ seen as group of symmetries  are the classical analogue of the non-commutative structure for the associated  quantum group.
%\beq
%\poi{D,D}_-= r D\ot D - D\ot D r
%\eeq
%In the left decomposition $\cD=G^*\bowtie G$, with $D=mv$ we have
%\beq
%\poi{m_1,m_2}_-=-\com{r, m_1m_2},
%& \quad \poi{m_1,v_2}_-=%\ell_1 r u_2, \quad
%\poi{v_1,m_2}_-=%\ell_2 r^t u_1,
%& \quad \poi{v_1,v_2}_-=-\com{r^t,u_1u_2}.
%\eeq

%\medskip
%In the previous section, we have dealt with two different phase spaces, $T^*\SU(2)\sim \ISO(3)$ and $\SL(2,\C)$. The previous theorem implies that the symmetries of these phase spaces are given by the Drinfeld double, that is in the respective cases $\cD_- = \ISO(3), \,\SL(2,\C)$ equipped with the relevant Poisson structure which makes them Poisson Lie groups.
\medskip

To have the explicit action of the rotations $v$ and the (deformed) translations $m$ on the phase space variables $\ell$ and $u$,  we split the action of the Drinfeld double $\cD_{\pi_-}$  on the Heisenberg double $\cD_{\pi_+}$ into the action of its subgroups, $G$ and $G^*$ on $\cD_{\pi_+}$. The left or right actions of $G$ and $G^*$ by multiplication on the Heisenberg double $\cD_{\pi_+}$ are given below where we have chosen the left decomposition of $D=\ell u \in \cD_{\pi_+}$. \\ %This explicit action will  of course depend on the factorization of $\cD_+$, which we pick to be the left decomposition.  \\%The action on the right decomposition can be o
%Given an element  $u\in G$  (resp. in $m\in G^*$), and $D\in\cD$, we shall consider the left and right action as given by the left or right multiplication. For example the left action of $G$ and $G^*$  on $\cD$ is
%\beq
%D\dr uD, \quad D\dr m D.
%\eeq
%In each case,  $\cD=\ISO(3), \SL(2,\C)$, we call  $G^*=\R^3, \SB(2,\C)$ the translation group, $G=\SO(3), \SU(2)$ being the rotation group.  Let us evaluate these actions on the configuration and momentum variables  in more details, according to the chosen factorization.
The left and right  (deformed) translations are given by the action of $G^*$ on $\cD$.
\beq\label{translation}
D\dr m\,D= m\, \ell u\,=  \,^{(m)} \ell\;  \,^{(m)} u \dr \left\{
\begin{array}{l}
\,^{(m)}  u= u \\
\,^{(m)} \ell = m \ell
\end{array}
\right.\,, \qquad D\dr D\, m = \ell u\, m = \ell ^{(m)} \; u^{(m)} \dr \left\{
\begin{array}{l}
\,^{(u)}m\, u^{(m)} = m\,u \\
\ell ^{(m)} = \ell \left(\,^{(u)}m\right)
\end{array}\right. \,.
\eeq
Note that to keep track on the side of the action, we use the (standard) notation $ \ell^{(m)} $ and $ u^{(m)}$  for the right action of $m$, as opposed to  $\,^{(m)} \ell$ and $\,^{(m)} u$
 for the left action. We use a similar notation   for the rotation $v$.  \\
The left and right  rotations are given by the action of $G$ on $\cD$.
\beq \label{rotation}
D\dr v\, D= v \, \ell u= \,^{(v)} \ell\; \,^{(v)} u \dr \left\{
\begin{array}{l}
\,^{(v)} u = v^{(\ell)} \,u \\
\,^{(v)} \ell \, v^{(\ell)}  =v  \ell
\end{array}
\right.\, , \qquad D\dr D\, v=\ell u\, v= \ell ^{(v)} \; u^{(v)} \dr \left\{
\begin{array}{l}
u^{(v)} = u\,v \\
\ell ^{(v)} = \ell
\end{array}
\right.\,.
\eeq
In the case $\cD=\ISO(3)$,  these transformations take a well known form. Indeed, if we denote  $m=(\textbf{1},{\bf a}),\, v=(g,{\bf 0}),\, \ell = (\one, \bX), \,  u= ({\bf R},{\bf 0})$, we recover that
\be
\begin{array}{llll}
\textrm{left translations: }&
\left\{
\begin{array}{l}
\,^{(m)}  u= u \\
\,^{(m)} \ell = m \ell \leadsto \bX\dr \bX+  {\bf a}
\end{array}
\right.
&
\textrm{right translations: }&
 \left\{  \begin{array}{l}
^{(u)}m\, u^{(m)} = u\,m  \leadsto \left\{ \tabl{l}{^{(u)}m= umu\mone, \\  u^{(m)}=u} \right. \\
\ell ^{(m)} = \ell \left(\,^{(u)}m\right)   \leadsto \bX\dr  \bX+ {\bf R}\rhd {\bf a}
\end{array}\right.
\\ &&&\\
\textrm{left rotations: } & \left\{ \begin{array}{l}
 \,^{(v)}u =v\, u\,,\,  \,^{(\ell)}v\, = v,     \\
 \,^{(v)}\ell = v \,\ell\, v\mone \leadsto \bX \dr g \rhd \bX \\
 \end{array}\right.
  & \textrm{right rotations:}  & \left\{ \begin{array}{l}
  u\,^{(v)} =u\, v  \\
\ell\,^{(v)}  = \ell\\
 \end{array}\right.
\end{array}
\ee
The interpretation of the $\SL(2,\C)$ case  is more complicated but some analogues of the above equations can be found in the appendix \ref{app:symmetries}.

In order to relate the constraints introduced in the next section to the above Poisson Lie symmetries, we express the infinitesimal  version of the  Poisson Lie symmetries \eqref{translation} and  \eqref{rotation} in terms of Poisson brackets on the \textit{Heisenberg double}. Indeed, later on we will build a lattice gauge theory with constraints which are functions over several copies of the $\SL(2,\C)$ phase space. We will have to show that the Poisson brackets from the symplectic structure $\poi{,}_{\pi_+}$ with the constraints generate gauge symmetries. Due to the different representations we have been using in the $\ISO(3)$ and $\SL(2,\C)$ cases (vector versus spinor representations), the formulae will not look exactly the same.

The proofs of the following propositions consists on one hand in determining the infinitesimal version of \eqref{translation} and  \eqref{rotation} and on the other hand calculating explicitly the Poisson brackets we will give to get a match. Most of these details are found in the appendix \ref{app:symmetries}.

In the following, we consider $m\sim\one +\delta m$ and $v\sim\one +i\vec \veps \cdot \vec J$, where $\vec J$ are the $\su(2)$ generators in the relevant representation.

\begin{prop} \label{translation ISO} \textbf{Translations for }$\ISO(3)$\\
 For $\cD_+= \ISO(3)$, we consider $\delta m = \vec \veps \cdot \vec E$ and we recall that we have the duality $\la E_i;J_j\ra=\delta_{ij}$. The infinitesimal right  translation  \eqref{translation} acting on $\ell$ and $u$ can then be rewritten as
 \beq
 \delta^{(m)}_R u \,=\,
\la \delta m_1;  u_1^{-1}\{u_1,u_2\}\, \ra_1 =0\,, \qquad
\delta^{(m)}_R \ell \,=\,
\la \delta m_1;  u_1^{-1}\{u_1,\ell_2\} \ra_1 = \ell\, u\, \delta m\, u\mone
 \eeq
The infinitesimal left  translation  acting on $\ell$ and $u$ can be rewritten as
  \beq
 \delta^{(m)}_L u  \, =  \la \delta m_1;  \{\tilde u_1,u_2\}\,  \tilde u_1^{-1}\ra_1\, = 0
%\tr_1 \delta M_1  u_1^{-1}\{u_1,u_2\}\,
\qquad
\delta^{(m)}_L \ell \,=\,  \la \delta m_1;  \{\tilde u_1,\ell_2\}\,  \tilde u_1^{-1}\ra_1\, = \delta m \, \ell
 %\tr_1 \delta M_1  u_1^{-1}\{u_1,\ell_2\}
 \eeq
\end{prop}
In the $\SB(2,\C)$ case, $\delta m$ cannot not contain the full information, as we have seen in the previous section when we used $\ell$ and $ \ell^\dagger$ to build a vector from $L=\ell \ell^\dagger$ . So we need to consider $\delta m^\dagger$ as well and to this goal we introduce $M=m \, m^\dagger$ and $\delta M = \delta m + \delta m^\dagger$. The infinitesimal deformed translations for the $\SL(2,\C)$ case are given by the following proposition.
\begin{prop} \label{translation SL}    \textbf{Deformed translations for }$\SL(2,\C)$\\
%For $\cD_+= \SL(2,\C)$,
We use $\tr_1$ for the trace on the first factor of the tensor product. The infinitesimal right and left translation acting on $\ell$ and $u$ can be rewritten as
\bes
&& \delta^{(m)}_R u \,=\,-i
\ka\mone \tr_1 \delta M_1 u_1^{-1}\{u_1,u_2\}\,  = \f14\,u\,[\delta M, \sigma_z-u^{-1}\sigma_z u] \,, \nn \\
&&\delta^{(m)}_R \ell \,=\,-i \ka\mone
\tr_1 \delta M_1  u_1^{-1}\{u_1,\ell_2\}
\,=\,
 \f12  \ell\left(
u\, \delta M\, u^{-1} +\f12[u\, \delta M\, u^{-1},\sigma_z]\right)\,.
 \ees
The infinitesimal left  translation acting on $\ell$ and $u$ can be rewritten as
  \bes
 \delta^{(m)}_L u  &=&  -\f{i}{4\ka}  \tr_1\delta M_1  \{\tilde u_1,u_2\}\,  \tilde u_1^{-1}, \,=\,  - \f{i}{4\ka}\tr_1\delta M_1  0\,=\, 0,
\\
\delta^{(m)}_L \ell &=& - \f{i}{4\ka} \tr_1 \delta M_1  \{\tilde u_1,\ell_2\}\,  \tilde u_1^{-1}
%\,=\, \la \delta M_1; r^\dagger \tu_1\ell_2\,  \tilde u_1^{-1}\ra_1
\,=\,\f{i}{4\ka} \sum_i \tr (\delta M  \tau_i ^\dagger)\sigma_i \ell \, = \, \delta m \, \ell,
 \ees

\end{prop}
We now address the rotations. As expected, they are generated by the variables $\ell$ and $\tell$ in the $\ISO(3)$ case.
\begin{prop} \label{rotation ISO} \textbf{Rotations for }$\ISO(3)$\\
\label{su2_1edge-iso3}
 For $\cD_+= \ISO(3)$, consider a $\SU(2)$ group element $v\sim \one + i\vec{\eps}\cdot\vec J = \one +V$  where $\vec J$ is  in the vector representation.
Then the variations of $\ell$ and $u$ under a left infinitesimal $\SU(2)$ transformations are given by
\be
\delta^{v}_L\ell =-\la V_1;\ell_1\mone \poi{\ell_1, \ell_2}\ra_1= V\ell - \ell V
\qquad
\delta^{v}_L u = -\la V_1;\ell_1\mone \poi{\ell_1, u_2}\ra_1= V\, u
\ee
The  variations of $\ell$ and $u$ under a right infinitesimal $\SU(2)$ transformations are given instead  by the Poisson brackets with the Hermitian matrix $\tell$.
\be
\delta^{v}_R\ell = \la V_1;\tell_1\mone \poi{\tell_1, \ell_2}\ra_1  =0,
\qquad
\delta^{v}_R u =  \la V_1;\tell_1\mone \poi{\tell_1, u_2}\ra_1= u V
\ee
\end{prop}
In the $\SB(2,\C)$ case, $\ell$ does not contain the full information about the flux, we need to consider also $\ell^\dagger$. Hence, we can use $L=\ell \ell^\dagger$ and $\tilde L^{op} = \tell^\dagger\tell$ to generate the left and right rotations.
\begin{prop}\label{rotation SL} \textbf{Rotations for }$\SL(2,\C)$\\
\label{su2_1edge-sl2c}
 For $\cD_+= \SL(2,\C)$, consider a $\SU(2)$ group element $v=\one + i\vec{\eps}\cdot\vsigma$  and the matrix $V$ %infinitesimally close to the identity, we parametrize it as follows:
\be
v=\one+i(V-\f12\tr V \one),
%\qquad\textrm{with}
\quad
V=\mat{cc}{2\eps_z & \eps_- \\ \eps_+ & 0}
=\eps_z(\one+\sigma_z)+\eps_-\sigma_++\eps_+\sigma_-\,.
\ee
Then the variations of $\ell$ and $u$ under a left infinitesimal $\SU(2)$ transformations are given by the Poisson brackets with the Hermitian matrix $L=\ell\ell^\dagger$.
\be
\delta^{v}_L\ell = -\f{\lambda^{-2}}{\ka}\,\{\tr VL,\ell\},
\qquad
\delta^{v}_L u =-\f{\lambda^{-2}}{\ka}\,\{\tr VL,u\}\,,
\ee
The generators of the rotations $\tr VL$  can be expanded explicitly as:
$$
\ka^{-1}\tr VL
\,=\,
\ka^{-1}(2\eps_z\lambda^2+\eps_-\lambda z+\eps_+\lambda \bz)
\,=\,
(2\eps_z(T_0+T_z)+\eps_-T_++\eps_+T_-)\,.
$$
The  variations of $\ell$ and $u$ under a right infinitesimal $\SU(2)$ transformations are given instead  by the Poisson brackets with the Hermitian matrix $\tL^{op}=\tell^\dagger\tell$.
\be
\delta_R^{(v)} u =  \f{\tlambda^{2}}\ka \poi{\tr V (\tL^{op})\mone, u}\, ,\quad
\delta_R^{(v)} \ell = \f{\tlambda^{2}}\ka \poi{\tr V (\tL^{op})\mone , \ell} = 0.
\ee
\end{prop}

%\begin{prop}
%For an observable $f$ on the phase space $\SL(2,\C)$, the vanishing of the Poisson brackets $\{\bT,f\}=0$ is equivalent to the vanishing of the Poisson brackets $\{\ell,f\}=0$ and $\{l,f\}=0$. Moreover, it automatically implies that $\{T_0,f\}=0$. These translate the invariance of $f$ under $\SU(2)$ transformations.
%\end{prop}
%\begin{proof}
%Since $2\ka T_\mu=\tr \ell\ell^\dagger\sigma_\mu$, one way is trivial. The reverse is proved by expanding the $T_i$'s explicitly in terms of $\lambda$, $z$ and $\bz$. First using $T_0^2=\bT^2+\ka^2$, we show that $\{\bT,f\}=0$ implies $\{T_0,f\}=0$. Then, we have:
%$$
%\{\lambda,f\}
%\,=\,
%\f1{4T_0}(2\lambda\{T_z,f\}+\bz\{T_+,f\}+z\{T_-,f\})\,.
%$$
%Then we conclude using
%$$
%\{z,f\}
%\,=\,
%\{\ka T_+,f\}-z\{\lambda,f\}\,,
%$$
%and similarly for $\bz$ using the bracket with $T_-$.
%
%\end{proof}

\section{Lattice with Heisenberg double and constraints}
\subsection{Graphs, ribbons and phase spaces}
\begin{figure}[h]
\centering
%\subfigure[A ribbon edge.
%]
{\includegraphics[scale=2]{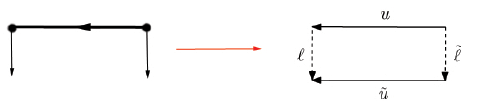}  %$\longrightarrow$
}
%\hskip1cm
%\subfigure[Part of a ribbon graph. There is a closed face, with holonomy $\tu_4^{-1} \tu_5^{-1} \tu_1^{-1} u_2 u_3$.]{
%\includegraphics[scale=.42]{RibbonFace.pdf}
%}
\caption{The ribbon is orientated such that   $\ell u = \tu \tell$. The solid lines are the strands and the dashed lines represent the thickened source and target points of each edge. Flipping the box is not allowed, only rotations in the plane are authorized. }\label{ribbon1}
\label{fig:RibbonEdge}
\end{figure}

In the standard LQG formalism, we associate to an edge of a graph the elements of the $T^*\SU(2)$
phase space, i.e. the holonomy $h$ and the fluxes $\bX$ and ${\tilde \bX}$ (related to one another by the holonomy). In the new formulation, we are going to associate to an edge an element $D$ of $\cD_+$. To take into account the fact that this group element can be decomposed into two ways, either $D=\ell u$ or $D=\tu \tell$, we are going to fatten the edge into a ribbon\footnote{
We would like to thank L. Freidel for sharing his idea regarding the possible use of the ribbon formalism in LQG.} with orientations on its boundary, as described in Fig. \ref{ribbon1}. This naturally encodes the fact that $\ell u = \tu \tell$ on each edge.

\begin{figure}[h]
\subfigure[]{\includegraphics[scale=.35]{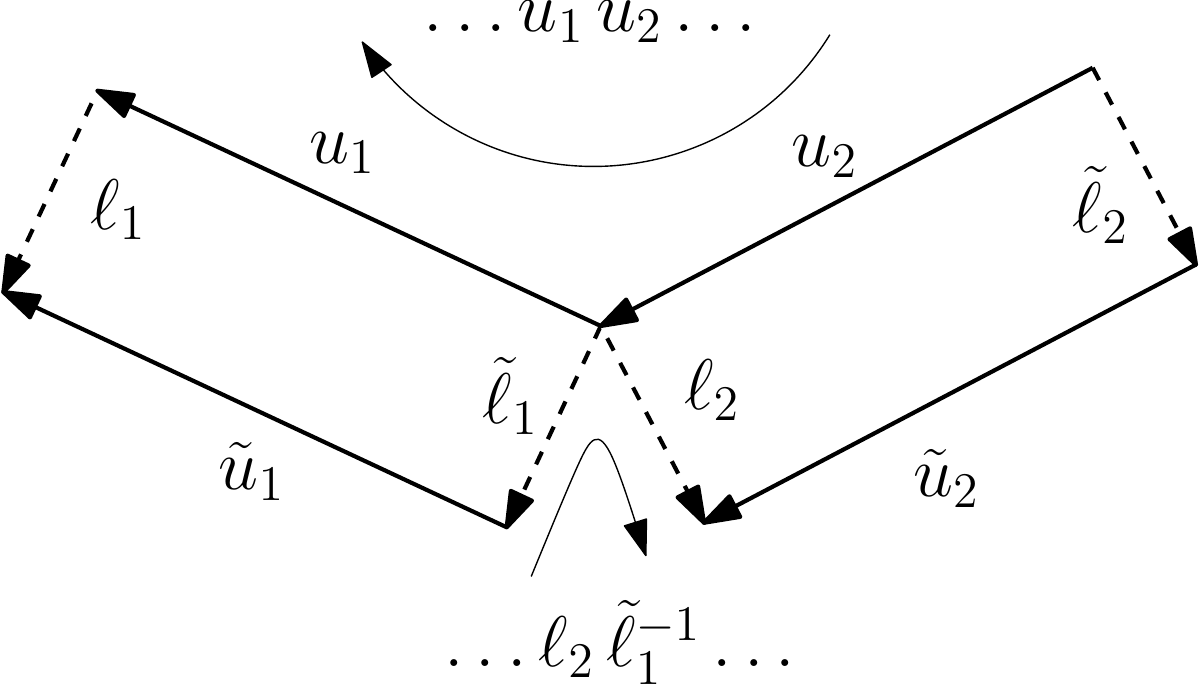} }
\subfigure[]{\includegraphics[scale=.35]{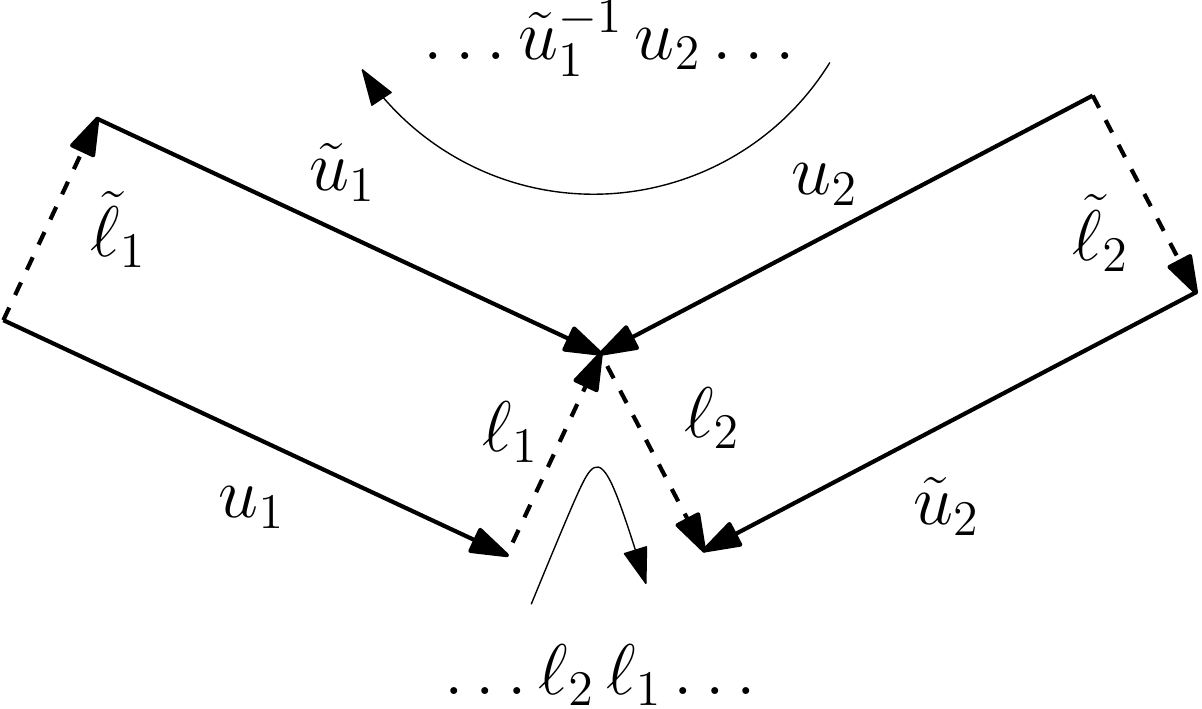} }
\subfigure[]{\includegraphics[scale=.35]{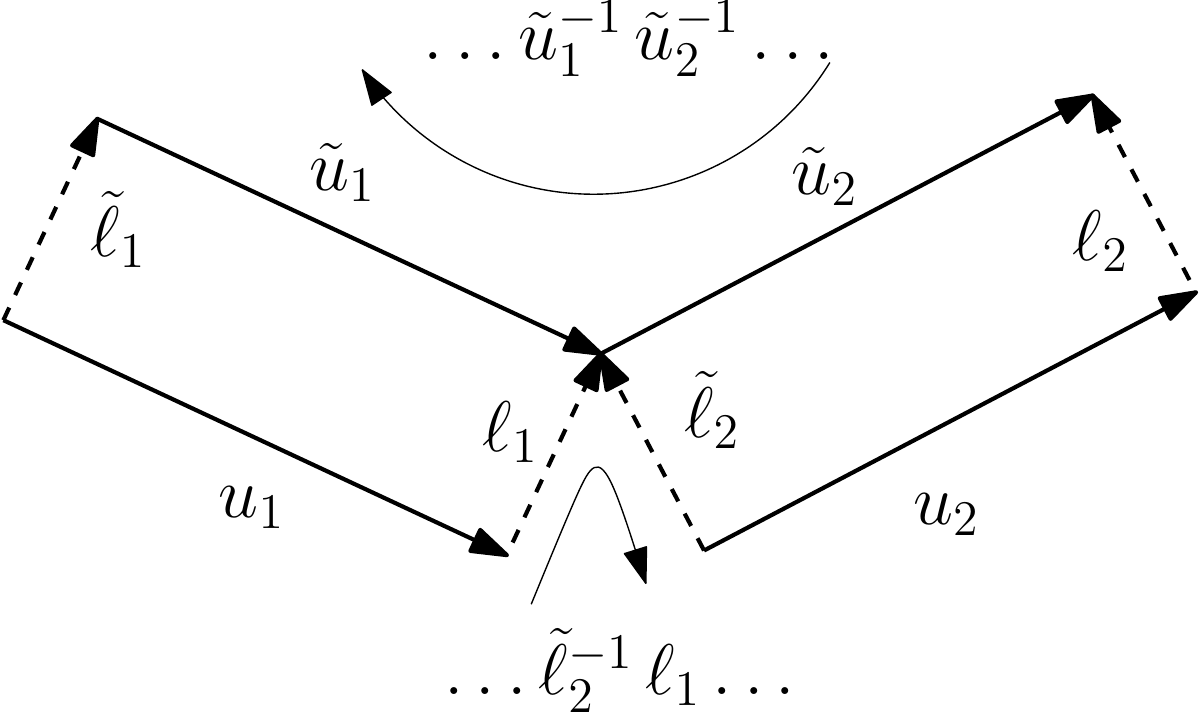} }
\subfigure[]{\includegraphics[scale=.35]{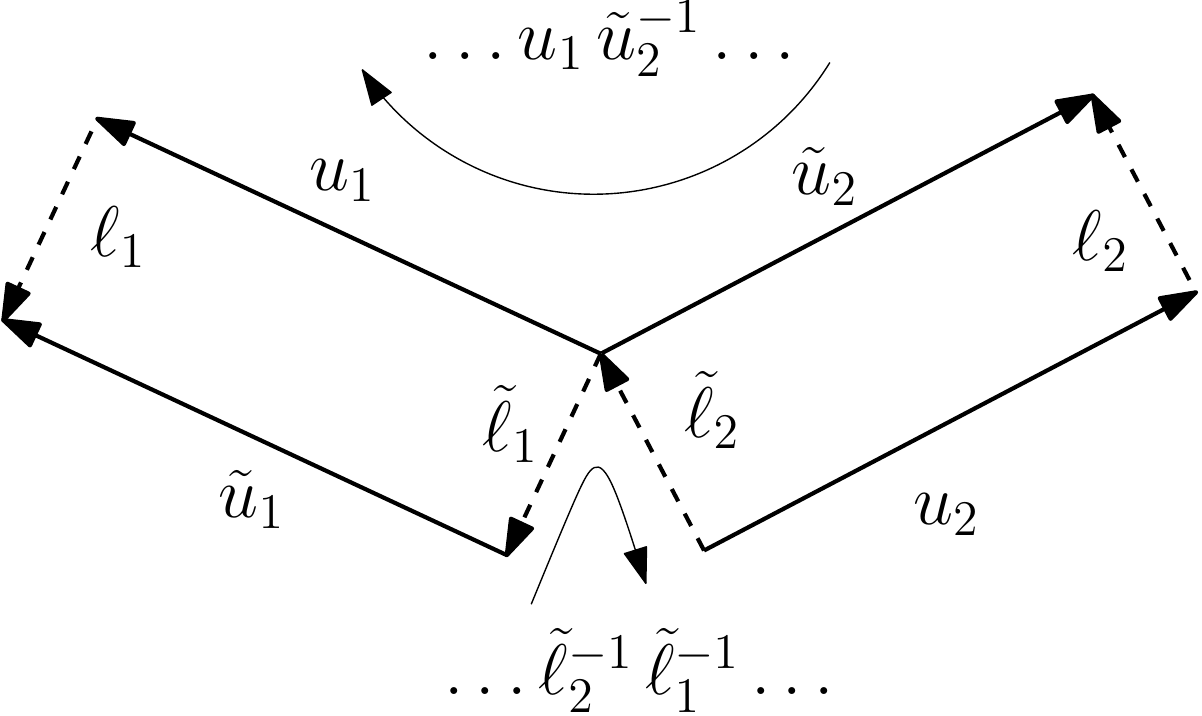} }
\caption{ The four possible ways to glue two ribbon edges together. The product of the $\SU(2)$ elements along the solid lines always only contains some $u$ and/or $\tu^{-1}$. The product of  elements along the dashed lines always only contains some $\ell$ and/or $\tell^{-1}$.}\label{fig:GluingTwoRibbonEdges}
\end{figure}

Note that since we deal with an orientable manifold, it is well known that a cell decomposition of a 2D surface can be represented as a ribbon graph. The 2D cell decomposition is therefore defined as a ribbon graph, with ribbon edges incident on ribbon vertices. Unlike the standard graph, we now have holonomies for both $\SU(2)$ and $\SB(2,\C)$. The $\SU(2)$ elements $u, \tu$ are on the side of the ribbon edges, while the $\SB(2,\C)$ elements $\ell, \tell$ lie where the ribbon edges are glued to the ribbon vertices. The boundaries of the ribbon vertices therefore carry "translations holonomies", i.e. $\ell$ and/or $\tell^{-1}$, as illustrated in Figure \ref{fig:GluingTwoRibbonEdges}.

With this formulation, we can easily generalize the construction from $T^*\SU(2)\sim \ISO(3)$ to $\SL(2,\C)$. This is not affecting the ribbon structure, the only thing that really changes is that $\ell$ is now an element of the non-Abelian group $\SB(2,\C)$.

%%%%%%%%%%
\subsection{The constraints}
%%%%%%%%%%

Now that we have detailed the Poisson structure on a single edge, we want to build a dynamics on a graph, made of first class constraints which generate the (infinitesimal) rotation  and translation transformations and which can be used to describe discrete hyperbolic geometries on the graph.

\begin{figure}[h]
\centering
%\subfigure[]{
\includegraphics[scale=.42]{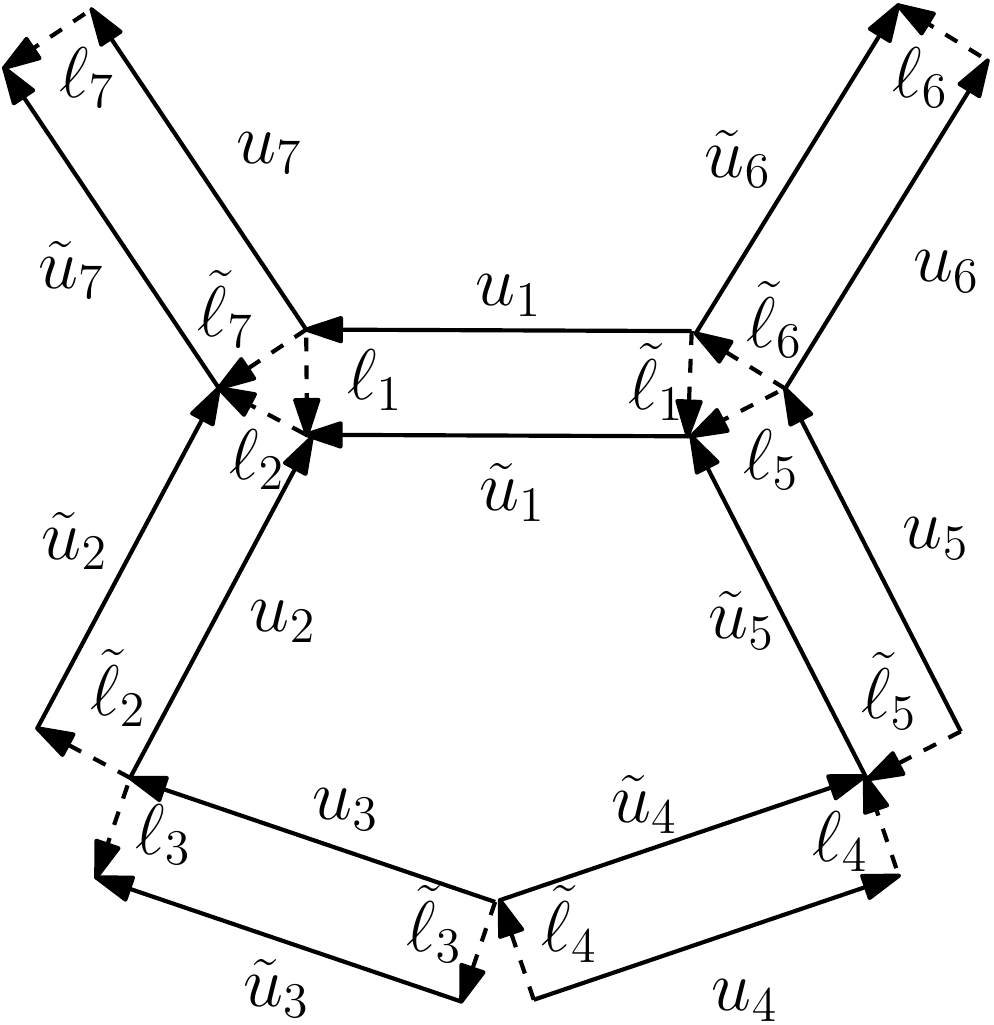}
\caption{Part of a ribbon graph. There is a closed face in bold lines, with holonomy $\tu_4^{-1} \tu_5^{-1} \tu_1^{-1} u_2 u_3$. There are two closed ribbon vertices with holonomies $\ell_1\ell_2\tell_7\mone$ and  $\ell_5\tell_1\mone \tell_6\mone$  in dashed lines.}\label{fig:RibbonFace}
\end{figure}

In the graph, there are two types of "faces": the faces of the cell decomposition, whose boundary edges are $u$ and $\tu$, and the ribbon vertices, whose boundary edges carry some $\ell$ and $\tell$, both represented in the Figure \ref{fig:RibbonFace}. Notice that they are four ways to glue two ribbon edges at a vertex as displayed in the Figure \ref{fig:GluingTwoRibbonEdges}.

As a consequence, the product of the $\SU(2)$ elements around a face only contains some $u$ and/or $\tu^{-1}$ (or some $u^{-1}$ and/or $\tu$ upon reversing the face orientation, but never some $u$ and $u^{-1}$ for instance). Similarly, the product of $\SB(2,\C)$ elements around a ribbon vertex only contains some $\ell$ and/or $\tell^{-1}$ (or some $\ell^{-1}$ and/or $\tell$ depending on the orientation, but never some $\ell$ and $\ell^{-1}$ for example).

In addition to the constraint $\ell u = \tu \tell$ on each edge, we propose the following set of constraints,
\begin{align}
\cG_v &\equiv \mathcal{L}_1 \dotsm \mathcal{L}_{N_v} = \one,& &\text{with $\mathcal{L}_i = \ell_i$ or $\tell^{-1}_i$},\label{gauss def}\\
\cC_f &\equiv \mathcal{U}_1 \dotsm \mathcal{U}_{N_f} = \one,& &\text{with $\mathcal{U}_i = u_i$ or $\tu^{-1}_i$},
\end{align}
where each $\mathcal{L}_i = \ell_i$ or $\tell^{-1}_i$ is the $\SB(2,\C)$ or $\R^3$ element on the leg $i =1,\dotsc,N_v$ around the ribbon vertex $v$ (while the order matters, the choice of the first edge does not), and $\mathcal{U}_i$ is the $\SU(2)$ element on the edge $i=1,\dotsc,N_f$ around the face $f$. For instance, on the top of the Figure \ref{fig:RibbonFace}, there are two closed ribbon vertices, one with $\tell^{-1}_7\ell_2\ell_1$ and the other with $\tell^{-1}_1 \ell_5 \tell^{-1}_6$, and there is one face with $\tu_4^{-1} \tu_5^{-1} \tu_1^{-1} u_2 u_3$.

We will call $\cG$ the {\em Gauss law} and $\cC$ the \textit{flatness constraint}. We now show that $\cG$ generates $\SU(2)$ transformations (hence the name) and that $\cC$ generates the (deformed) translations.

%%%%%%%%%%
\subsection{The Gauss law $\cG_v$ as generator of $\SU(2)$ transformations}
%%%%%%%%%%

Consider for simplicity a $N$-valent ribbon vertex whose incident edges are oriented inward. The Gauss law then reads
\be
\cG = \ell_1\,\dotsm \,\ell_N.
\ee
In the $\ISO(3)$ case, where $\ell=(\one,\bf{X})$, this product of matrices leads to the usual constraint of vanishing total angular momentum,
\beq
\cG = \one \Leftrightarrow \sum  \bX_i = {\bf 0}.
\eeq
Hence we recover the usual the Gauss constraint and we know that it generates the local (at each vertex) $\SU(2)$ transformations.
\medskip

Let us consider the new case, when $\ell_i\in\SB(2,\C)$. The product $\prod_i \ell_i$ of the triangular matrices $\ell_i$ is easy to perform,
\be
\cG = \ell_1\dotsm \ell_N = \begin{pmatrix} \prod_{i=1}^N \lambda_i &0\\ \sum_{i=1}^N \bigl[\prod_{j=1}^{i-1} \lambda_j^{-1}\bigr] z_i \bigl[\prod_{k=i+1}^{N} \lambda_k\bigr] & \prod_{i=1}^N \lambda_i^{-1} \end{pmatrix},
\ee
and therefore the constraints are
\beq
\left.\begin{array}{c}\cG=\ell_1\dotsm\ell_N = \one\\ \cG^{-1\dagger}=l_1\dotsm l_N= \one \end{array}\right\} \quad \Leftrightarrow \quad \left\{\begin{array}{c} \prod_{i=1}^N \lambda_i = 1\\  \sum_{i=1}^N \bigl[\prod_{j=1}^{i-1} \lambda_j^{-1}\bigr] z_i \bigl[\prod_{k=î+1}^N \lambda_k\bigr]=0\\ \sum_{i=1}^N \bigl[\prod_{j=1}^{i-1} \lambda_j^{-1}\bigr] \overline{z}_i \bigl[\prod_{k=î+1}^N \lambda_k\bigr]=0. \end{array}\right.
\eeq
The brackets between the matrix elements of $\cG$ and an arbitrary function $f$ are
\begin{align}
\{ \cG_{11},f\} &= \left\{ \prod_{i=1}^N \lambda_i, f\right\} = \prod_{i=1}^N \lambda_i \sum_{k=1}^N \lambda_k^{-1}\,\{\lambda_k,f\}\\
\{ \cG_{22},f\} &= \left\{ \prod_{i=1}^N \lambda_i^{-1}, f\right\} = -\prod_{i=1}^N \lambda_i^{-1} \sum_{k=1}^N \lambda_k^{-1}\,\{\lambda_k,f\} = - \prod_{i=1}^N \lambda_i^2\,\{\cG_{11},f\}\\
\{ \cG_{21},f\} &= \prod_{j=1}^N \lambda_j^{-1} \sum_{i=1}^N \left\{ \lambda_i z_i \prod_{k=i+1}^N \lambda_k^2, f \right\} + \sum_{i=1}^N \lambda_i z_i \prod_{k=i+1}^N \lambda_k^2 \left\{ \prod_{j=1}^N \lambda_j^{-1}, f \right\},
\end{align}
and similarly with $\cG^\dagger$. It is convenient to express those brackets in terms of the transformation generated by $\cG\cG^\dagger$,
\be \label{su2Constraint}
\prod_{k=1}^N \lambda_k^{-2} \left\{ \tr V \cG\cG^\dagger, f\right\} \equiv \epsilon_z \delta_z f +\epsilon_- \delta_- f +\epsilon_+ \delta_+ f,
\ee
where $V = \left(\begin{smallmatrix} 2\epsilon_z & \epsilon_-\\ \epsilon_+ &0 \end{smallmatrix}\right)$. A direct calculation leads to
\be
\tr V \cG\cG^\dagger = 2\epsilon_z \prod_{i=1}^N \lambda_i^2 +\epsilon_- \sum_{i=1}^N \lambda_i z_i \prod_{j=i+1}^N \lambda_j^2 + \epsilon_+ \sum_{i=1}^N \lambda_i \bar{z}_i \prod_{j=i+1}^N \lambda_j^2,
\ee
therefore
\begin{align}
\delta_z f &= \prod_{k=1}^N \lambda_k^{-2} \left\{ 2\prod_{i=1}^N \lambda_i^2,f \right\} = 2 \sum_{i=1}^N \lambda_i^{-2} \{\lambda_i^2,f\} = 4 \sum_{i=1}^N \lambda_i^{-1} \{\lambda_i, f\}, \label{deltazf} \\
\delta_-f &= \prod_{k=1}^N \lambda_k^{-2} \left\{\sum_{i=1}^N \lambda_i z_i \prod_{j=i+1}^N \lambda_j^2, f\right\},\qquad
\delta_+f = \prod_{k=1}^N \lambda_k^{-2} \left\{\sum_{i=1}^N \lambda_i \bar{z}_i \prod_{j=i+1}^N \lambda_j^2, f\right\}. \label{delta-f}
\end{align}
We get
\begin{align}
\{\cG_{11}, f\} &= \frac{1}{4} \left[\prod_{k=1}^N \lambda_k \right]\,\delta_zf = \frac{1}{4} \cG_{11}\,\delta_zf\\
\{\cG_{21}, f\} &= \left[\prod_{k=1}^N \lambda_k\right] \delta_-f -\frac{1}{4} \left[\sum_{i=1}^N \prod_{k=1}^{i-1} \lambda_k^{-1} z_i \prod_{j=i+1}^N \lambda_j\right]\ \delta_zf = \cG_{11} \delta_-f - \frac{1}{4}\, \cG_{21}\,\delta_zf.
\end{align}

Notice that in the simple case $N=1$, the bracket \eqref{su2Constraint} generates the $\SU(2)$ transformations on the phase space $\SL(2,\C)$. In order to prove that $\cG, \cG^\dagger$ generates $\SU(2)$ rotations on $N>1$ legs, we have to match the brackets with some linear combination of brackets with the generators $\ell_k \ell^\dagger_k$ of $\SU(2)$ transformations on the leg $k$.

\begin{prop} The Gauss constraint generates $\SU(2)$ transformations with braided parameter $V^{(k-1)}$ on the leg $k=1,\dotsc,N$, i.e.
\be
\prod_{k=1}^N \lambda_k^{-2} \left\{ \tr V \cG\cG^\dagger, f\right\} = \sum_{k=1}^N \lambda_k^{-2}\,\tr V^{(k-1)}\,\{ \ell_k\ell_k^\dagger, f\},
\ee
where $V^{(k)}$ is defined by induction,  $V^{(k)} = \lambda_{k}^{-2} \ell_k^\dagger V^{(k-1)} \ell_k$, and $V^{(0)} = V = \left(\begin{smallmatrix} 2\epsilon_z &\epsilon_- \\ \epsilon_+ & 0\end{smallmatrix} \right)$.
\end{prop}

{\bf Proof.} The braided parameter is found to be
\be
V^{(k)} = \bigl[\prod_{i=1}^k \lambda_i^{-2}\bigr] \ell_k^\dagger \dotsm \ell_1^\dagger V \ell_1\dotsm \ell_k = \begin{pmatrix} 2 \epsilon_z^{(k)} & \epsilon_-^{(k)} \\ \epsilon_+^{(k)} & 0 \end{pmatrix},
\text{ with}\qquad \left\{ \begin{split} \epsilon_z^{(k)} &= \epsilon_z +\frac{1}{2}\sum_{i=1}^k \bigl[\prod_{j=1}^i \lambda_j^{-2}\bigr] \bigl(\epsilon_- \lambda_i z_i +\epsilon_+ \lambda_i \bar{z}_i\bigr)\\
\epsilon_{\pm}^{(k)} &= \bigl[\prod_{i=1}^k \lambda_i^{-2}\bigr]\, \epsilon_\pm
\end{split} \right.
\ee
It can be noted that $V^{(N)} = V$ on-shell, i.e. when $\cG=\cG^\dagger=\one$. First consider the $\SU(2)$ transformations along $z$,
\be
\sum_{k=1}^N \lambda_k^{-2}\, \tr V^{(k-1)}\,\{ \ell_k\ell_k^\dagger, f\}_{|\epsilon_\pm=0} = 2\epsilon_z \sum_{k=1}^N \lambda_k^{-2}\, \{\lambda_k^2,f\} = \epsilon_z \delta_z f,
\ee
by direct comparison with \eqref{deltazf}. Next we set $\epsilon_z=\epsilon_+=0$ and focus on the variations with $\epsilon_-$,
\be
\sum_{k=1}^N \lambda_k^{-2}\, \tr V^{(k-1)}\,\{ \ell_k\ell_k^\dagger, f\}_{|\epsilon_z=\epsilon_+=0} = \epsilon_- \sum_{k=2}^N\lambda_k^{-2} \sum_{i=1}^{k-1} \bigl[\prod_{j=1}^{i} \lambda_j^{-2}\bigr] \lambda_i z_i\,\{\lambda_k^2,f\} + \epsilon_- \sum_{k=1}^N\lambda_k^{-2}  \bigl[\prod_{j=1}^{k-1} \lambda_j^{-2}\bigr]\,\{\lambda_k z_k, f\}.
\ee
The first term of the right hand side is
\be
\begin{aligned}
\epsilon_- \sum_{k=2}^N\lambda_k^{-2} \sum_{i=1}^{k-1} \bigl[\prod_{j=1}^{i} \lambda_j^{-2}\bigr] \lambda_i z_i\,\{\lambda_k^2,f\}
&= \epsilon_- \prod_{j=1}^N \lambda_j^{-2} \sum_{i=1}^{N-1} \lambda_i z_i \bigl[\prod_{p=i+1}^N \lambda_p^2\bigr] \sum_{k=i+1}^N \lambda_k^{-2} \{\lambda_k^2,f\}\\
&= \epsilon_-\prod_{j=1}^N \lambda_j^{-2} \sum_{i=1}^{N} \lambda_i z_i \, \left\{ \prod_{k=i+1}^N \lambda_k^2, f\right\},
\end{aligned}
\ee
where the empty product is 1. The second term writes
\be
\epsilon_- \sum_{k=1}^N\lambda_k^{-2}  \bigl[\prod_{j=1}^{k-1} \lambda_j^{-2}\bigr]\,\{\lambda_k z_k, f\} = \epsilon_- \prod_{j=1}^N \lambda_j^{-2} \sum_{k=1}^N \Bigl[\prod_{p=k+1}^N \lambda_p^2\Bigr]\,\{\lambda_k z_k, f\}.
\ee
Therefore we get
\be
\sum_{k=1}^N \lambda_k^{-2}\, \tr V^{(k-1)}\,\{ \ell_k\ell_k^\dagger, f\}_{|\epsilon_z=\epsilon_+=0} = \epsilon_- \Bigl[ \prod_{j=1}^N \lambda_j^{-2} \Bigr] \sum_{i=1}^N \left\{ \lambda_i z_i \prod_{p=i+1}^N \lambda_p^2, f\right\} = \epsilon_- \delta_-f,
\ee
in agreement with \eqref{delta-f}. The variation with $\epsilon_+$ works similarly.

%%%%%%%%%%%%%%%%
\subsection{The flatness constraints as generator of the (deformed) translations}
%%%%%%%%%%%%%%%%

We consider for simplicity a face whose $N$ boundary edges have the same orientation as the face so that the flatness constraint reads
\be
\cC = u_N\,\dotsm\,u_1.
\ee
Again, in the $\ISO(3)$ case, we know that this constraint generates the Abelian translational symmetry. So we focus instead on the $\SL(2,\C)$ case.

Consider the transformation generated by the following bracket with $\cC$ on an arbitrary function $f$,
\be
\delta_M f \equiv \tr \delta M\, \cC^{-1} \left\{ \cC, f\right\}, \qquad \text{with $\delta M = \left(\begin{smallmatrix} 2\epsilon_3 &\epsilon_-\\ \epsilon_+ &-2\epsilon_3 \end{smallmatrix}\right)$}.
\ee
It is equivalent to a translation on all the edges around the face (the translation on a single edge is generated by $\tr \delta M\,u^{-1} \{u, \cdot\}$, as stated in the Proposition \ref{translation SL}).

\begin{prop}
The flatness constraint generates $\SB(2,\C)$ tranformations with braided parameter $\delta M^{(k-1)}$ on the edge $k=1,\dotsc,N$,
\be
\tr \delta M\,\cC^{-1} \left\{\cC, f\right\} = \sum_{k=1}^N \tr \delta M^{(k-1)}\,u_k^{-1} \left\{u_k, f\right\}.
\ee
where $\delta M^{(k-1)} = u_{k-1} \dotsm u_1 \delta M u_1^{-1} \dotsm u_{k-1}^{-1}$.
\end{prop}

{\bf Proof.} The calculation is straightforward,
\be
\begin{aligned}
\tr \delta M\,\cC^{-1} \left\{\cC, f\right\} &= \sum_{k=1}^N \tr u_{k-1} \dotsm u_1 \delta M\, \cC^{-1} u_N\dotsm u_{k+1} \{u_k,f\}\\
&= \sum_{k=1}^N \tr \Bigl[u_{k-1} \dotsm u_1 \delta M\,u_1^{-1} \dotsm u_{k-1}^{-1}\Bigr] u_k^{-1} \{u_k,f\} = \sum_{k=1}^N \tr \delta M^{(k-1)}\,u_k^{-1} \left\{u_k, f\right\}.
\end{aligned}
\ee

%%%%%%%%%%%%%%%%%
\subsection{A first class constraint algebra}
%%%%%%%%%%%%%%%%%
Interestingly, the proof that the constraints form a first class algebra will not depend on the choice of phase space, $\ISO(3)$ or $\SL(2\,C)$. Using  Section \ref{symmetries}, the left rotation
given by an element  $v\in\SU(2)$ acting at a ribbon vertex reads
\be \label{SU(2)OnAVertex}
\begin{aligned}
&\ell_1 u_1
&\quad\mapsto\quad&
v \ell_1 u_1 = (v \ell_1 (v^{(\ell_1)})^{-1})((v^{(\ell_1)})u_1 )= \,^{(v_1)}\ell_1\, ^{(v_1)}u_1\,, \textrm{ with } v_{1} \equiv  v^{(\ell_1)}\\
&\ell_2 u_2
&\quad\mapsto\quad&
(v^{(\ell_1)}) \ell_2 u_2 = ((v^{(\ell_1)})  \ell_2 (v^{(\ell_1\ell_2)}) ^{-1})((v^{(\ell_1\ell_2)})u_2 )=\,^{(v_{2})}\ell_2\,^{(v_{2})}u_2\,, \textrm{ with } v_{2} \equiv  v^{(\ell_1\ell_2)}\\
&\ell_N u_N
&\quad\mapsto\quad&
v_{N-1} \ell_N u_N = (v_{N-1} \ell_N v_N^{-1})(v_Nu_N )=\,^{(v_N)}\ell_N\,^{(v_N)}u_N\,, \textrm{ with } v_{N} \equiv  v^{(\ell_1\ell_2..\ell_N)}.
\end{aligned}
\ee
%where $v_1$ is the modified transformation $v$ transited through $u_1$, $v_2$ the modified transformation $v$ transited through $u_1$ and $u_2$ and so on.
%
Note that we have used the fact that there is a (possibly trivial) right action $\rhd$ of $G^*=\SB(2,\C), \R^3$ on $G=\SU(2)$, so that we have
\beq
v\lhd \ell_1 = v\ell_1 \equiv v^{(\ell_1)}, \quad v^{(\ell_1)}\, \ell_2=(v\lhd \ell_1)\lhd \ell_2 =  v\lhd (\ell_1 \ell_2) \equiv  v^{(\ell_1\ell_2)}, \quad v\in G, \, \ell_i\in G^*.
\eeq
Let us compute the transformation of the Gauss law under $v$ (again in the case the incident edges are oriented inward),
\be
\,^{(v_1)}\ell_1\dots \, ^{(v_N)}\ell_N
\,=\,
(v \ell_1 v_1^{-1})(v_1 \ell_2 v_2^{-1})\dots (v_{N-1} \ell_N v_N^{-1})
\,=\,
v\,\ell_1\dots\ell_N\,v_N^{-1}\,.
\ee
Thus when the Gauss law is satisfied, $\ell_1\dots\ell_N=\one$, we know that $v_N=v$ which implies that the transformed triangular matrices still satisfy the Gauss law, $\,^{(v_1)}\ell_1\dots \, ^{(v_N)}\ell_N
\,=\,\one$.

As the Gauss law generates $\SU(2)$ transformations, we expect the brackets of the Gauss law with itself to vanish on-shell. One can directly check that those constraints form a first class system,
\be
\{ \cG_{1}, \cG_{2}\}= -[r, \cG_{1}\cG_{2}],\qquad
\{ \cG_{1}^{-1\dagger}, \cG_{2}^{-1\dagger}\}= -[r^\dagger, \cG_{1}^{-1\dagger}\cG_{2}^{-1\dagger}],\qquad
\{ \cG_{1}, \cG_{2}^{-1\dagger}\}= -[r^\dagger, \cG_{1}\cG_{2}^{-1\dagger}].
\ee

Let us check the brackets in the case one edge is outgoing, and to simplify we consider a bivalent vertex (this generalizes straightforwardly, though quite unconveniently in terms of notations). We have $\cG = \ell_\alpha \tell_\beta^{-1}$ where $\alpha, \beta$ are the two edges. The bracket reads
\be
\begin{aligned}
\left\{\cG_1, \cG_2\right\} = \left\{ \ell_{\alpha1} \tell_{\beta1}^{-1}, \ell_{\alpha2} \tell_{\beta2}^{-1} \right\} &= -\ell_{\alpha1}\ell_{\alpha2} [r^\dagger, \tell_{\beta1}^{-1} \tell_{\beta2}^{-1}] - [r, \ell_{\alpha1}\ell_{\alpha2}] \tell_{\beta1}^{-1}\tell_{\beta2}^{-1}\\
&= -[r, \ell_{\alpha1}\tell_{\beta1}^{-1}\ \ell_{\alpha2}\tell_{\beta2}^{-1}]= -[r, \cG_{1}\cG_{2}].
\end{aligned}
\ee
We have used that $r-r^\dagger$ is the Casimir to go from the first to the second line.

Furthermore, the brackets between the Gauss constraints associated to two ribbon vertices $v, v'$ connected by an edge $e$ vanish. Indeed, if $\cG_v$ contains $\ell_e$, then $\cG_{v'}$ contains $\tell_e$ (or vice versa), and moreover $\{\ell, \tell\} = 0$. This way, we can conclude that the algebra generated by the Gauss law is first class, whatever the edge orientations are.
\medskip

Now we consider the bracket between the Gauss law at a vertex $v$ and the flatness constraint around a face $f$ such that $v$ is a vertex of $f$. There are several situations depending on the orientations of the two edges of $f$ that meet at $v$. Let us look at the case where the edge $\alpha$ is incoming at $v$ while the edge $\beta$ is outgoing. The only non-trivial part of the bracket has the following form
\be
\left\{ \ell_{\alpha1} \tell_{\beta 1}^{-1}, u_{\beta 2} u_{\alpha 2} \right\} = \ell_{\alpha1} \{\tell_{\beta1}^{-1} ,u_{\beta2}\} u_{\alpha2} + u_{\beta2} \{\ell_{\alpha1}, u_{\alpha2}\} \tell_{\beta1} = 0.
\ee
The fact that the bracket vanishes identically reflects the fact that the flatness constraint is $\SU(2)$ invariant. Indeed, if $D_\alpha$ transform on the left (at $v$) with $v\in\SU(2)$, then
\be
u_\alpha \mapsto v^{(\ell_1)} \, u_\alpha,
\ee
and $D_\beta$ transforms on the right (since the edge is oriented outward) with the modified parameter $(v^{(\ell_1)})\mone$ that has transited through $D_\alpha$. More generally we get
\be
\left\{ \cG_1, \cC_2\right\} =0.
\ee

The bracket of the flatness constraint with itself on the same face gives (in the case it only contains some $u$, $\cC = u_N \dotsm u_1$)
\be
\left\{\cC_1, \cC_2 \right\} = \left[ r^\dagger, \cC_1 \cC_2 \right].
\ee
The last case to check involves the two faces $f, f'$ shared by an edge. However, if $f$ contains the variable $u$ associated to that edge then $f'$ contains $\tu$ and not $u$ (or vice versa), and due to $\{u,\tu\} = 0$, the bracket between the two flatness constraint vanishes.

%%%%%%%%%%%%%%%%%%%%%%%%%%%%%%%%%%
\section{Geometric meaning of the  constraints}
%%%%%%%%%%%%%%%%%%%%%%%%%%%%%%%%%%

\subsection{Gauss constraint and cosine laws}
\subsubsection{$\ISO(3)$ case and the "flat" cosine law}
In the flat case, the   Gauss law we have introduced in \eqref{gauss def}, realized  on a 3-valent vertex with incident edges, boils down to
\beq
\ell_1\ell_2\ell_3 = \one \leadsto \bX_1+\bX_2+\bX_3= \vec 0.
\eeq
The 3-valent vertex is combinatorially dual to a triangle and $\bX_e $ is interpreted as the normal to the edge dual to $e$, in the plane spanned by the triangle. This provides the well known geometric interpretation  that the triangle dual to the vertex geometrically closes: this is the \textit{closure constraint}.

\begin{figure}[h]
\includegraphics[scale=.4]{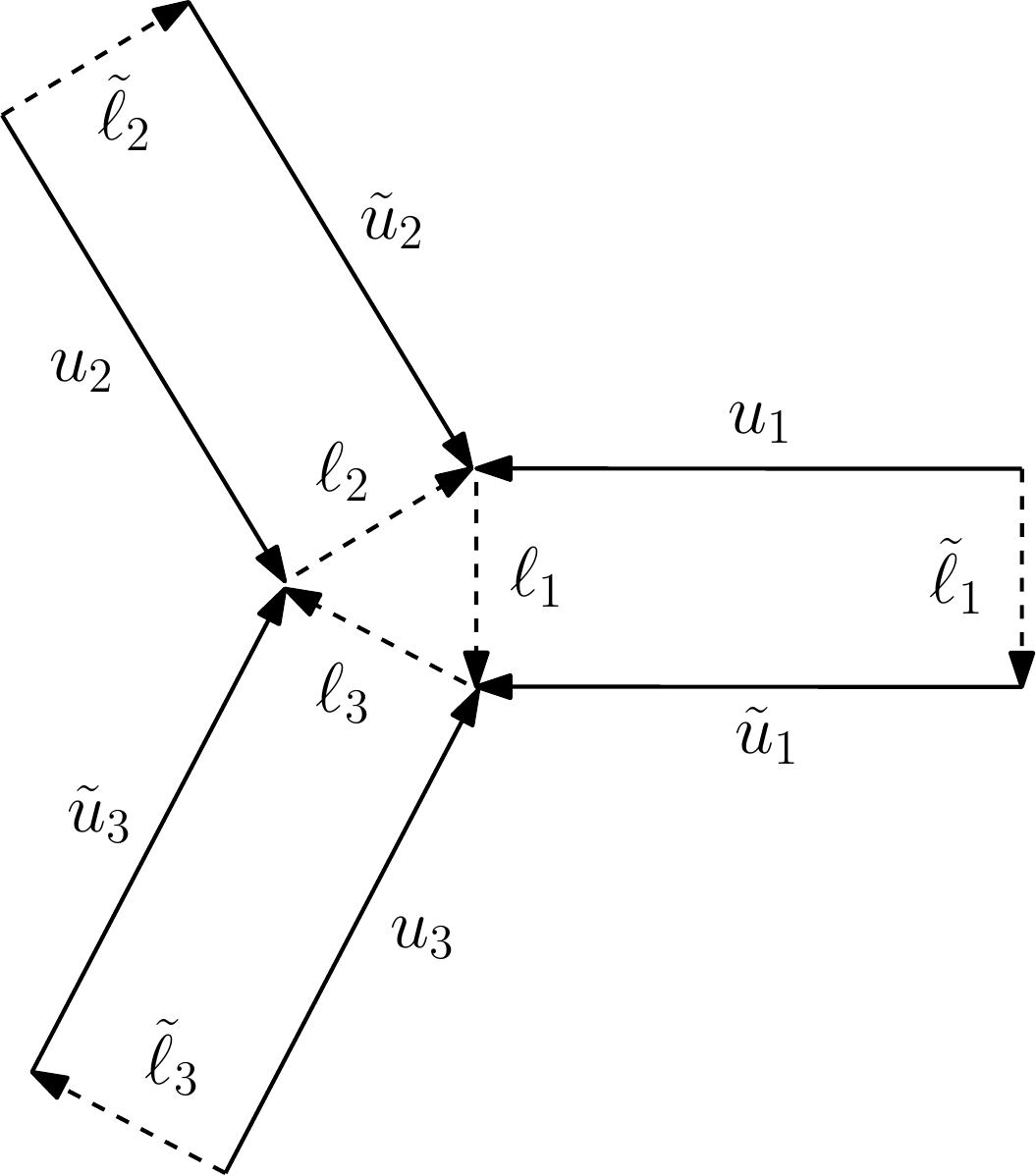}
\caption{A 3-valent ribbon vertex with its incident edges inward. The Gauss law is $\cG = \ell_1 \ell_2 \ell_3 = \one$.}\label{fig:3ValentVertex}
\end{figure}

Equivalently, the Gauss law corresponds to the three $\SU(2)$ invariant equations,
\be \label{ggauss}
\tr  \ell_j^\dagger \ell_i^\dagger \ell_i \ell_j = \tr  (\ell_k \ell_k^\dagger)\mone \, \Leftrightarrow\,  (\bX_i + \bX_j)^2 = \bX_k^2, \textrm{ for } i,j,k=1,2,3 \textrm{ all different.}
\ee
By expanding the square, it comes
\be
|\bX_i|^2+|\bX_j|^2 - 2\,|\bX_i|\,|\bX_j|\,\cos \phi_{ij} = |\bX_k|^2, \textrm{ with } \cos \phi_{ij} \equiv - \bX_i\cdot \bX_j/(|\bX_i| |\bX_j|).
\ee
These relations describe the angles $\phi_{ij}$ of the triangles in terms of the edge lengths $|\bX_e|$. We have just rederived the well known fact that the Gauss constraint encodes in the flat case the closure constraint that is equivalent to the three \textit{cosine laws}.

Trading $\ISO(3)$ for $\SL(2,\C)$ the expressions \eqref{ggauss} will generalize to the curved case. The difficulty is however to identify the analogue of the normal vectors $\bX_e$.

\subsubsection{$\SL(2,\C)$ case and the hyperbolic cosine law}

%%%%%%%%%%%%%%%%%%
%\subsubsection{Boosts and the Cartan decomposition}
%%%%%%%%%%%%%%%%%%

Beside the Iwasawa decomposition, the Cartan decomposition of $\SL(2,\C)$ is also available.  It will provide the key to identify the analogue of the normal vectors.  This decomposition  states that any element can be written as the product of a boost and a rotation. When dealing with the momentum variable $\ell$, we have
$$
\ell = B h\mone, h\in\SU(2),
\quad
B= e^{-\vsigma\cdot \vb}\,,
$$
where $B\in\SL(2,\C)$ is a pure boost in the Lorentz group, that is uniquely characterized as a 2$\times$2 Hermitian matrix satisfying $B=B^\dagger$ and $\det B=1$.
The main advantage is that the set of pure boosts is stable under conjugation by $\SU(2)$ group elements, unlike triangular matrices. Moreover, $B$ and $\ell$ define the same vector  $\bT=\tr\ell\ell^\dagger\vec{\sigma}/(2\kappa)$, since
\be\nonumber
L=DD^\dagger = \ell\ell^\dagger = BB^\dagger.
\ee

\medskip

It will also be useful to have the explicit formula for the boost $B=\cosh b \one -\sinh b\,(\hb\cdot\vsigma)$ in its spinorial form, with  rapidity $2b$ and boost direction $\hb$ expressed in terms of $\lambda$ and $z$:
\bes
&&B= \mat{cc}{\cosh(b)-\sinh(b)\, \hat{b}_z&- \sinh(b)\, \hat{b}_-\\ - \sinh(b)\, \hat{b}_+& \cosh(b)+\sinh(b)\, \hat{b}_z }
 \\
&& \textrm{with}
\quad
\cosh(2b)=\f{1+\lambda^2(\lambda^2+|z|^2)}{2\lambda^2},
\quad
\hb_+\sinh(2b)=-\lambda z, \quad \sinh(2b) \hb_z=\f{\lambda^{-2}+|z|^2-\lambda^2}{2}. \nn
\ees

Let us also remark that $hBh^{-1}$, which is still a pure boost and which enters the definition of $\bTop=\f{1}{2\kappa} \tr(\ell^\dagger \ell \vec{\sigma})=\f{1}{2\kappa} \tr(hB^{\dagger} Bh^{-1} \vec{\sigma})$, has the simple expression
\be
hBh^{-1}=\f{1}{\sqrt{2+2\ka T_0}}(\one + \ell^\dagger \ell)=\f{1}{\sqrt{2+2\ka T_0}}(\one + \Lop)
\ee

%\subsubsection{Hyperbolic cosine law}

We want to show now  that the Gauss law on a 3-valent vertex contains exactly the information to construct a \textit{hyperbolic} triangle.  Note that unlike  the flat case, a hyperbolic triangle $t$ is totally specified by its three angles or its three lengths.%, the equivalence being encoded in the hyperbolic cosine law as recalled in the appendix \ref{cosine law}.

\begin{figure}
\includegraphics[scale=.4]{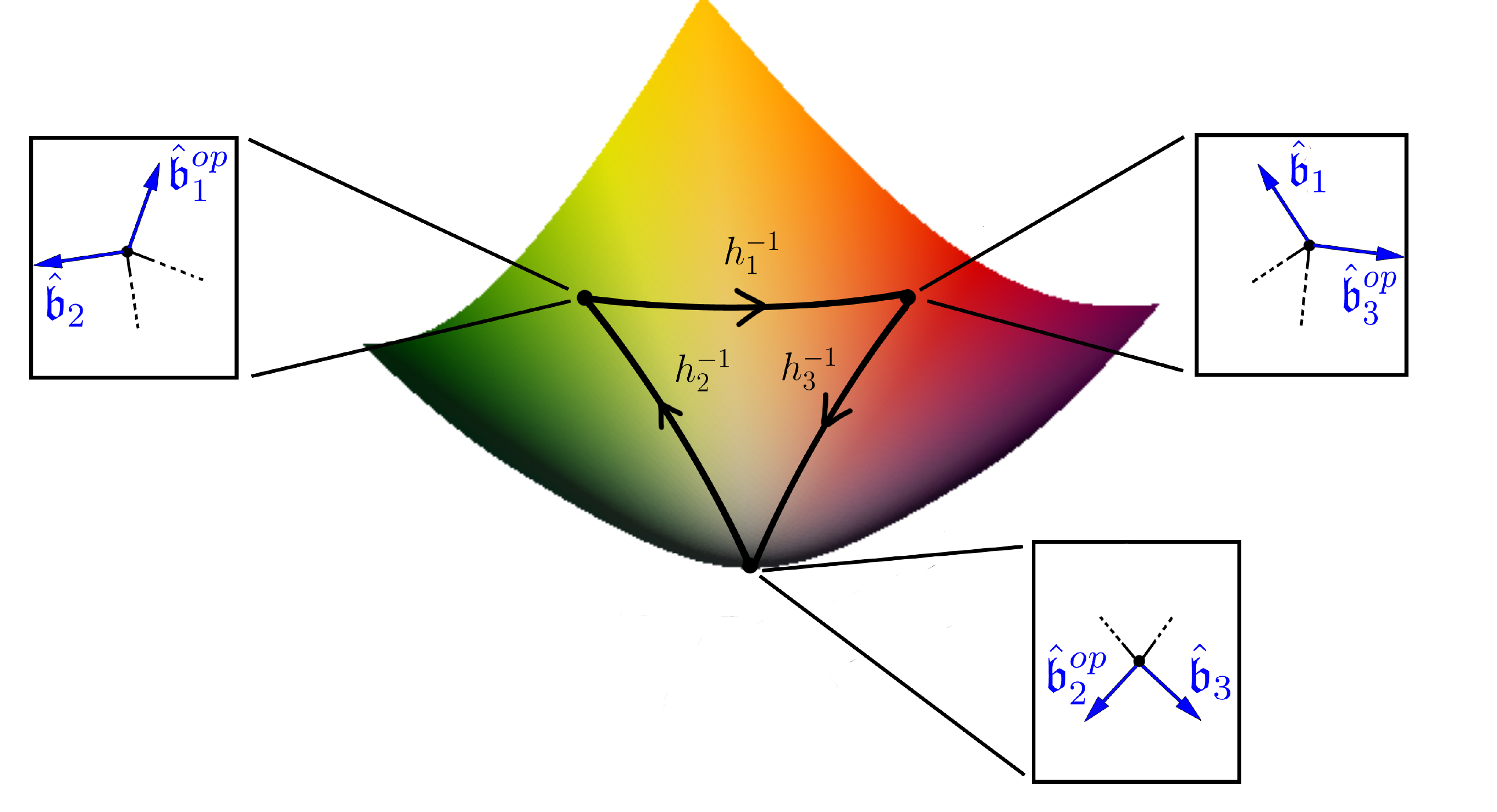}
\caption{This represents the way the vectors $\hfb_e, \hfb_e^{op}$ are assigned to the dashed lines of the ribbon vertex.}\label{fig:RibbonVertexZoom}
\end{figure}

We consider three incoming edges, so that $\cG = \ell_1 \ell_2 \ell_3$, as in the Figure \ref{fig:3ValentVertex}. We use the Cartan decomposition $\ell = B h^{-1}$, with $B$ a boost and $h$ a rotation, to get
\be
\ell_1 \ell_2 \ell_3 = \fB_1\,\fB_2\,\fB_3\,H^{-1} = \one,
\ee
with
\be\label{boost braided}
\fB_1 = B_1,\qquad \fB_2 = h_1^{-1}\, B_2\, h_1,\qquad \fB_3 = h_1^{-1} h_2^{-1}\, B_3\, h_2 h_1, \qquad H = h_3 h_2 h_1.
\ee
We take one $\fB_i$ to the right hand side, say $\fB_1 \fB_2 = H \fB_3^{-1}$, and multiply on the left by the adjoint equation to get rid of the rotation $H$, which gives
\beq
\ell_2^\dagger \ell_1^\dagger \ell_1 \ell_2 =   (\ell_3 \ell_3^\dagger)\mone  \Leftrightarrow \fB_2 \fB_1^2 \fB_2 = (\fB_3^{-1})^2
\eeq
By taking the trace, we get the following three equations,
\be
\tr \fB_1^2\,\fB_2^2 = \tr \fB_3^2,\qquad \tr \fB_2^2\,\fB_3^2 = \tr \fB_1^2,\qquad \tr \fB_3^2\,H^{-1}\,\fB_1^2\,H = \tr\fB_2^2.
\ee
By writing
\be
\fB = \cosh b/2\ \one + \sinh b/2\ \hfb\cdot \vec{\sigma},
\ee
where $\hfb$ is the normalized direction of the boost, we get explicitly
\be \label{TraceGaussLaw}
\begin{aligned}
&\cosh b_1\,\cosh b_2 + \sinh b_1\,\sinh b_2\ \hfb_1\cdot R(h_1^{-1}) \hfb_2 = \cosh b_3,\\
&\cosh b_2\,\cosh b_3 + \sinh b_2\,\sinh b_3\ \hfb_2\cdot R(h_2^{-1}) \hfb_3 = \cosh b_1,\\
&\cosh b_3\,\cosh b_1 + \sinh b_3\,\sinh b_1\ \hfb_3\cdot R(h_3^{-1}) \hfb_1 = \cosh b_2.
\end{aligned}
\ee
Here $R(h)$ denotes the rotation $h$ in the vector representation, i.e. a 3D rotation. There is a nice way to rewrite the scalar products $\hfb_i\cdot R(h_i^{-1}) \hfb_j$ in terms of the vector variables $\bT, \bTop$. First note that $\hfb$ is the normalized vector $\bT$, and $R(h) \hfb$ is the normalized $\bTop$,
\begin{align}
&\sinh b\ \hfb = \frac12\,\tr B^2 \vec{\sigma} = \frac12\,\tr \ell \ell^\dagger \vec{\sigma} = \bT,\\
&\sinh b\ R(h)\hfb = \frac12\,\tr B^2\,h^{-1} \vec{\sigma} h = \frac12\,\tr \ell^\dagger \ell \vec{\sigma} = \bTop,
\end{align}
i.e. the rotation $h$ maps $\bT$ to $\bTop$. This suggests to define $\hfb^{op} = \bTop/|\bTop|$, such that
$
\hfb^{op} = R(h) \hfb,
$
and
\be
\cos \phi_{ij} \equiv - \hfb^{op}_i\cdot \hfb_j,\qquad (ij) = (12), (23), (31),
\ee
so that $\phi_{ij}$ is the angle between $i$ and $j$ in the hyperbolic triangle defined by the lengths $\kappa b_1, \kappa b_2, \kappa b_3$. Indeed, the equations \eqref{TraceGaussLaw} take the form of the hyperbolic cosine law
\be
\cosh b_i\ \cosh b_j - \sinh b_i\ \sinh b_j\ \cos \phi_{ij} = \cosh b_k,
\ee
where $i,j,k$ are such that $\epsilon_{ijk}=1$. The ribbon vertex corresponds to a triangle whose boundary is composed of oriented dashed edges carrying $\ell_1, \ell_2, \ell_3$. The angles $\phi_{ij}$ are associated to the corners of this triangle. On the dashed line $i$, we associate $\hfb_i$ to its target vertex and $\hfb_i^{op}$ to its source vertex, and the rotation $h_i^{-1}$ maps the former to the latter. This is summarized in the Figure \ref{fig:RibbonVertexZoom}. $\hfb^{op}_i$ is interpreted as the normal vector to the edge $i$ at the meeting point of $i$ and $j$, and $\hfb$ as the normal vector at the meeting point of $i$ and $k$ (still with the convention $\epsilon_{ijk}=1$).
% {\bf [V] Not clear to me in which 3D hyperplane??}.

\medskip

If one edge $e$ incident to the ribbon vertex is outgoing, we get $\tell^{-1}_e$ in the Gauss law. By writing $\tell^{-1} = \tilde{B} \tilde{h}^{-1}$, we define
\begin{align}
&-\sinh \tilde{b}\ \hat{\tilde{\mathfrak{b}}} = \frac12\,\tr \tilde{B}^{-2}\ \tilde{h}^{-1}\, \vec{\sigma}\,\tilde{h} = \frac12\,\tr \tell \tell^\dagger \vec{\sigma} = \tilde{\bT},\\
&-\sinh \tilde{b}\ R(\tilde{h}^{-1})\hat{\tilde{\mathfrak{b}}}^{op} = \frac12\,\tr \tilde{B}^{-2} \vec{\sigma} = \frac12\,\tr \tell^\dagger \tell \vec{\sigma} = \tilde{\bT}^{op},
\end{align}
and then find that the angles $\phi_{ij}$ are evaluated through the scalar products with $\hat{\tilde{\mathfrak{b}}}_e$ instead of $\hfb_e$ at the target vertex of the dashed line $\tell_e$ and with $\hat{\tilde{\mathfrak{b}}}^{op}_e$ instead of $\hfb^{op}_e$ at the source vertex of the dashed line. Moreover, the relation between both is now
\be
\hat{\tilde{\mathfrak{b}}}^{op} = R(\tilde{h}^{-1})\, \hat{\tilde{\mathfrak{b}}}.
\ee

\subsection{Dihedral angles between hyperbolic triangles and the flatness constraint}
\begin{figure}
\includegraphics[scale=.45]{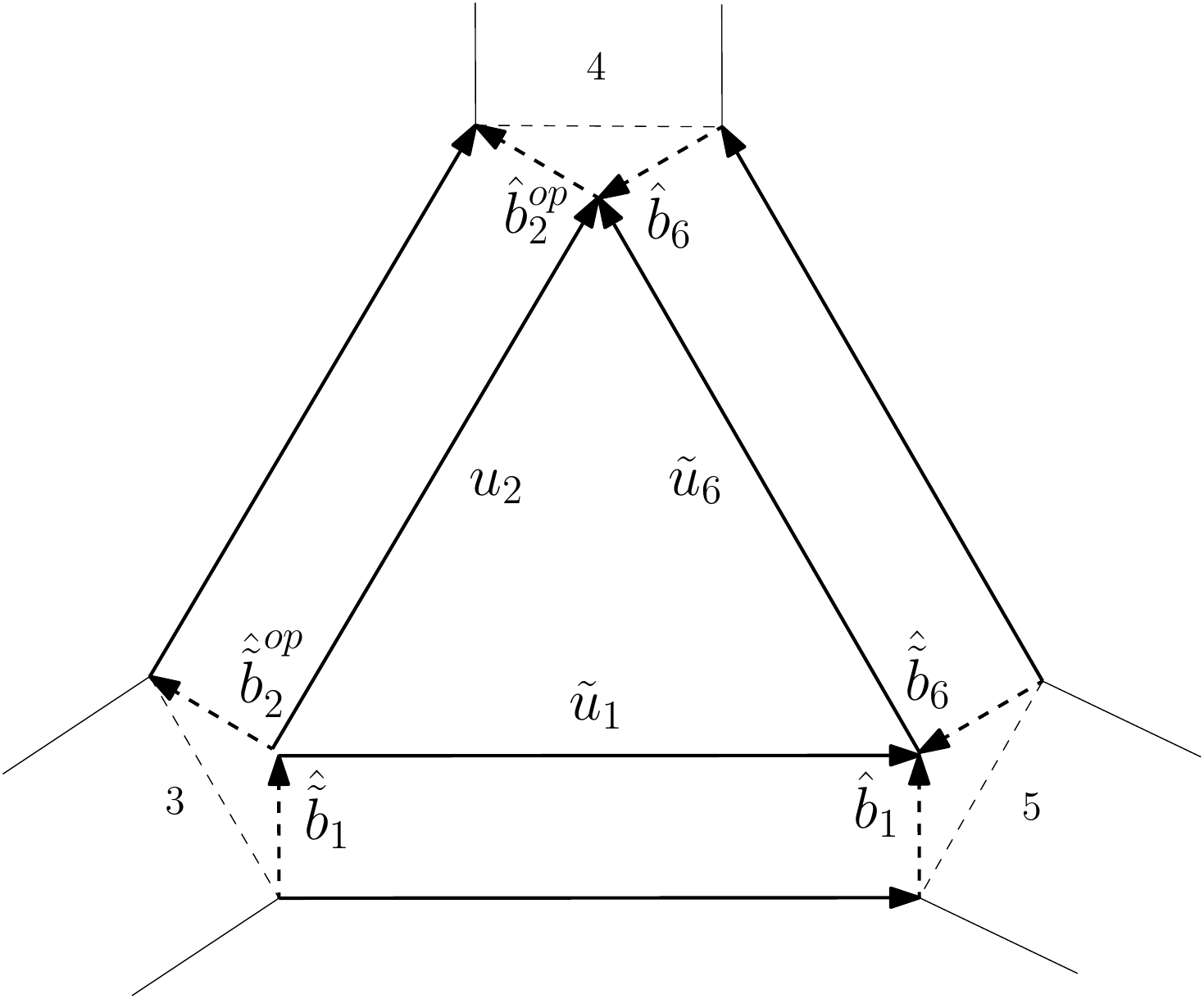}
\caption{This is a face with three edges, and the flatness constraint $\cC = u_1 \tu^{-1}_1 \tu^{-1}_6$. It is dual to a vertex shared by three triangles, which meet along the edges dual to $e_1, e_2, e_6$. We have indicated the vectors $\hfb_e, \hfb^{op}_e, \hat{\tilde{\mathfrak{b}}}_e, \hat{\tilde{\mathfrak{b}}}_e^{op}$ which are relevant to the 2d angles.}\label{fig:3ValentFace}
\end{figure}

%%%%%%%%%%%%%%%%%%%%%%
In the flat case, the geometric interpretation of the flatness constraint was described in details in \cite{valentin1}. We focus here on the new curved case, and show that similar results can be obtained. That is, we want to show that the flatness constraint enables to evaluate the extrinsic curvature (measured by dihedral angles between triangles) as the one of a homogeneous 3D hyperbolic geometry made by gluing hyperbolic triangles. For simplicity, we restrict our attention to the case of three triangles glued together. In the dual picture, this corresponds to three ribbon vertices, connected by three edges $e_1, e_2, e_6$ which represent the edges of the triangles meeting at a common point. This point is represented as the face formed by those three ribbon edges. The notations and orientations are fixed by the Figure \ref{fig:3ValentFace}.

The 2D angles around the face are
\be
\cos \phi_{26} = -\hfb_2^{op} \cdot \hfb_6,\qquad \cos \phi_{16} = - \hfb_1\cdot \hat{\tilde{\mathfrak{b}}}_6,\qquad \cos\phi_{12} = - \hat{\tilde{\mathfrak{b}}}_1\cdot \hat{\tilde{\mathfrak{b}}}_2^{op}.
\ee
If the gluing of triangles can be embedded in 3D hyperbolic space, the dihedral angle between the two triangles which meet along the edge dual to $e_1$ is
\be \label{Costheta}
\cos\theta_1 = \frac{\cos\phi_{26} - \cos\phi_{12}\ \cos\phi_{16}}{\sin \phi_{12}\ \sin\phi_{16}}.
\ee
This would fix the extrinsic curvature as a function of the intrinsic geometry, determined by the 2D angles or equivalently the lengths.

On the other hand, we have an independent notion of dihedral angles on our phase space. Introduce the normals to the two triangles hinged on the edge dual to $e_1$, at the point where the three triangles meet,
\be
N_{12} = \frac{\hat{\tilde{\mathfrak{b}}}_1 \times \hat{\tilde{\mathfrak{b}}}_2^{op}}{|\hat{\tilde{\mathfrak{b}}}_1 \times \hat{\tilde{\mathfrak{b}}}_2^{op}|} = \frac{\hat{\tilde{\mathfrak{b}}}_1 \times \hat{\tilde{\mathfrak{b}}}_2^{op}}{\sin\phi_{12}}
,\qquad N_{16} = \frac{\hat{\tilde{\mathfrak{b}}}_6\times \hfb_1}{|\hat{\tilde{\mathfrak{b}}}_6\times \hfb_1|} = \frac{\hat{\tilde{\mathfrak{b}}}_6\times \hfb_1}{\sin\phi_{16}}.
\ee
The dihedral angle is the scalar product between these two normals, but since they are defined in the frame of two different triangles (one uses $\hat{\tilde{\mathfrak{b}}}_1$ and the other uses $\hfb_1$), it is necessary to transport $N_{16}$ along $e_1$ using the rotation $R(\tu_1^{-1})$. This leads to the definition
\be
\cos \Theta_1 = - N_{12}\cdot R(\tu_1^{-1}) N_{16}.
\ee
Using the fact that $R(\tu_1^{-1}) \hfb_1 = - \hat{\tilde{\mathfrak{b}}}_1$, a direct calculation shows that
\be \label{CosTheta}
\begin{aligned}
\cos \Theta_1 &= -\frac{\left(\hat{\tilde{\mathfrak{b}}}_1 \times \hat{\tilde{\mathfrak{b}}}_2^{op}\right)\cdot \left(R(\tu_1^{-1})\hat{\tilde{\mathfrak{b}}}_6\times R(\tu_1^{-1})\hfb_1\right)}{\sin\phi_{12}\ \sin\phi_{16}},\\
&= -\frac{\bigl(\hat{\tilde{\mathfrak{b}}}_1\cdot R(\tu_1^{-1})\hat{\tilde{\mathfrak{b}}}_6\bigr) \bigl(\hat{\tilde{\mathfrak{b}}}_2^{op} \cdot R(\tu_1^{-1})\hfb_1\bigr) - \bigl(\hat{\tilde{\mathfrak{b}}}_2^{op}\cdot R(\tu_1^{-1})\hat{\tilde{\mathfrak{b}}}_6\bigr) \bigl(\hat{\tilde{\mathfrak{b}}}_1\cdot R(\tu_1^{-1})\hfb_1\bigr)}{\sin\phi_{12}\ \sin\phi_{16}},\\
&= -\frac{\cos\phi_{16}\ \cos\phi_{12} + \hat{\tilde{\mathfrak{b}}}_2^{op}\cdot R(\tu_1^{-1})\hat{\tilde{\mathfrak{b}}}_6}{\sin\phi_{12}\ \sin\phi_{16}}
\end{aligned}
\ee

Now we are in position to relate the two above notions of dihedral angles to the flatness constraint. The latter reads $\cC = u_2 \tu_1^{-1} \tu_6^{-1}$. We consider the matrix element of $\cC$ in the vector representation contracted with $\hfb^{op}_2$ and $\hfb_6$,
\be
H_1 = -\hfb_2^{op} \cdot R(\cC) \hfb_6 + \hfb_2^{op} \cdot \hfb_6.
\ee
$H_1=0$ on-shell and it is one of the three independent components of $\cC-\one$. Using $R(\tu_6^{-1}) \hfb_6 = - \hat{\tilde{\mathfrak{b}}}_6$ and $R(u_2^{-1}) \hfb_2^{op} = -\hat{\tilde{\mathfrak{b}}}_2^{op}$, we get
\be
\begin{aligned}
H_1 &= -\hat{\tilde{\mathfrak{b}}}_2^{op}\cdot R(\tu_1^{-1}) \hat{\tilde{\mathfrak{b}}}_6 - \cos\phi_{26},\\
&= \sin\phi_{12}\,\sin\phi_{16}\ \cos\Theta_1 + \cos\phi_{12}\,\cos\phi_{16} - \cos\phi_{26},\\
&= \sin\phi_{12}\,\sin\phi_{16} \bigl(\cos \Theta_1 - \cos \theta_1\bigr).
\end{aligned}
\ee
To get to the second line, we have used \eqref{CosTheta}, and to get the last line we have recognized $\cos\theta_1$ as defined in \eqref{Costheta}.

Therefore we see that the constraint $\cC=\one$ forces the holonomy around the face, $u_2 \tu_1^{-1} \tu_6^{-1}$, to know about the hyperbolic dihedral angles, and identifies the extrinsic curvature as measured by $\cos\Theta_1$ with the geometry of 3D hyperbolic space determined by the intrinsic geometry of the triangles (measured by $\cos\theta_1$ as a function of the 2D angles).

%%%%%%%%%%%
\section{Solutions of the constraints}
%%%%%%%%%%%
We show in this section that the theory is topological, % without needing to specify if we deal with  $\ISO(3)$ or $\SL(2,\C)$. Since it is well known that in  flat case $\ISO(3)$ the theory is topological, we only focus on the new curved case $\SL(2,\C)$.  It is topological
in the sense that the reduced phase space, i.e. the set of solutions of the constraints, only depends on the topology and not on the choice of cell decomposition. We proceed in a way completely analogous to the way the moduli space of flat connections is identified as the set of solutions in BF theory (2D Yang-Mills at weak coupling). In fact since both cases $\ISO(3)$ and $\SL(2,\C)$ are expressed in the same formalism, the proof for the flat case extends in a direct manner to the curved case, once we have re-expressed the variables in terms of the Heisenberg double variables. Let us consider the  $\SL(2,\C)$ case since it is new.

Just as in the flat case, first we have to   gauge fix the  ($\SU(2)$) rotational  and ($\SB(2,\C)$) deformed translational symmetries (up to global residual symmetries), then solve the constraints on as many faces and vertices as possible, until we arrive at equations which only depend on the topology.

We consider the graph dual to a triangulation of a surface of genus $g$, with $V, E, F$ denoting the sets of vertices, edges and faces. The set $E$ of edges can be partitioned into 3 sub-sets: a spanning tree $\mathcal{T}$ in the graph (with $|V|-1$ edges), a set of edges $\mathcal{T}^*$ dual to a spanning tree in the triangulation (with $|F|-1 = |E|-|V| +1 -2g$ edges), and a set of $2g$ ``crossing'' edges $CE$.

The gauge fixing of the $\SU(2)$ symmetry consists of setting $u_e=\id$ for $e\in\mathcal{T}$. This implies
\be \label{GaugeFixingU}
\forall\,e\in \mathcal{T},\qquad u_e=\tilde{u}_e = \id,\qquad \ell_e = \tell_e\,.
\ee

The gauge fixing of the $\SB(2,\C)$  symmetry is similar in the dual: we take $\ell_e = \id$ for $e\in\mathcal{T}^*$ so that
\be \label{GaugeFixingL}
\forall\,e\in \mathcal{T}^*,\qquad \ell_e=\tell_e = \id,\qquad u_e = \tu_e\,.
\ee

These two gauge fixings are consistent with one another because the condition \eqref{GaugeFixingU} is left invariant under a $\SB(2,C)$ transformation (the transformed $u^{(m)}$ is defined by $u\, m = \, ^{(u)}m\, u^{(m)}$ which means $^{(u)}m=m$ and $u^{(m)}=\one$ when $m=\one$), and the condition \eqref{GaugeFixingL} is left invariant by $\SU(2)$ transformations. They can be seen graphically as a retract of the ribbon lines carrying the $\SU(2)$ elements for the edges in $\mathcal{T}$ (the solid lines in the Figure \ref{fig:RibbonEdge} are reduced to two points) and a thinning of the ribbon edges in $\mathcal{T}^*$ so that they become regular edges (and the dashed lines in the Figure \ref{fig:RibbonEdge} disappear).

We choose arbitrary roots for $\mathcal{T}$ and the spanning tree dual to $\mathcal{T}^*$ in the triangulation. This induces a natural orientation of their edges from the leaves towards the root. There is then a canonical procedure to solve the constraints.

Let us start with the flatness constraints. The vertices of the spanning tree in the triangulation are dual to faces of the ribbon graph and we denote $F_{\mathcal{T}^*}$ this set of faces. We can solve the set of constraints $\{\mathcal{C}_f\}_{f\in F_{\mathcal{T}^*}}$ except for the face dual to the root. Indeed, each edge in $\mathcal{T}^*$ is shared by exactly two faces, one being the source and the other the target with respect to the orientation induced by the root. Moreover, each such edge carries a single $\SU(2)$ element since $u_e=\tu_e$, which appears in the flatness constraints of the source and target faces only. Moreover, $u_e$ only appears once in those constraints. We can therefore use the constraint at the source face to express $u_e = f(\{u_\alpha\})$ as a product of other $\SU(2)$ elements. This expression is then injected in the constraint at the target face. We say that this way $u_e$ has been eliminated. Starting from the faces dual to the leaves, down to the root face, we see that all $\SU(2)$ elements carried by the edges in $\mathcal{T}^*$ are eliminated, since they are expressed as products of the only $\SU(2)$ elements we are left with, i.e. those on the edges in $CE$. Graphically, this elimination process simply consists in removing the edges of $\mathcal{T}^*$ which merges all faces together until only one remains, with the ribbon edges of $CE$ on the boundary. It is called a \emph{deletion} process.

We proceed similarly for the Gauss constraints. Each edge $e\in \mathcal{T}$ carries a single $\SB(2,\C)$ element since $\ell_e = \tell_e$, which appears in exactly two Gauss constraints, those at the source and target vertices $e$ is incident to. We solve the Gauss constraints along $\mathcal{T}$, from its leaves down to its root vertex. At each step we use the constraint at the source vertex to express $\ell_e$ as a product of other elements which is then injected in the constraint at the target vertex. All $\{\ell_e\}_{e\in\mathcal{T}}$ are thus expressed as functions on $\{\ell_e\}_{e\in CE}$. One is left with the constraint at the root vertex on the remaining variables. Graphically, this process contracts the ribbon edges of $\mathcal{T}$ so as to merge the ribbon vertices they are incident to. This is called a \emph{contraction} process.

\begin{figure}
\includegraphics[scale=.7]{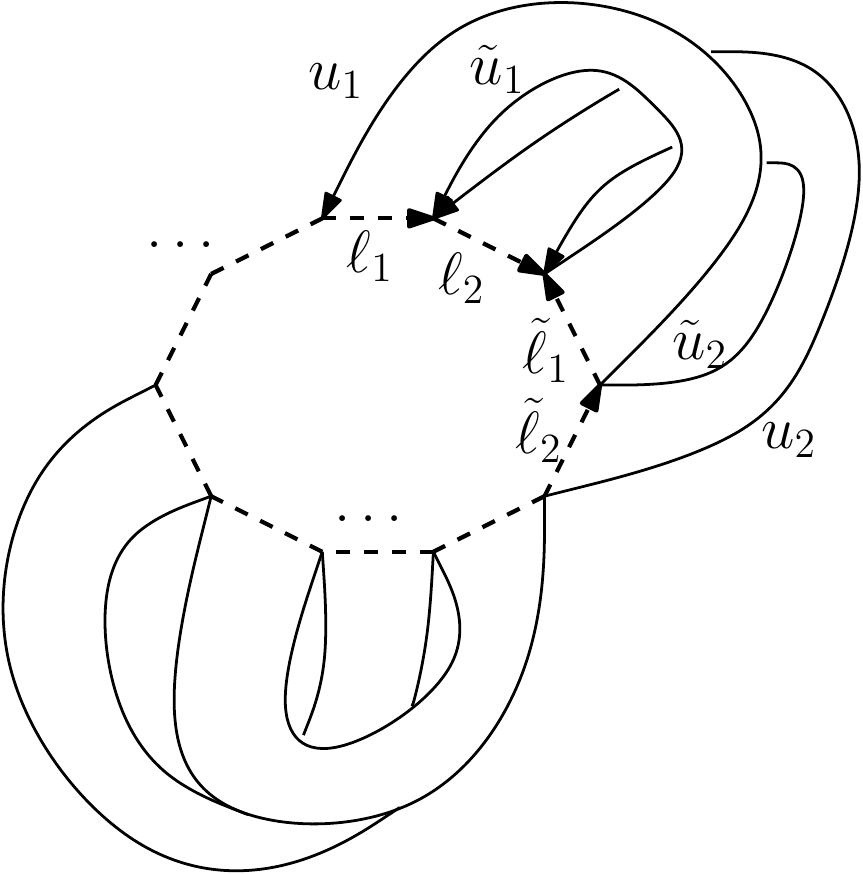}
\caption{The rosette graph, with a single ribbon vertex, $2g$ ribbon edges and a single face.}\label{fig:Rosette}
\end{figure}

The sequence of contractions and deletions we described above is well known and turns the ribbon graph we started with in a ``rosette'' which has a single ribbon vertex, $2g$ ribbon edges and a single face, as shown in the Figure \ref{fig:Rosette}. From this rosette it is trivial to read the remaining constraints,
\begin{align}
\mathcal{C} &= u_1 \tu_2^{-1} \tu_1^{-1} u_2 \dotsm u_{2g-1} \tu_{2g}^{-1} \tu_{2g-1}^{-1} u_{2g} = \id,\\
\mathcal{G} &= \tell_{2g}^{-1} \tell_{2g-1}^{-1} \ell_{2g} \ell_{2g-1} \dotsm \tell_2^{-1} \tell_1^{-1} \ell_2 \ell_1= \id.
\end{align}

We end up with them no matter which triangulation we started with and they clearly depend on the genus only, meaning that we have successfully defined a topological theory.

\section*{Conclusion \& Outlook}
We have constructed  topological models describing flat and hyperbolic discrete geometries. In the flat case, we have recalled the standard discretization of $BF$ theory and we have reformulated it in the less standard formalism of Heisenberg double. This reformulation has allowed us to construct  in a direct manner  a new phase space, namely $\SL(2,\C)$, as a deformation of the Heisenberg double of the flat case.
%In this new phase space, the momentum space is curved.
The new $\SL(2,\C)$ phase space is constructed from the classical $r$-matrix of $\sl(2, \C)$  and is interpreted as a deformation of the flat $\ISO(3)$ case, with deformed notions of rotations and translations. This phase space can be viewed as a $\SU(2)$ configuration space and a curved momentum space $\SB(2,\C)$ (identified as the hyperboloid of time-like normalized vectors in Minkowski space-time).

Considering a ribbon decomposition of the 2D surface, provided with one copy of $\SL(2,\C)$ on each edge, we introduced closure constraints on vertices and flatness constraints on faces. On the one hand, these constraints generate rotations and $\SB(2,\C)$ translations; on the other hand, they have a natural geometrical meaning: on the trivalent graph dual to a 2D triangulation, the closure constraint at each vertex (dual to a triangle) imposes the hyperbolic cosine law on the triangle, and the flatness constraint on each face (dual to a point) ensures that these triangles are consistently glued in the sense that the triangulation can be embedded in a homogeneous hyperbolic geometry. From this perspective, we have showed how this new model describes discrete hyperbolic geometries. We have further proved that this model is topological, in a way similar to discretized BF theory.

This new phase space and model open up many different routes to explore.

\medskip

\textbf{Link with continuum limit:} In the flat case, we know that we have considered  a discretization of the $BF$ theory with $\Lambda=0$. We would like to show that similarly the new model corresponds to a discretization of $BF$ theory with $\Lambda<0$ (of the form $BF+\Lambda B^3$). This would ensure that we are indeed describing 3D Riemannian gravity with negative cosmological constant. It would require understanding how to derive the $\SL(2,\C)$ phase space from the continuous triad and connection fields of BF theory.
%A possible route to explore this is to use the recent results of \cite{marc}.
This is currently under investigation.

\medskip

\textbf{Quantization:} %Link with LQG defined in terms of $\UQ$/Turaev-Viro based on $\UQ$:
Once again in the flat case, we know that the quantization of this model gives rise to LQG which can be precisely related to the Ponzano-Regge model. In the $\SL(2,\C)$ case, we would expect to obtain LQG  and the Turaev-Viro model defined in terms of $\UQ$ (with $q$ real though). Some elements of the quantization process already exist. For example in \cite{yakushin}, the quantum analogue of the components of $\ell$ are given, with $q=e^{ \hbar\ka}$
\bes
\ell \dr \hat \ell = \left(\begin{array}{cc} K\mone&0\\ -(q^{\f12}-q^{-\f12})J_+&K \end{array}\right) \quad l \dr \hat l = \left(\begin{array}{cc} K&(q^{\f12}-q^{-\f12})J_-\\0&K\mone \end{array}\right).
\ees
Hence it seems natural to identify $\lambda\dr K^{-1}=q^{-J_z/2}$, $z\dr -(q^{\f12}-q^{-\f12})J_+$, $\ov{z}\dr -(q^{\f12}-q^{-\f12})J_-$, where we recognize the usual  $\UQ$ generators.
The quantum version $R$ of the classical $r$-matrix becomes
\beq
r= \frac{i \ka}{4}
\left(\begin{array}{cccc}
1&&&\\ &-1&&\\&4&-1&\\&&&1
\end{array}\right)
\quad\leadsto\quad
R=q^{-\f14}\left(\begin{array}{cccc} q^{\f12}\\ &1 \\ &q^{\f12}-q^{-\f12}&1\\&&&q^{\f12}\end{array}\right)\sim \one -i \hbar r.
\eeq
The quantum version of the Poisson bracket is then given by
\bes
R^\dagger \hat \ell_1 \hat \ell_2=  \hat \ell_2 \hat \ell_1R^\dagger, \quad R^\dagger \hat l_1 \hat l_2=  \hat l_2 \hat l_1R^\dagger,  \quad R^\dagger \hat l_1 \hat \ell_2=  \hat \ell_2 \hat l_1R^\dagger,
\ees
which give then  the  $\UQ$ commutation relations
\be \label{commutationSU2q}
K J_{\pm}K\mone=q^{\pm\demi} J_\pm, \quad [J_+, J_-]= \f{K^{2}-K^{-2}}{q^{1/2}-q^{-1/2}}.
\ee
%Furthermore, if we look at the classical Gauss constraint, as in Figure \ref{fig:3ValentVertex},   and consider the components, say $\cG_{11}$,  and $G_{12}$ we have
%\beq
%\cG_{11}= \lambda_1\Lambda_2\Lambda_3 =1, \quad G_{12}= z\lambda_2\lambda_3 + \lambda_1\mone z_2 \lambda_3+ \lambda_1\mone \lambda_2\mone z_3 = 1.
%\eeq
%We recognize the classical version of the $\UQ$ invariance which would read
%\beq
%K\ot K\ot K= \one, \quad J_+\ot K\ot K+ K\mone \ot J_+ \ot K+ K\mone \ot K\mone\ot J_+.
%\eeq
%All these elements seem to indicate that the quantum theory behind our model will be LQG defined in terms of $\UQ$ as a gauge group.
The precise quantization procedure is currently under investigation.

\medskip

\textbf{Braiding:} In the quantum group case such as for $\UQ$, the tensor product becomes "non-commutative" which leads to the notion of braiding. This braiding is characterized by the quantum $R$-matrix. In the construction we have proposed here, there are different braidings involved. We have seen that both the translations and the rotations have a different type of braiding in Proposition \ref{translation ISO} and \ref{translation SL}. We have also seen that the vector $\hfb$ was  constructed from the boosts, using  a different braiding (cf \eqref{boost braided}).  It would be interesting to see how these braidings relate together and to the $\UQ$ braiding used to construct the observables in the LQG context \cite{ours1, ours2}. A priori the braiding of the symmetries is likely to be related to the braiding of the Drinfeld double $U_q(\SL(2,\C))$.  The quantum analogue of the braiding  for the $\hfb$  is less clear.
\medskip

\textbf{Link with Chern-Simons:} It is well known that a $BF$ theory with a cosmological constant can be rewritten as a Chern-Simons theory \cite{witten, catherine, bernd}. The discretization of the phase space approach to Chern-Simons theory is well-understood and is given by the Fock-Rosly formalism \cite{fock}. In the flat case $\Lambda=0$, the link between the discretized $BF$ theory and the Fock-Rosly formalism was given in \cite{karim-cat}.  It would be interesting to see how their approach can be generalized to our construction.

\medskip

\textbf{Curved twisted geometries:} Spin networks defined with gauge group $\SU(2)$ can be interpreted as the quantization of a special type of discrete geometries, the \textit{twisted geometries} \cite{twisted}. The generalization of this framework to the curved case would be extremely interesting. Another useful tool, related to the twisted geometries, is their parametrization in terms of spinor variables \cite{un0, un1, twisted1}. The quantum group  version  of this approach has been described in details in \cite{ours2}. It would be interesting to explore what are the classical analogue of these spinor operators and how they could be  used to parametrize the $\SL(2,\C)$ phase space. Essentially one would like to define the variables that transforms as spinors or vectors under the action of the non trivial Poisson Lie group $\SU(2)$. This is currently under investigation.

\medskip

\textbf{Other signatures and signs for $\Lambda$: } For simplicity we have focused here on the Euclidian case with $\Lambda\leq 0$. The  other combinations are also of interest. For example, the Euclidian case with  $\Lambda>0$ would be interesting to explore to get discrete spherical geometries. It would lead supposedly to the quantum group $\UQ$ with $q$ root of unity. In this case, it is well known that this quantum group is not a Hopf algebra but a quasi-Hopf algebra. Hence, at the classical level, we would expect to deal with quasi-bialgebras \cite{quasi}. This is currently under investigation.

%\textbbf{q-deformed lattice gauge theory:} In this article, we have used standard techniques of lattice gauge theory, generalized to the case of a deformed phase space. Once the quantization procedure will be understood, our construction will provide a general setting to deal with $q$-deformed lattice gauge theories. In particular the Yang-Mills case could be discussed in this framework. A preliminary attempt had been made in \cite{rotator}, but the Gauss constraint at each vertex were missing. We expect that our approach will allow to complete this work. It would be interesting to see if our theory, hopefully describing $BF$ theory with $\Lambda\neq0$, could be seen as a limiting case of a $q$-deformed lattice Yang-Mills case, just as when $\Lambda=0$.

\medskip

\textbf{Insights for the 4D case?}
The fact that the cosmological constant is implemented using a quantum group in a 3D space-time has been used as a motivation to do the same thing for a 4D space-time \cite{yetter, roche, winston, muxin0}. The proof of this conjecture is a long standing problem. Thanks to our model we have now some new tools to explore if this proposal is actually justified. Indeed, we can expect that unlocking the link between  our model with the continuum model in 3D will shed some new light on how the kinematical space is affected by $\Lambda\neq 0$.

%%%%%%%%%%%%%%%%%%%%%%%%%%%%%%
\section*{Acknowledgements}
We would like to thank L. Freidel for sharing his idea regarding the use of the ribbon formalism in LQG.
We are also grateful to B. Schroers for an enlightening discussion on our approach and its relation to the Fock-Rosly formalism.
M. Dupuis also thanks C. Meusburger and K. Noui for interesting discussions.
M. Dupuis and F. Girelli acknowledge financial support from the Government of Canada through
respectively a Banting fellowship and a NSERC Discovery grant.

%%%%%%%%%%%%%%%%%%%%%%%%%%%%%%
%%%%%%%%%%%%%%%%%%%%%%%%%%%%%%
%%%%%%%%%%%%%%%%%%%%%%%%%%%%%%
\appendix

\section{The $\ISO(3)$ case} \label{ISO(3)}
 \paragraph*{Generators and $r$-matrix:}
 The $\so(3)$ generators (in the spin 1 representation) can be written as $3\times 3$ anti-symmetric matrices:
 \be
 j_1=\left(\tabl{ccc}{0&0&0\\ 
 0&0&-1\\
 0&1&0}\right), \quad j_2=\left(\tabl{ccc}{0&0&1\\ 
 0&0&0\\
 -1&0&0}\right), \quad j_3= \left(\tabl{ccc}{0&-1&0\\ 
 1&0&0\\
 0&0&0}\right).
 \ee
 They satisfy the following commutation relations: $[ j_i, j_j]=\epsilon_{ij}^{\; \; k} j_k$. The generators of the Abelian Lie algebra $\R^3$ are 3-vectors:
 \be
 \bee^1 =\left(\tabl{c}{1\\ 
 0\\
 0}\right), \quad \bee^2= \left(\tabl{c}{0\\ 
 1\\
 0}\right), \quad \bee^3=\left(\tabl{ccc}{0\\ 
 0\\
 1}\right).
 \ee

We can write the generators of the classical double of $\so(3)$ using $4\times4$ matrices:
 \be
 J_i=\left(\tabl{cc}{\; j_i \; &{\bf 0} \\
 0\, 0\, 0 & 0}\right), \quad E^i=\left(\tabl{cc}{\; 0_{3\times 3} \; & {\bee}^i \\
 0\, 0\, 0 & 0}\right).
 \ee
 Then the $r$-matrix is given by
 \be \label{r iso3}
r=\sum_i E^i \otimes J_i=\left( \tabl{cccc}{0&0&0&J_1\\
0&0&0&J_2\\
0&0&0&J_3\\
0&0&0&0}\right) , \quad r^t=\sum_i J_i\otimes E^i =\left( \tabl{cccc}{0&E^3&-E^2&0\\
-E^3&0&E^1&0\\
E^2&-E^1&0&0\\
0&0&0&0}\right).
 \ee
They are $16\times16$ matrices written by blocks of $4\times4$ matrices.   If $\so(3)$ with generators $J_i$ is equipped with the trivial cocycle $\delta=0$ then the Abelian Lie algebra  $\R^3$ is its dual Lie algebra. Its associate cocycle is $\delta(E^k)=\epsilon_{ij}^k E^i\otimes E^j$. To completely precise the structures of the classical double $\fd_0(\so(3))$, we need to introduce the $\fd_0$-invariant bilinear form which characterizes the fact that $E^i$ is the dual of $J_i$: $\la E^i, J_j\ra= \delta^i_j,  \; \la J_i, J_j \ra= \la E^i, E^j\ra=0$.

 \medskip

 \paragraph*{The Heisenberg double and Poisson bracket structure:}

 Let $D \in \ISO(3)$, its left decomposition is given by
 \be
 D= ( {\bf R},  \bX) = \ell u, \quad \ell= \left ( \tabl{cc}{ \id_{3\times 3} & \bX  \\
 0\, 0\, 0 & 1}\right) , \; u=\left( \tabl{cc}{ {\bf R} & \bf{0} \\
 0\, 0\, 0 & 1}\right),
 \ee
where $ \bX$ is a 3-vector and $\textbf{R}$ is a $3\times 3$ rotation matrix. Introducing the notation $D_1=D\otimes \id$ and $D_2=\id \otimes D$, its commutation relations, for $\ISO(3)$ seen as a Heisenberg double, can be written as
 \be
 \{ D_1, D_2\}= -r D_1 D_2 + D_1 D_2 r^t.
 \ee
 which imply for the  rotation and translation variables, $u$ and $\ell$ respectively,
 \be
 \{ \ell_1, \ell_2\}=-[r, \ell_1 \ell_2], \quad \{\ell_1, \, u_2\}=-\ell_1 r u_2, \quad \{u_1, \, \ell_2\}=\ell_2 r^\dagger u_1, \quad \{u_1, u_2\}=-[r^\dagger, u_1 u_2].
 \ee
 This gives as back  the usual Poisson brackets of the flux and holonomy variables of LQG,
 \be
\{ X_i, X_j\}= \epsilon_{ij}^k X_k, \qquad \{ X_i, \textbf{R}\}=-j_i \textbf{R}, \qquad \{ \textbf{R}, \textbf{R}\}=0.
 \ee
 For the right Iwasawa decomposition of $\ISO(3)$, $D=\tilde{u} \tilde{\ell}$, we get,
 \be
 \{\tilde{\ell}_1, \tilde{\ell}_2\}=-[r, \tilde{\ell}_1\tilde{\ell}_2], \quad \{\tilde{\ell}_1, \tilde{u}_2\}=\tilde{u}_2 r \tilde{\ell}_1, \quad \{ \tilde{u}_1, \tilde{\ell}_2\}=-\tilde{u}_1 r^t\tilde{\ell}_2, \quad \{\tilde{u}_1, \tilde{u}_2\}=- [r^t, \tilde{u}_1\tilde{u}_2],
 \ee
 or equivalently,
 \be
 \{\tilde{X}_i, \tilde{X}_j\}=\epsilon_{ij}^{\;\; k}\tilde{X}_k, \qquad \{\tilde{X}_i, \tilde{\bR}\}= \tilde{\bR} j_i, \qquad \{\tilde{\bR}, \, \tilde{\bR}\}=0.
 \ee

  \section{The $ \SL(2,\C)$ case}\label{poi formulae}

 \paragraph*{Generators and $r$-matrix:}
The $\su(2)$ generators (in the spin 1/2 representation) can be represented by the Pauli matrices $\vec{\sigma}$ satisfying $[ \sigma_i, \sigma_j]= 2i \epsilon_{ij}^k \sigma_k$. Explicitly,
\be
\sigma_x= \left( \tabl{cc}{ 0&1 \\
1 & 0}\right), \quad \sigma_y= \left( \tabl{cc}{ 0&-i \\
i & 0}\right), \quad \sigma_x= \left( \tabl{cc}{ 1&0 \\
0& -1}\right).
\ee
%Note that $\sigma_i$ used in this paper corresponds to half the usual Pauli matrix found in the literature.

The generators of $\sb(2, \C)$ which generate the triangular matrix $\ell$ are modified boosts defined by
\be
\kappa^{-1}\tau^i= i \sigma_i - \f i2 [ \sigma_z, \sigma_i]= i \sigma_i + \epsilon_{zi}^k \sigma_k
\ee
$$
%\kappa \mone \,\tau_i%=i\sigma_i -\f i2[\sigma_z,\sigma_i]
%=i\sigma_i+\epsilon_{zik} \sigma_k, \quad
\kappa \mone \, \tau^x
=i\sigma_x +\sigma_y= \mat{cc}{0&0\\2i&0}, \quad \kappa \mone \,\tau^y=i\sigma_y -\sigma_x=i\tau^x= \mat{cc}{0&0\\-2&0}, \quad \kappa \mone \,\tau^z=i\sigma^z =\mat{cc}{i&0\\0&-i}.
$$
They satisfy the commutation relations,  $[\tau^i,\tau^j]=2i\ka (\delta _{jz}\tau^i - \delta _{iz}\tau^j)$.

\medskip

The $r$-matrix then explicitly reads as,
\be
r=\f14\sum_i  \tau^i\ot \sigma_i \,=\,
\frac{i \ka}{4}
\left(\begin{array}{cccc}
1&&&\\ &-1&&\\&4&-1&\\&&&1
\end{array}\right) \quad \textrm{ and } \quad r^t=\f14\sum_i \sigma_i\ot \tau^i \,=\,
\frac{i \ka}{4}
\left(\begin{array}{cccc}
1&&&\\ &-1&4&\\&&-1&\\&&&1
\end{array}\right).
\ee
If $\su(2)$ with generators $\sigma_i$ is equipped with the cocycle  $ \delta(\sigma_k)=2i\, \ka \left( \delta^i_k \delta^j_z-\delta^j_k \delta^i_z\right) \sigma_i \otimes \sigma_j$, then the Lie algebra $\sb(2, \C)$ is its dual algebra. Its associated cocycle is $\delta_*(\tau^k)=2i\epsilon_{ij}^k \tau^i\otimes \tau^j$. To precise completely the structures of the classical double $\fd(\su(2))$, we need to introduce the $\fd$-invariant bilinear form which characterizes the fact that $\tau^i$ is the dual of $\sigma_i$: $\la \tau^i, \sigma_j\ra= \delta^i_j,  \; \la \sigma_i, \sigma_j \ra= \la \tau^i, \tau^j\ra=0$.
%In this case, the $\fd$-bilinear form can be taken as $i \tr..$.

\medskip

 \paragraph*{The Heisenberg double and Poisson bracket structure:}
 The phase space $\SL(2, \C)$ can be seen as a Heisenberg double. Then its symplectic structure is given by \eqref{SL(2,C) poi} for its left Iwasawa decomposition where $D=\ell u \, \in\, \SL(2, \C)$.
 We  expand the $\SL(2,\C)$ Poisson brackets \eqref{lu poi}, \eqref{lu poi 2} explicitly in terms of the  components of the momentum variables $\ell$, $l=(\ell^\dagger)^{-1}$ and configuration variables $u$.
\begin{itemize}
\item
The brackets between components of the triangular matrix $\ell$ and $\ell^\dagger$:
\be
\poi{\lambda,z}= \frac{i\ka}{2}\lambda z,
\qquad
\poi{\lambda,\ov{z}}= - \frac{i\ka}{2}\lambda\ov{z},
\qquad
\poi{\ov{z},z}= -{i\ka}\left( \lambda^2- \lambda^{-2} \right)\,,
\ee
\item The brackets between components of the unitary matrix $u$:
\beq
\begin{array}{llll}
\poi{\alpha, \beta} = -\f{i\ka}{2} \alpha \beta, & \poi{ \alpha, \bar{\beta}}=-\f{i\ka}{2} \alpha \bar{\beta}, & \poi{\alpha, \bar{\alpha}}=i \ka |\beta|^2, \\
\\
\poi{\bar{\alpha}, \beta}=\f{i \ka}{2} \bar{\alpha} \beta, & \poi{\bar{\alpha}, \bar{\beta}}=\f{i \ka}{2} \bar{\alpha} \bar{\beta}, & \poi{\beta, \bar{\beta}}=0\,,
\end{array}
\eeq
\item The brackets between $\ell$ or $l$ and $u$:
\beq
\begin{array}{llll}
\poi{\lambda, \alpha}= -\f{i\ka}{4} \lambda \alpha, & \poi{\lambda, \beta} = \f{i \ka}{4} \lambda \beta, & \poi{\lambda, \bar{\alpha}}= \f{i\ka}{4} \lambda \bar{\alpha}, & \poi{\lambda, \bar{\beta}} =- \f{i \ka}{4} \lambda \bar{\beta},  \\
\\
\poi{\lambda^{-1}, \alpha}= \f{i\ka}{4} \lambda^{-1} \alpha, & \poi{\lambda^{-1}, \beta} = -\f{i \ka}{4} \lambda^{-1} \beta, & \poi{\lambda^{-1}, \bar{\alpha}}=- \f{i\ka}{4} \lambda^{-1} \bar{\alpha}, & \poi{\lambda^{-1}, \bar{\beta}} =\f{i \ka}{4} \lambda^{-1} \bar{\beta},  \\
\\
\poi{z, \alpha}= -\f{i\ka}{4} (z \alpha+ 4 \lambda^{-1} \beta), & \poi{z, \beta} = \f{i \ka}{4} z \beta, & \poi{z, \bar{\alpha}}= \f{i\ka}{4} z \bar{\alpha}, & \poi{z, \bar{\beta}} =- \f{i \ka}{4}( z \bar{\beta}-4\lambda^{-1}\bar{\alpha}),  \\
\\
\poi{\bar{z}, \bar{\alpha}}= \f{i\ka}{4}(\bar{ z} \bar{\alpha}+4 \lambda^{-1} \bar{\beta}), & \poi{\bar{z}, \bar{\beta}} =-\f{i \ka}{4} \bar{z} \bar{\beta}, & \poi{\bar{z}, \alpha}=- \f{i\ka}{4} \bar{z} \alpha, &   \poi{\bar{z}, \beta} =- \f{i \ka}{4} (-\bar{z} \beta +4\lambda^{-1} \alpha), \\
\end{array}
\eeq
\end{itemize}
Let us now use the relation between the right and left Iwasawa decompositions to determine the Poisson brackets between the $\ell$, $u$ variables of the left Iwasawa decomposition and the $\tell$, $\tu$ variables of the right Iwasawa decomposition.
Starting from $\ell u = \tilde{u}\tell$, we get rid of $\tilde{u}$ by considering $\tell^\dagger \tell = u\mone \ell ^\dagger \ell u$, which gives
\be
%\begin{pmatrix} \tlambda &\overline{\tz} \\ 0 &\tlambda^{-1} \end{pmatrix} \begin{pmatrix} \tlambda & 0\\ \tz & \tlambda^{-1} \end{pmatrix} =
\begin{pmatrix} \tlambda^2+|\tz|^2 &\tlambda^{-1}\overline{\tz} \\ \tlambda^{-1} \tz & \tlambda^{-2} \end{pmatrix} = \begin{pmatrix} \overline{\alpha} &\overline{\beta} \\ -\beta &\alpha \end{pmatrix} \begin{pmatrix} \lambda^2+|z|^2 &\lambda^{-1}\overline{z} \\ \lambda^{-1} z & \lambda^{-2} \end{pmatrix} \begin{pmatrix} \alpha &-\overline{\beta} \\ \beta &\overline{\alpha} \end{pmatrix}.
\ee
We deduce that
\be
\begin{aligned}
\tlambda^{-2} &= \lambda^{-2} +|\beta|^2 \bigl(\lambda^2-\lambda^{-2}+|z|^2\bigr) -\lambda^{-1} \bigl(\overline{\alpha}\beta\,\overline{z} + \alpha \overline{\beta}\,z\bigr)\\
\tz &= \tlambda\ \Bigl( -\alpha\beta\bigl(\lambda^2-\lambda^{-2}+|z|^2\bigr)+ \lambda^{-1}\bigl(\alpha^2z-\beta^2 \overline{z}\bigr)\Bigr).
\end{aligned}
\ee
This allows to evaluate the Poisson brackets between $\tell$ and $\ell$ and $u$,
\be
\left\{ \tell^{-1}_1,u_2\right\} = u_2\,r\,\tell^{-1}_1, \qquad \left\{ \tell_1, \ell_2\right\} = 0.
\ee
From the first equality of the above equation, we can in particular infer that
\be
\{\tilde{l}_1, u_2\}=-\tilde{l}_1u_2 r^\dagger.
\ee
Finally, using $\tu = \ell u \tell^{-1}$, we get
\be
\left\{\tu_1, u_2 \right\} = 0,\qquad \left\{\tu_1, \ell_2 \right\} = -r^\dagger\,\tu_1 \ell_2.
\ee
In terms of the components, we have
\begin{itemize}
\item The brackets between $\tell$ and $u$:
\begin{alignat}{4} \label{tell u}
&\{\tlambda, \alpha\} = -i\frac{\kappa}{4}\,\alpha \tlambda,& \qquad &\{\tlambda, \overline{\alpha}\} = i\frac{\kappa}{4}\,\overline{\alpha}\tlambda,& \qquad &\{\tlambda, \beta\} = -i\frac{\kappa}{4}\,\beta \tlambda,& \qquad &\{\tlambda, \overline{\beta}\} = -i\frac{\kappa}{4}\,\overline{\beta} \tlambda, \\
&\{\tz, \alpha\} = -i\frac{\kappa}{4}\,\alpha \tz,& \qquad &\{\tz, \overline{\alpha}\} = i\frac{\kappa}{4}\,\overline{\alpha}\tz - i\kappa \tlambda^{-1} \beta,& \qquad &\{\tz, \beta\} = -i\frac{\kappa}{4}\,\beta \tz,& \qquad &\{\tz, \overline{\beta}\} = i\frac{\kappa}{4}\,\overline{\beta} \tz+i\kappa \tlambda^{-1} \alpha.
\end{alignat}
\item The brackets between $\tell^\dagger$ and $u$:
\beq
\begin{array}{llll}
\poi{\overline \tz, \alpha}=-\f{i\ka}4(-4\ov{\beta}\tlambda\mone+\overline \tz\alpha),&\poi{\overline \tz, \ov \beta}=\f{i\ka}4 \overline \tz\beta, & \poi{\overline \tz,  \beta}=-\f{i\ka}4 (4\ov \alpha\tlambda\mone+\beta \overline \tz ),& \poi{\overline \tz, \ov \alpha} =\f{i\ka}4 \overline \tz\ov \alpha.
\end{array}
\eeq
\end{itemize}

%%%%%%%%%%%%%%%%%%%%%%%%%%%%%%%
\section{Infinitesimal symmetries}\label{app:symmetries}
\subsection*{Translations}
%%%%%%%%%%%%%%%%%%%%%%%%%%%%%%%%
We have the translations realized as
\beq\label{translation bis}
D\dr m\,D= m\, \ell u\,=  \,^{(m)} \ell\;  \,^{(m)} u \dr \left\{
\begin{array}{l}
\,^{(m)}  u= u \\
\,^{(m)} \ell = m \ell
\end{array}
\right.\,, \qquad D\dr D\, m = \ell u\, m = \ell ^{(m)} \; u^{(m)} \dr \left\{
\begin{array}{l}
\,^{(u)}m\, u^{(m)} = m\,u \\
\ell ^{(m)} = \ell \left(\,^{(u)}m\right)
\end{array}\right.
\eeq
\textbf{$\ISO(3)$ case,   translation:}  In the case of the right translation for $\ISO(3)$, we have $\,^{(u)}m=u\,m\, u\mone $ and $u^{(m)}=u$. Setting $m=\one+\delta m=\one +\vec\veps\cdot \vec E$, and plugging it into \eqref{translation bis}, we obtain
\beq \label{infinitesimal translation} \textrm{Infinitesimal left translations: }\left\{ \begin{array}{l}
 \delta^{(m)}_L  u= 0 \\
 \delta^{(m)}_L \ell = \delta m \,\ell
\end{array}
\right.\,, \quad \textrm{Infinitesimal right translations: } \left\{
\begin{array}{l}
 \delta^{(m)}_R u = 0 \\
 \delta^{(m)}_R \ell = \ell\, (u\, \delta m\, u\mone )
\end{array}\right.
\eeq
Let us calculate now the different Poisson brackets we proposed in Proposition \ref{translation ISO}.
 \bes
 && \delta^{(m)}_R u \,=\, \la \delta m_1;  u_1^{-1}\{u_1,u_2\}\, \ra_1  \,=\, \la \delta m_1;  0\, \ra_1\,=\, 0\,,  \\
&&\delta^{(m)}_R \ell \,=\, \la \delta m_1;  u_1^{-1}\{u_1,\ell_2\} \ra_1  \,=\, \la \delta m_1;  u_1^{-1}\ell_2 r^t u_1\ra_1 \,= \, \sum _i \la \vec \veps\cdot \vec E\ot \one ;  u^{-1}J_i u \ot \ell E_i \ra_1 \,= \, \ell\, (u\, \delta m\, u\mone) \\
&& \delta^{(m)}_L u  \, = \, \la \delta m_1;  \{\tilde u_1,u_2\}\,  \tilde u_1^{-1}\ra_1\, = \, \la \delta m_1;  0 \ra_1\,=\, 0\\
&& \delta^{(m)}_L \ell \,=\,  \la \delta m_1;  \{\tilde u_1,\ell_2\}\,  \tilde u_1^{-1}\ra_1\, = \, \la \delta m_1;  \ell_2 r^t\tu_1,  \tilde u_1^{-1}\ra_1\, = \,  \sum _i \la \vec \veps\cdot \vec E\ot \one ;  J_i  \ot \ell E_i   \ra_1 \, = \, \delta m \, \ell
 \eeq

\medskip

\textbf{$\SL(2,\C)$ case,  deformed translation:}  We write again  $m=\one + \delta m \in \SB(2,\C)$, and furthermore introduce $M= m m^\dagger$, as well as  $\delta M=\delta m + \delta m^\dagger$. We can actually express $\delta m$ in terms of $\delta M$.
\beq \label{cool1}
\delta m = \mat{cc}{\veps_z & 0 \\ \veps_+ & -\veps_z},
\quad
\delta M = \mat{cc}{2\veps_z & \veps_- \\ \eps_+ & -2\veps_z},
\quad
\f12[\delta M,\sigma_z]= \mat{cc}{0 & -\veps_- \\ \veps_+ & 0}\,,
\quad  \delta m = \f12\left(
\delta M +\f12[\delta M,\sigma_z]
\right)\,.
\eeq
We focus first on the right infinitesimal deformed translations. We  deduce the twisted  infinitesimal translation $\,^{(u)} \delta m$, coming from $\,^{(u)}m$ in \eqref{translation bis}, using  $\,^{(u)}M=u\, M \, u \mone$,
\be
\,^{(u)} \delta m = \f12\left(
u\,\delta M\, u^{-1} +\f12[u\,\delta M \, u^{-1},\sigma_z]\right).
\ee
This allows to recover each of the transformations.
%\be
%u^{(m)}= \left(\,^{(u)}m\right)^{-1}\, u\, m
%\,=\,
%(\one-\,^{(u)} \delta m)\, u \, (\one+\delta m)
%\,\sim\,
%u(\one +\delta m - u^{-1}\, (\delta^{(u)} m)\, u )
%\,\sim\,
%u(\one +\f14[\delta M, \sigma_z-u^{-1}\sigma_z u])\,,
%\ee
%which gives therefore that
\bes
\delta^{(m)}_R u &=& u^{(m)}-u \sim
(\one-\,^{(u)} \delta m)\, u \, (\one+\delta m) - u
\,=\,
\f14\,u\,[\delta M, \sigma_z-u^{-1}\sigma_z u]\,,  \label{rbab}\\
%We similarly compute the infinitesimal version of  $\ell^{(m)}=\ell \left(\,^{(u)}m\right)$.
 \label{rbob}
\delta^{(m)}_R \ell &=& \ell (\,^{(u)}\delta m)
\,=\,
\ell \f12\left(
u\, \delta M\, u^{-1} +\f12[u\, \delta M\, u^{-1},\sigma_z]\right)\,.
\ees
We would like now to relate the expression in terms of the Poisson brackets given in \eqref{translation SL} to these infinitesimal transformations. For this we use once again the explicit expression Poisson brackets. %We recall that in the $\SL(2,\C)$ case, the duality map is given by the trace, $\la A,B\ra=(2i)\mone \tr (AB)$.
\beq
\tr_1 \delta M_1  u_1^{-1}\{u_1,u_2\}
&=&
\tr_1 \delta M_1  u_1^{-1}(r^\dagger u_1u_2-u_1u_2r^\dagger) =
%\tr_1 \delta M_1  u_1^{-1}r^\dagger u_1u_2
%-\tr_1 \delta M_1   u_2r^\dagger
%\,=\,
 \sum_i \left(
\tr (u\delta Mu^{-1}\tau_i^\dagger ) \, (\sigma_i u)
-\tr (\delta M\tau_i^\dagger)\,( u\sigma_i)
\right)\,.
\eeq
Then coming back to the expression of $\tau_i$ in terms of the Pauli matrices,
$
\tau_i^\dagger = -i\ka(\sigma_i +\f{1}2[\sigma_z,\sigma_i])\,,
$
and using the identity  $
\sum_i \tr(A\sigma_i)\,\sigma_i
\,=\,
  2A-\tr A\,\one\,,
$ holding for arbitrary  2$\times$2 matrices $A$,
we have:
\be \label{cool2}
\sum_i \tr(A\tau_i^\dagger)\,\sigma_i
\,=\,
%-2iA-i[A,\sigma_z]+i\tr A
%\,=\,
%-2iA-i[\sigma_z,A]+i\tr A
i\ka(-2A-[A,\sigma_z]+\tr A)\,.
\ee
Putting the pieces together, we finally get:
\beq
\ka \mone\tr_1 \delta M_1  u_1^{-1}\{u_1,u_2\}
&=&
-\f i 2\left(
(u\delta M u^{-1}+\f12[u\delta M u^{-1},\sigma_z])\,u
-u \,(\delta M +\f12[\delta M, \sigma_z ])
\right)\nn
=\,i\f u4[\delta M,\sigma_z-u^{-1}\sigma_z u]\,,\\ \nn
\ka \mone \tr_1 \delta M_1  u_1^{-1}\{u_1,\ell_2\}
&=&
-\ka \mone\tr_1 \delta M_1  u_1^{-1}\ell_2 r^\dagger u_1
\,=\,
-\,\f1{4\ka} \ell \sum_i \tr(u\delta M u^{-1}\tau_i^\dagger)\,\sigma_i \nn\\
&=&
\,i \f\ell2
\left(
u\delta M u^{-1}+\f12[u\delta M u^{-1}, \sigma_z]
\right)
\,,
\ees
which match respectively  \eqref{rbab} and  \eqref{rbob}, modulo the factor $-i$. Let us consider now the infinitesimal left translations, which are much easier to determine, since we are using the left decomposition. By inspection of \eqref{translation bis}, we have directly that
\beq
\delta_L^{(m)}u= 0, \quad \delta_L^{(m)}\ell = \delta m \,\ell.
\eeq
We now evaluate directly the Poisson brackets from Proposition \ref{translation SL} (TO CHECK again)
  \bes
 \delta^{(m)}_L u  &=&  -\f{i}{4\ka}  \tr_1\delta M_1  \{\tilde u_1,u_2\}\,  \tilde u_1^{-1}, \,=\,  - \f{i}{4\ka}\tr_1\delta M_1  0\,=\, 0,
\\
\delta^{(m)}_L \ell &=& - \f{i}{4\ka} \tr_1 \delta M_1  \{\tilde u_1,\ell_2\}\,  \tilde u_1^{-1}
%\,=\, \la \delta M_1; r^\dagger \tu_1\ell_2\,  \tilde u_1^{-1}\ra_1
\,=\,\f{i}{4\ka} \sum_i \tr (\delta M  \tau_i ^\dagger)\sigma_i \ell \, = \, \delta m \, \ell,
 \ees
where we used \eqref{cool1} and \eqref{cool2} in the last equation.

%%%%%%%%%%%
\subsection*{Rotations}
%%%%%%%%%%%%%
The action of the rotations is given by
\beq \label{rotation bis}
D\dr D\, v=\ell u\, v= \ell ^{(v)} \; u^{(v)} \dr \left\{
\begin{array}{l}
u^{(v)} = u\,v \\
\ell ^{(v)} = \ell
\end{array}
\right.\,, \qquad
D\dr v\, D= v \, \ell u= \,^{(v)} \ell\; \,^{(v)} u \dr \left\{
\begin{array}{l}
\,^{(v)} u = v^{(\ell)} \,u \\
\,^{(v)} \ell \, v^{(\ell)}  =v  \ell
\end{array}
\right.
\eeq

\textbf{$\ISO(3)$ case,   rotation:} A quick inspection of \eqref{rotation bis} with $v=\one +i \vec \veps\cdot J= \one +V$ in mind  leads to
\beq \label{rotation bis ISO}
\textrm{Infinitesimal left rotations: }\left\{
\begin{array}{l}
\delta_L^{(v)} u = V \,u \\
\delta_L^{(v)} \ell   = V\ell-\ell V
\end{array}
\right.\,,
 \qquad
\textrm{Infinitesimal right rotations: }
 \left\{
\begin{array}{l}
\delta_R^{(v)}u = u\,V \\
\delta_R^{(v)}\ell =0
\end{array}
\right.\,.\eeq
Let us calculate now the different Poisson brackets we proposed in Proposition \ref{rotation ISO}.
\bes
\delta^{v}_L\ell &=&-\la V_1;\ell_1\mone \poi{\ell_1, \ell_2}\ra_1= \la V_1;\ell_1\mone (r\ell_1\ell_2-\ell_1\ell_2 r)\ra_1= \sum_i \la \vec \veps \cdot \vec J;(\ell\mone E_i\ell \ot J_i\ell- E_i \ot \ell J_i )\ra_1= V\ell - \ell V,
\\
\delta^{v}_L u &=& -\la V_1;\ell_1\mone \poi{\ell_1, u_2}\ra_1 \,=\,   \la V_1;\ell_1\mone \ell_1r u_2\ra_1 \,=\,  \sum_i \la  \vec \veps \cdot \vec J; E_i \ot J_i u\ra_1 \,=\, V\, u.
\\
\delta^{v}_R\ell &=& \la V_1;\tell_1\mone \poi{\tell_1, \ell_2}\ra_1 \, =\,  \la V_1;0\ra_1 \, =\,0,
\\
\delta^{v}_R u &=&  \la V_1;\tell_1\mone \poi{\tell_1, u_2}\ra_1\,=\, \la V_1;\tell_1\mone u_2r\tell_1\ra_1\,=\, \sum_i \la \vec \veps \cdot \vec J; E_i\ot u \,J_i\ra_1\,=\,u V,
\ees

\textbf{$\SL(2,\C)$ case,   rotation:}  We are going to plug in $v=\one + i{\vveps}\cdot\vsigma$
%$$
%v=\one + i{\vveps}\cdot\vsigma=\mat{cc}{1+i\veps_z & i\veps_- \\ i\veps_+ & 1-i\veps_z},
%\textrm{ with }
%\overline{\veps_-}=\veps_+\,,
%$$
into \eqref{rotation bis} to get the action of the left infinitesimal rotation. To study the transformation of  $\ell$  under  such infinitesimal  rotation, it is actually convenient to look at how  $L$ transforms since $L\dr vLv^{-1}$. %We can then easily re-construct $\ell^{(v)}$ from $L^{(v)}$.
Recalling the expressions for $\ell$ and $L$,
$$
\ell=\mat{cc}{\lambda & 0 \\ z & \lambda^{-1}},
\qquad
L=\mat{cc}{\lambda^2 & \lambda\,\bz \\ \lambda\,z & \lambda^{-2}+|z|^2}\,,
$$
we explicitly parameterize the $\SU(2)$ matrix $v$ and we get the parameters $\,^{(v)} \lambda $ and $\,^{(v)}z$ of $\,^{(v)}\ell$ as functions of $\ell$ and $v$:
$$
v=\mat{cc}{\alpha & -\bbeta \\ \beta & \balpha}
 \quad\Rightarrow\quad
\left|
\begin{array}{lcl}
\, ^{(v)}\lambda{}^2&=&|\alpha|^2\lambda^2+|\beta|^2(\lambda^{-2}+|z|^2)-\balpha\bbeta \lambda z -\alpha\beta \lambda \bz \\
\, ^{(v)}\lambda  \,^{(v)}z &=& \balpha^2\lambda z -\beta^2 \lambda \bz +\balpha \beta (\lambda^2-\lambda^{-2}-|z|^2)
\end{array}
\right.
$$
Replacing $v$ by its infinitesimal version, we get at leading order in $\veps$ the components of $\delta_L^{(v)}\ell$:
\be \label{yes good}
\delta_L^{(v)}\lambda=\lambda^{(v)} - \lambda
\sim \f{i}2(\veps_-z - \veps_+\bz)\,,
\qquad
\delta_L^{(v)} z=z^{(v)} -z
\sim -2i\veps_z z
-\f{i}{2\lambda}z(\veps_-z - \veps_+\bz)
+\f{i}{\lambda}\veps_+(\lambda^2-\lambda^{-2}-|z|^2)
\,.
\ee
The twisted $\SU(2)$ transformation $v^{(\ell)}$ is given by  $v^{(\ell)}\,=\,(^{(v)}\ell)^{-1}v\ell$ and  its infinitesimal version is
\be
 v^{(\ell)}=\one + i\vec{\veps'}\cdot\vsigma
\quad \textrm{ with }
\eps_z' =\eps_z +\f{z\veps_-+\bz\veps_+}{2\lambda},
\qquad
\eps_+'=\lambda^{-2}\veps_+\,.
\ee
That $ v^{(\ell)}$ is not simply obtained from $v$ by conjugation by $\ell$ comes from the fact that triangular matrices are not stable under conjugation by $\SU(2)$ matrices and vice-versa.  This allows to recover the components of $\delta_L^{(v)} u\sim i (\vec{\veps'}\cdot\vsigma)u$.
%We can do the same for the holonomy $u$, whose variation should match $\delta u \sim i (\vec{\veps'}\cdot\vsigma)u$ at leading order in $\veps$:
\be \label{yes good 1}
%\delta u \sim i (\vec{\veps'}\cdot\vsigma)u
%\quad\Longrightarrow\quad
\begin{array}{lclcl}
\delta_L^{(v)} \alpha &=& i(\veps'_z\alpha+\veps'_-\beta)
&=& i(\veps_z \alpha +\f12\lambda^{-1}(\veps_-z+\veps_+\bz)\alpha+\lambda^{-2}\veps_-\beta) \\
\delta_L^{(v)} \beta &=& i(-\veps'_z\beta+\veps'_+\alpha)
&=& i(-\veps_z \beta -\f12\lambda^{-1}(\veps_-z+\veps_+\bz)\beta+\lambda^{-2}\veps_+\alpha)
\end{array}
\ee
We can recover these transformations thanks to the Poisson brackets given in Proposition \ref{rotation SL} by brute force. We first recall that
\be
\ka^{-1}\tr VL
\,=\,
\ka^{-1}(2\eps_z\lambda^2+\eps_-\lambda z+\eps_+\lambda \bz)
\,=\,
(2\eps_z(T_0+T_z)+\eps_-T_++\eps_+T_-)\,.
\ee
We have then from Proposition \ref{rotation SL}
\be
\delta_L^{(v)}\ell = -\lambda^{-2}\,\{\ka^{-1}\tr VL,\ell\},
\qquad
\delta_L^{(v)} u = -\lambda^{-2}\,\{\ka^{-1}\tr VL,u\}\,,
\ee
and by considering each component, we prove they match \eqref{yes good} and \eqref{yes good 1}.
\bes
\delta_L^{(v)} z
&=&
-\ka^{-1}\lambda^{-2}\,\{2\veps_z\lambda^2+\veps_-\lambda z+\veps_+\lambda \bz\,,\,z\}
\,=\,
-2i\veps_z z
+i\veps_+\lambda^{-1}(\lambda^2-\lambda^{-2}-\f12|z|^2)
-i\veps_-\lambda^{-1}z^2 \\
\delta_L^{(v)} \lambda
&=&
-\lambda^{-2}\ka^{-1}\,\{2\veps_z \lambda^2+\veps_- \lambda z+\veps_+ \lambda \bz\,,\,\lambda\}
\,=\,
\f{1}{\ka\lambda^2}\,(\veps_-\{\lambda,\lambda z\}+\veps_+\{\lambda,\lambda \bz\})
\,=\,
\f{i}2(\veps_- z-\veps_+ \bz)\,,\\
\delta_L^{(v)}\alpha
&=&
-{\ka\mone \lambda^{-2}}\,\{2\veps_z\lambda^2+\veps_-\lambda z+\veps_+\lambda \bz\,,\,\alpha\}\,=\,
 i(\veps_z \alpha +\f12\lambda^{-1}(\veps_-z+\veps_+\bz)\alpha+\lambda^{-2}\veps_-\beta) \\
\delta_L^{(v)}\beta
&=&
-{\ka\mone \lambda^{-2}}\,\{2\veps_z\lambda^2+\veps_-\lambda z+\veps_+\lambda \bz\,,\,\beta\}\,=\, i(-\veps_z \beta -\f12\lambda^{-1}(\veps_-z+\veps_+\bz)\beta+\lambda^{-2}\veps_+\alpha)
\ees
The right rotations take a simple form as one can see in \eqref{rotation bis}. Their infinitesimal version is therefore given by
\beq
\delta_R^{(v)} \ell = 0, \quad \delta_R^{(v)} u = i u(\vec \veps \cdot \vec \sigma).
\eeq
We want to show how to recover these expressions from the Poisson brackets in \eqref{rotation SL}. The action on the $\ell$ variable is direct since $\poi{\tell,\ell}=0$.
\bes
\delta_R^{(v)} \ell &=& \f{\tlambda^{2}}\ka \poi{\tr V (\tL^{op})\mone , \ell} = 0.
\ees
We calculate the action on $u$ component by component, using \eqref{tell u}.
\bes
\delta_R^{(v)} u &=&  \f{\tlambda^{2}}\ka \poi{\tr V (\tL^{op})\mone, u} \leadsto \left\{ \begin{array}{c}\delta^{(v)}_R\alpha = i (\veps_z\alpha - \veps_+\overline{\beta}) \\
\delta^{(v)}_R\beta = i(\veps_z \beta+\veps_+ \overline{\alpha})
 \end{array} \right. \, \leadsto  \delta_R^{(v)} u = i u(\vec \veps \cdot \vec \sigma).
\ees


\begin{thebibliography}{99}
\bibitem{carlip}
S.~Carlip,
\textit{Quantum gravity in 2+1 dimensions}, Cambridge, UK: Univ. Pr. (1998)

\bibitem{louapre}
 L.~Freidel and D.~Louapre,
\textit{Ponzano-Regge model revisited II: Equivalence with Chern-Simons},
  gr-qc/0410141.

  \bibitem{karim}
   K.~Noui,
  \textit{Three Dimensional Loop Quantum Gravity: Particles and the Quantum Double},
  J.\ Math.\ Phys.\  {\bf 47} (2006) 102501
  [gr-qc/0612144].

 \bibitem{karim-cat}
  C.~Meusburger and K.~Noui,
 \textit{The Hilbert space of 3d gravity: quantum group symmetries and observables},
  Adv.\ Theor.\ Math.\ Phys.\  {\bf 14} (2010) 1651
  [arXiv:0809.2875 [gr-qc]].

 \bibitem{ale-karim}
  K.~Noui and A.~Perez,
  \textit{Three-dimensional loop quantum gravity: Physical scalar product and spinfoam models},
  Class.\ Quant.\ Grav.\  {\bf 22} (2005) 1739
  [gr-qc/0402110].

\bibitem{witten}
 E.~Witten,
  \textit{(2+1)-Dimensional Gravity as an Exactly Soluble System},
  Nucl.\ Phys.\ B {\bf 311} (1988) 46.

 \bibitem{viro}
   V.~G.~Turaev and O.~Y.~Viro,
  \textit{State sum invariants of 3 manifolds and quantum 6j symbols},
  Topology {\bf 31} (1992) 865.

\bibitem{CS-TV}
J.~Roberts, \textit{Skein theory and Turaev-Viro invariants}, Topology,  34,  4,  (1995) 771-787.

\bibitem{pranzetti}
  K.~Noui, A.~Perez and D.~Pranzetti,
  \textit{Canonical quantization of non-commutative holonomies in 2+1 loop quantum gravity},
  JHEP {\bf 1110} (2011) 036.
  [arXiv:1105.0439 [gr-qc]].

\bibitem{ours1}
M.~Dupuis and F.~Girelli,
 \textit{Quantum hyperbolic geometry in loop quantum gravity with cosmological constant},
  Phys.\ Rev.\ D {\bf 87} (2013) 121502
  [arXiv:1307.5461 [gr-qc]].

\bibitem{ours2}
M.~Dupuis and F.~Girelli,
 \textit{Observables in loop quantum gravity with a cosmological constant},
  [arXiv:1311.6841 [gr-qc]].

\bibitem{valentin1}
  V.~Bonzom and L.~Freidel,
  \textit{The Hamiltonian constraint in 3d Riemannian loop quantum gravity},
  Class.\ Quant.\ Grav.\  {\bf 28} (2011) 195006
  [arXiv:1101.3524 [gr-qc]].

\bibitem{valentin2}
  V.~Bonzom and E.~R.~Livine,
 \textit{ A New Hamiltonian for the Topological BF phase with spinor networks},
  arXiv:1110.3272 [gr-qc].

  \bibitem{frolov1}
S.A. Frolov,
{\it Gauge-invariant Hamiltonian formulation of lattice Yang-Mills theory and the Heisenberg double},
Mod.Phys.Lett. A10 (1995) 2619-2632 [arXiv:hep-th/9501143]


\bibitem{frolov2}
S.A. Frolov,
{\it Hamiltonian lattice gauge models and the Heisenberg double},
Mod.Phys.Lett. A10 (1995) 2885-2896 [arXiv:hep-th/9502121]

 \bibitem{yvette} Y. Kosmann-Schwarzbach,  \textit{Lie bialgebras, Poisson Lie groups and dressing transformations}.  (1997).  Lecture Notes in Physics  \textbf{495}, (1997),  104-170.


\bibitem{chari}
V.~Chari and A.~Pressley,
\textit{A guide to quantum groups},
  Cambridge, UK: Univ. Pr. (1994)

%%%%%%%%%%%% main text

\bibitem{lu} J. H. Lu, \textit{Multiplicative and Affine Poisson Structures on Lie Groups}, PhD thesis, University of California, Berkeley, 1990.

\bibitem{marmo}   G.~Marmo, A.~Simoni and A.~Stern,
  ``Poisson lie group symmetries for the isotropic rotator,''
  Int.\ J.\ Mod.\ Phys.\ A {\bf 10} (1995) 99
  [hep-th/9310145].





%%%%%%%%%%%%%%% conclusion

\bibitem{yakushin}
  A.~Stern and I.~Yakushin,
  ``Deformation quantization of the isotropic rotator,''
  Mod.\ Phys.\ Lett.\ A {\bf 10} (1995) 399
  [hep-th/9312125].

\bibitem{fock}
V.~V.~Fock and A.~A.~Rosly,
 \textit{Poisson structure on moduli of flat connections on Riemann surfaces and r matrix},
  Am.\ Math.\ Soc.\ Transl.\  {\bf 191} (1999) 67
  [math/9802054 [math-qa]].



\bibitem{catherine}
C.~Meusburger and B.~J.~Schroers,
  \textit{Quaternionic and Poisson-Lie structures in 3d gravity: The Cosmological constant as deformation parameter},
  J.\ Math.\ Phys.\  {\bf 49} (2008) 083510
  [arXiv:0708.1507 [gr-qc]].

\bibitem{bernd}
  B.~J.~Schroers,
  \textit{Quantum gravity and non-commutative spacetimes in three dimensions: a unified approach},
  Acta Phys.\ Polon.\ Supp.\  {\bf 4} (2011) 379
  [arXiv:1105.3945 [gr-qc]].


\bibitem{twisted} L. Freidel and S. Speziale, \textit{Twisted geometries: A geometric parametrisation of SU(2) phase space},  Phys. Rev. D \textbf{82}, 084040 (2010).

\bibitem{un0}
  F.~Girelli and E.~R.~Livine,
 \textit{Reconstructing quantum geometry from quantum information: Spin networks as harmonic oscillators},
  Class.\ Quant.\ Grav.\  {\bf 22} (2005) 3295
  [gr-qc/0501075].

\bibitem{un1}
L.~Freidel and E.~R.~Livine,
\textit{The Fine Structure of $\SU(2)$ Intertwiners from $\U(N)$ representations},
  J.\ Math.\ Phys.\  {\bf 51} (2010) 082502
  [arXiv:0911.3553 [gr-qc]].
%
%\bibitem{un2}
% L.~Freidel and E.~R.~Livine,
%\textit{$\U(N)$ coherent states for Loop Quantum Gravity},
%  J.\ Math.\ Phys.\  {\bf 52} (2011) 052502
%  [arXiv:1005.2090 [gr-qc]].

\bibitem{twisted1}
M.~Dupuis, S.~Speziale and J.~Tambornino,
 \textit{Spinors and Twistors in Loop Gravity and spinfoams},
  arXiv:1201.2120 [gr-qc].
 \bibitem{twisted2}
 L.~Freidel and S.~Speziale,
\textit{From twistors to twisted geometries},
  Phys.\ Rev.\ D {\bf 82} (2010) 084041.
%  [arXiv:1006.0199 [gr-qc]].





\bibitem{quasi}
M. Bangoura,  Y. Kosmann-Schwarzbach,
\textit{The Double of a Jacobian Quasi-Bialgebra}, Letters in Mathematical Physics \textbf{28}, 13-29, 1993.


%%%%%%%%% lattice gauge theory
% \bibitem{rotator}
%G. Bimonte, A. Stern and P. Vitale,
%{\it $\SU_q(2)$ lattice gauge theory},
%Phys. Rev. D \textbf{54} (1996) 1, 1054



%%%%%%%%% 4d quantum groups spinfoams
\bibitem{yetter}
D.~Yetter, \textit{Generalized Barrett-Crane Vertices and Invariants of Embedded Graphs}, 	 arXiv:math/9801131 [math.QA]

\bibitem{roche}
  K.~Noui and P.~Roche,
  \textit{Cosmological deformation of Lorentzian spinfoam models},
  Class.\ Quant.\ Grav.\  {\bf 20} (2003) 3175
  [gr-qc/0211109].

\bibitem{winston}
W.~J.~Fairbairn and C.~Meusburger,
  \textit{Quantum deformation of two four-dimensional spinfoam models},
  J.\ Math.\ Phys.\  {\bf 53} (2012) 022501
  [arXiv:1012.4784 [gr-qc]].

\bibitem{muxin0}
M.~Han, \textit{4-dimensional Spin-foam Model with Quantum Lorentz GroupÓ}, J. Math. Phys. \textbf{52} (2011) 072501.



\end{thebibliography}
\end{document}